%% file: specialEdition_NQHRRC_arxiv.tex
\documentclass[12pt]{iopart}

\usepackage{citesort}
\usepackage{epsfig}
\usepackage{graphicx}
\usepackage{dcolumn}
\usepackage{bm}
\expandafter\let\csname equation*\endcsname\relax  
\expandafter\let\csname endequation*\endcsname\relax
\usepackage{amsmath, amsthm, amssymb}

\newcommand{\be}[1]{\begin{equation}\label{#1}}
\newcommand{\ee}{\end{equation}}
\newcommand{\vlowk}{V_{{\rm low}\,k}}

\newcommand{\fm}{\, \text{fm}}

\newcommand{\mev}{\, \text{MeV}}

\newcommand{\bra}[1]{\langle#1|}
\newcommand{\ket}[1]{|#1\rangle}
\newcommand{\braket}[2]{\langle#1|#2\rangle}

\newcommand{\matrELred}[3]{\langle#1||#2||#3\rangle}

\newcommand{\commut}[1]{\left[#1\right]}
\newcommand{\elemA}[2]{\ensuremath{{}^{#1}}\textrm{#2}}

\newcommand{\leftsub}[2]{\vphantom{#2}_{#1}{#2}}
\newcommand{\cut}[1]{}

\newcommand{\op}[1]{\hat{#1}}
\newcommand{\elem}[2]{\ensuremath{{}^{#2}\text{#1}}}

\begin{document}

\title{Unified {\em ab initio} approaches to nuclear structure and reactions}

\author{Petr Navr\'atil$^1$, Sofia Quaglioni$^2$, Guillaume Hupin$^{3,4}$, Carolina Romero-Redondo$^2$, Angelo Calci$^1$}
\address{$^1$TRIUMF, 4004 Wesbrook Mall, Vancouver, British Columbia, V6T 2A3, Canada}
\address{$^2$Lawrence Livermore National Laboratory, P.O.\ Box 808, L-414, 
Livermore, California 94551, USA}
\address{$^3$Institut de Physique Nucl\'eaire, Universit\'ee Paris-Sud, IN2P3/CNRS, F-91406 Orsay Cedex, France}
\address{$^4$CEA, DAM, DIF, F-91297 Arpajon, France}
\ead{navratil@triumf.ca}
\vspace{10pt}

\input{SE_abstract}
%
%
%
%
%

\input{SE_1-intro-general}

\input{SE_2-intro-formalism}

\input{SE_hamiltonian}

\input{SE_SRG}

\input{SE_NCSM_eigenstates}

\input{SE_binary-RGM}

\input{SE_ternary-RGM}
\input{SE_NCSMC}

\input{SE_R-matrix}
\input{SE_convergence}

\input{SE_transitions}
\input{SE_3-intro-section3}
\input{SE_scatt-s-shell}
\input{SE_scatt-p-shell}

\input{SE_6helium}

\input{SE_radiative-capt}

\input{SE_fusion}

\input{SE_conclusions}

\section*{References}
\bibliography{biblio}
\end{document}

%% file: SE_abstract.tex
\begin{abstract}
The description of nuclei starting from the constituent nucleons and the realistic interactions among them has been a long-standing goal in nuclear physics. In addition to the complex nature of the nuclear forces, with two-, three- and possibly higher many-nucleon components, one faces the quantum-mechanical many-nucleon problem governed by an interplay between bound and continuum states. In recent years, significant progress has been made in {\em ab initio} nuclear structure and reaction calculations based on input from QCD-employing Hamiltonians constructed within chiral effective field theory. After a brief overview of the field, we focus on {\it ab initio} many-body approaches - built upon the No-Core Shell Model - that are capable of simultaneously describing both bound and scattering nuclear states, and present results for resonances in light nuclei, reactions important for astrophysics and fusion research. In particular, we review recent calculations of resonances in the $^{6}$He halo nucleus, of five- and six-nucleon scattering, and an investigation of the role of chiral three-nucleon interactions in the structure of $^9$Be. Further, we discuss applications to the $^7$Be$(p,\gamma)^8$B radiative capture. Finally, we highlight our efforts to describe transfer reactions including the $^3$H$(d,n)^4$He fusion.
\end{abstract}

%% file: SE_1-intro-general.tex
\section{Introduction}
\label{sec:intro}

Understanding the structure and the dynamics of nuclei as many-body systems of protons and neutrons interacting through the strong (as well as electromagnetic and weak) forces is one of the central goals of nuclear physics. One of the major reasons why this goal has yet to be accomplished lies in the complex nature of the strong nuclear force, emerging form the underlying theory of Quantum Chromodynamics (QCD). At the low energies relevant to the structure and dynamics of nuclei,  QCD is non-perturbative and very difficult to solve. The relevant degrees of freedom for nuclei are nucleons, i.e., protons and neutrons, that are not fundamental particles but rather complex objects made of quarks, antiquarks and gluons. Consequently, the strong interactions among nucleons is only an ``effective" interaction emerging non-perturbatively from QCD. Our knowledge of the nucleon-nucleon (NN) interactions is limited at present to models. The most advanced and most fundamental of these models rely on a low-energy effective field theory (EFT) of the QCD, chiral EFT~\cite{Weinberg1991}. This theory is built on the symmetries of QCD, most notably the approximate chiral symmetry. However, it is not renormalizable and has an infinite number of terms. Chiral EFT involves unknown parameters, low-energy constants (LECs) fitted to experimental data. It predicts higher-body forces, in particular a three-nucleon (3N) interaction that plays an important role in nuclear structure and dynamics. 

{\it Ab initio} calculations in nuclear physics start from the fundamental forces among nucleons, typically the chiral EFT interactions, and aim at predicting the properties of nuclei. This is a very challenging task because of the complex nature of nuclear forces and because of our limited knowledge of these forces. The high-level strategy is to solve the non-relativistic many-nucleon Schr\"{o}dinger equation with the inter-nucleon interactions as the only input. This can be done exactly for the lightest nuclei ($A{=}3,4$) \cite{FRIAR19934,Nogga199719,Barnea2001565,PhysRevC.64.044001}. However, using new methods and well-controlled approximations, {\it ab initio} calculations have recently progressed tremendously and become applicable to nuclei as heavy as nickel and beyond. 

This progress has been in particular quite dramatic concerning the description of bound-state properties of light and medium mass nuclei. For light nuclei, the Green's Function Monte Carlo Method (GFMC) \cite{Pudliner1995,PhysRevC.56.1720,PhysRevC.62.014001,doi:10.1146/annurev.nucl.51.101701.132506,Wiringa2002,PhysRevC.66.044310,Pieper2004} has been applied up to $A\leq 12$. The No-Core Shell Model (NCSM) \cite{Navratil2000a,Navratil2000,Navratil2009,Barrett2013} with its importance-truncated extension \cite{Roth2007,Roth2009} up to oxygen isotopes \cite{Hergert2013}. Other NCSM extensions, e.g., symmetry-adapted NCSM \cite{PhysRevLett.111.252501} and no-core Monte-Carlo shell model \cite{PhysRevC.86.054301} under active development. Very recently, methods such as the Coupled Cluster (CCM) \cite{PhysRevLett.94.212501,PhysRevLett.104.182501,Binder2013,PhysRevLett.109.032502,0034-4885-77-9-096302,PhysRevC.91.064320,Hagen2007169,Hagen:2015yea,PhysRevLett.111.122502,PhysRevC.90.064619}, the Self-Consistent Green's Function (SCGF) \cite{Cipollone2013} and its Gorkov generalization \cite{PhysRevC.89.061301}, the newly developed In-Medium Similarity Renormalization Group (IM-SRG) method \cite{PhysRevLett.106.222502,Hergert2013,Hergert2013a,PhysRevC.90.041302,PhysRevLett.113.142501} achieved high accuracy and predictive power for nuclei up to the calcium region with a full capability to use chiral NN+3N interactions. Further, there has been progress in Monte Carlo methods such as the Nuclear Lattice EFT \cite{PhysRevLett.109.252501,PhysRevLett.112.102501} as well as the Auxiliary-Field Monte Carlo (AFDMC) method and the GFMC that are now also able to use chiral EFT NN+3N interactions \cite{PhysRevC.90.054323,Lynn:2015jua}. 

As to the inclusion of continuum degrees of freedom, for $A{=}3,4$ systems there are several successful exact methods, e.g., the Faddeev~\cite{PhysRevC.63.024007}, Faddeev-Yakubovsky ~\cite{PhysRevC.70.044002,PhysRevC.79.054007}, Alt-Grassberger and Sandhas (AGS)~\cite{PhysRevC.75.014005,PhysRevLett.98.162502}, and Hyperspherical Harmonics (HH)~\cite{0954-3899-35-6-063101,Marcucci2009} methods.  For $A>4$ nuclei, concerning calculations of nuclear resonance properties, scattering and reactions, there has been less activity and the No-Core Shell Model with Resonating-Group Method (NCSM/RGM) \cite{Quaglioni2008,Quaglioni2009} and in particular the No-Core Shell Model with Continuum (NCSMC) method \cite{Baroni2013,Baroni2013a} highlighted in this paper are cutting edge approaches. Still the field is rapidly evolving also in this area. The GFMC was applied to calculate $n{-}^4$He scattering \cite{Nollett2007,Lynn:2015jua}, the Nuclear Lattice EFT calculations were applied to the $^4$He-$^4$He scattering \cite{Elhatisari:2015iga}, and the $p{-}^{40}$Ca scattering was calculated within the CCM with the Gamow basis \cite{Hagen2012a}. The CCM with the Gamow basis was also used to investigate resonances in $^{17}$F~\cite{PhysRevLett.104.182501} and in oxygen isotopes~\cite{Hagen2012}. Further, the {\it ab initio} Gamow NCSM with a capability to calculate resonance properties is under development \cite{PhysRevC.88.044318,Barrett:2015wza}.

Let us stress that a predictive {\it ab initio} theory of nuclear structure and nuclear reactions is needed for many reasons: 

(i) Nuclear structure plays an important role in many precision experiments testing fundamental symmetries and physics beyond the Standard Model. Examples include the determination of the $V_{ud}$ matrix element of the Cabbibo-Kobayashi-Maskawa matrix and its unitarity tests, the conserved vector current hypothesis tests, neutrino oscillations experiments, neutrino-less double beta decay experiments, searches for right-handed, scalar and other currents not present in the Standard Model. Realistic nuclear structure is of great importance here and {\it ab initio} nuclear theory of light and medium mass nuclei can provide a significant help.

(ii) A predictive nuclear theory would greatly help our understanding of nuclear reactions important for astrophysics. Typically, capture, transfer or other reactions take place in the Cosmos at energies much lower than those accessible by experiments. A well-known example is provided by the triple-alpha and $^{12}$C($\alpha,\gamma$)$^{16}$O radiative capture reactions. The ratio of the thermonuclear reaction yields for these two processes determines the carbon-to-oxygen ratio at the end of helium burning with important consequences for the production of all species made in subsequent burning stages in the stars. At stellar energies ($\approx300$ keV) radiative capture rates are too small to be measured in the laboratory. Thus, measurements are performed at higher energies (see, e.g., the recent experiment of Ref.~\cite{PhysRevLett.97.242503}) and extrapolations to the low energy of interest using theory are unavoidable. Theoretical extrapolation are, however, challenging due to the influence of several resonances. A fundamental theory would be of great use here. 

(iii) {\it Ab initio} theory of medium mass nuclei helps to shed light on the shell evolution of the neutron rich nuclei that impact our understanding of the r-process and the equation of state \cite{PhysRevLett.105.032501,doi:10.1146/annurev-nucl-102313-025446}. 

(iv) Low-energy fusion reactions represent the primary energy-generation mechanism in stars, and could potentially be used for future energy generation on Earth. Examples of these latter reactions include the $^3$H($d,n$)$^4$He fusion used at the international project ITER in France and at the National Ignition Facility in the USA. Even though there have been many experimental investigations of the cross section of this reaction, there are still open issues. A first-principles theory will provide the predictive power to reduce the uncertainty, e.g., in the reaction rate at very low temperatures; in the dependence on the polarization induced by the strong magnetic fields.

(v) Nuclear reactions are one of the best tools for studying exotic nuclei, which have become the focus of the new generation experiments with rare-isotope beams. These are nuclei for which most low-lying states are unbound, so that a rigorous analysis requires scattering boundary conditions. In addition, much of the information we have on the structure of these short-lived systems is inferred from reactions with other nuclei. A predictive {\it ab initio} theory will help to interpret and motivate experiments with exotic nuclei.

In addition to all of the above, an accurate many-body theory for light and medium mass nuclei provides a feedback about the quality of the inter-nucleon interactions, e.g., those derived from the QCD-based chiral EFT, used in the calculations and ultimately helps to improve our knowledge of the NN interactions, and in particular of the still-not-completely-understood 3N interactions. 

In this paper, we focus on an {\it ab initio} description of both bound and unbound nuclear states in a unified framework. In particular, we discuss in detail the NCSM/RGM~\cite{Quaglioni2008,Quaglioni2009} and the very recent NCSMC~\cite{Baroni2013,Baroni2013a}.

Our approach to the description of light nuclei is based on combining the {\it ab initio} NCSM \cite{Navratil2009,Barrett2013} and the Resonating-Group Method (RGM) \cite{Wildermuth1977,Tang1978,Fliessbach1982,Langanke1986,Lovas1998} into new many-body approaches (the first version called {\it ab initio} NCSM/RGM) \cite{Quaglioni2008,Quaglioni2009} capable of treating bound and scattering states in a unified formalism, starting from fundamental inter-nucleon interactions. The NCSM is an {\it ab initio} approach to the microscopic calculation of ground and low-lying excited states of light nuclei. The RGM is a microscopic cluster technique based on the use of $A$-nucleon Hamiltonians, with fully anti-symmetric many-body wave functions using the assumption that the nucleons are grouped into clusters. Although most of its applications are based on the use of binary-cluster wave functions, the RGM can be formulated for three (and, in principle, even more) clusters in relative motion \cite{Tang1978}. The use of the harmonic oscillator (HO) basis in the NCSM results in an incorrect description of the wave function asymptotics (for bound states due to technical limitations on the size of the HO expansion) and a lack of coupling to the continuum. By combining the NCSM with the RGM, we complement the ability of the RGM to deal with scattering and reactions with the use of realistic interactions, and a consistent {\it ab initio} description of nucleon clusters, achieved via the NCSM. 

The state-of-the-art version of this approach is the NCSMC \cite{Baroni2013,Baroni2013a}. It is based on an expansion of the $A$-nucleon wave function consisting of a square-integrable NCSM part and an NCSM/RGM cluster part with the proper asymptotic behaviour. In this way, the NCSM description of short- and medium-range many-nucleon correlations is combined with the NCSM/RGM description of clustering and long-range correlations. This approach treats bound and unbound states on the same footing and provides a superior convergence compared to both the NCSM and the NCSM/RGM. Using the NCSMC method we can predict not only the bound ground- and excited-state observables of light nuclei, but also resonances and cross sections of nuclear reactions as well as electromagnetic and weak transitions among bound and unbound states.

In Sec.~\ref{sec:formalism}, we present the formalism of the binary-cluster as well as the three-body cluster NCSM/RGM, of the binary-cluster NCSMC, and introduce the formalism for the calculation of electric dipole transitions in the NCSMC. In Sec.~\ref{sec:s-p-shell-res}, we discuss NCSMC results for $A{=}5,6$ nuclei and for $^9$Be with the chiral EFT NN+3N interactions. In Sec.~\ref{sec:He6-res}, we show the application of the three-body cluster NCSM/RGM to the description of the Borromean halo nucleus $^6$He. In Sec.~\ref{sec:capture}, we review our first application of the NCSM/RGM formalism to a reaction important for astrophysics, the $^7$Be($p,\gamma$)$^8$B radiative capture. In Sec.~\ref{sec:fusion}, we discuss our past as well as new results for the $^3$H($d,n$)$^4$He fusion. Finally, we give conclusions and an outlook in Sec.~\ref{sec:concl}.

%% file: SE_2-intro-formalism.tex
\section{Unified {\em ab initio} description of bound and scattering states}
\label{sec:formalism}

As mentioned in the Introduction, {\em ab initio} many-body approaches making use of expansions on square-integrable basis functions have been quite successful in explaining the properties of many well-bound systems. At the same time, microscopic cluster approaches, in which the wave function is represented by the continuous motion of two or more clusters, are naturally adapted to the description of clustering, scattering and reaction observables. In general, both approaches, taken separately, tend to have significant limitations owing to the  fact that in most practical calculations one must severely restrict the number of basis states in the trial function. In this section, we discuss how these seemingly very different views can be combined into a unified approach to structure and reactions, particularly in the context of {\em ab initio} calculations.
To build such a unified theory we start from an accurate microscopic Hamiltonian, described in Sec.~\ref{sec:ham}. We then make use of the similarity renormalization group (SRG) approach~\cite{Wegner1994,Bogner2007,SzPe00} to soften this Hamiltonian, as described in Sec.~\ref{sec:srg}. The NCSM approach we use to obtain the square-integrable eigenstates is briefly outlined in Sec.~\ref{sec:eigenstates}. A unified description of structure and dynamics can be achieved by means of the RGM, discussed in Secs.~\ref{sec:rgm2} and \ref{sec:rgm3}, but is more efficiently obtained working within the NCSMC, presented in~\ref{sec:ncsmc}.

%% file: SE_hamiltonian.tex
\subsection{Hamiltonian}
\label{sec:ham}

{\em Ab initio} approaches start from the microscopic Hamiltonian for the $A$-nucleon system 
\begin{equation}\label{eq:ham_Ham}
\hat{H} = \hat T_{\rm int} +\hat V
\end{equation}
composed of the intrinsic kinetic energy operator $\hat T_{\rm int}$ and the nuclear interaction $\hat V = \hat V^{NN}+\hat V^{3N}+\dots$, which describes the strong and electro-magnetic interaction among nucleons. The interaction $\hat{V}$ generally consist of realistic NN and 3N contributions that accurately reproduce few-nucleon properties, but in principle can also contain higher many-nucleon contributions. More specifically, the Hamiltonian can be written as
\begin{equation}
\hat{H}=\frac{1}{A}\sum_{i<j=1}^A\frac{(\hat{\vec{p}}_i-\hat{\vec{p}}_j)^2}{2m}+\sum_{i<j=1}^A \hat{V}^{NN}_{ij}+\sum_{i<j<k=1}^A \hat{V}^{3N}_{ijk}+\dots,
\label{H}
\end{equation} 
where $m$ is the nucleon mass and $\vec{p}_i$ the momentum of the $i$th nucleon.
The electro-magnetic interaction is typically described by the Coulomb force, while the determination of the strong interaction poses a tremendous challenge.  

\begin{figure}
\centering\includegraphics[width=1.0\textwidth]{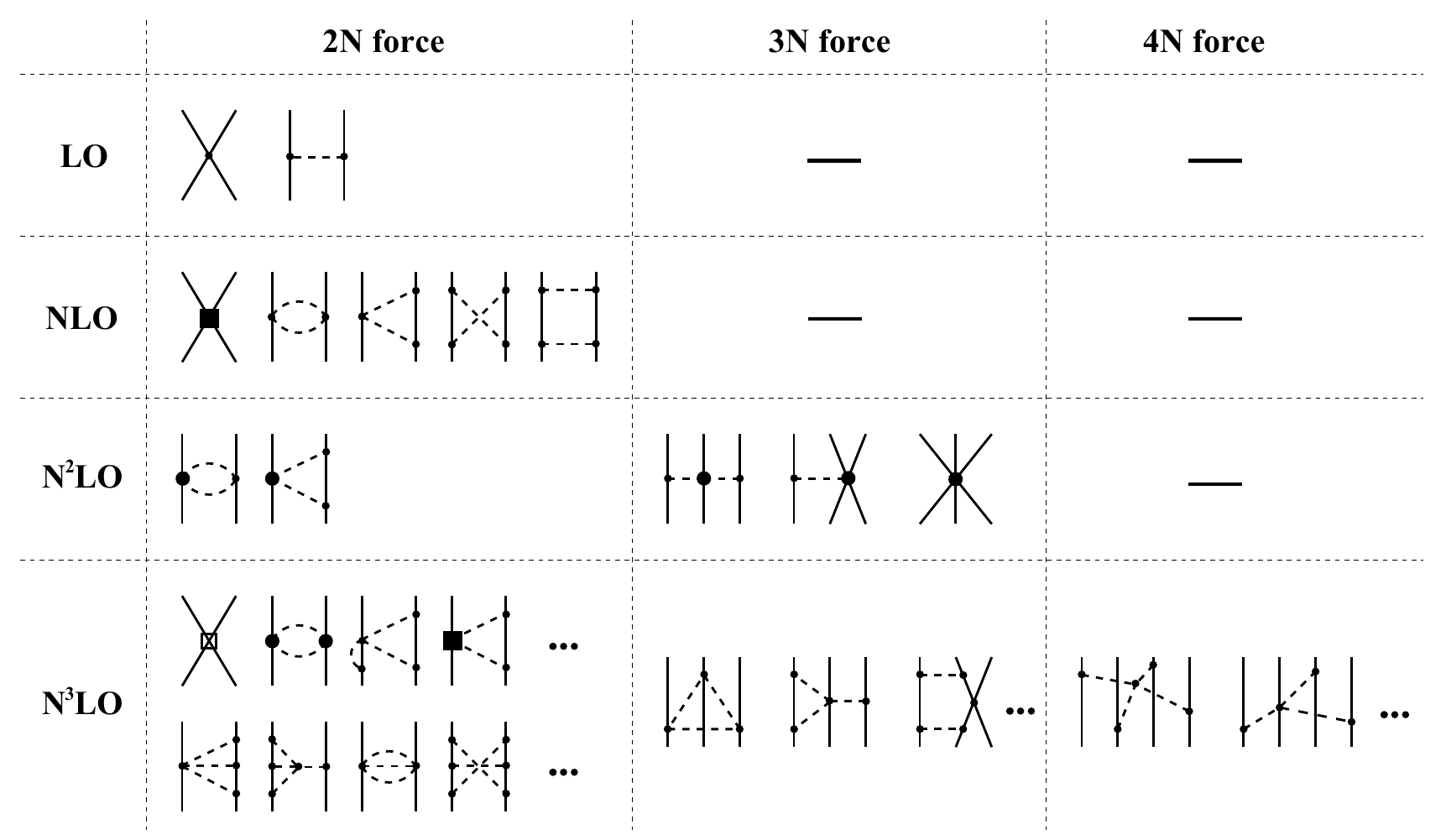}

\caption{{Hierarchy of nuclear forces in chiral EFT~\cite{Kalantar-Nayestanaki2012}:} 
{\small The interaction diagrams up to N$^{3}$LO arranged by the particle rank of the interaction. Dashed lines represent pions and the solid lines nucleons. Small dots, large
solid dots, solid squares and open squares denote vertices at increasing expansion orders. Figure from Ref.~\cite{Kalantar-Nayestanaki2012}.
}
\label{fig:XEFTDiagram}}
\end{figure}

According to the Standard Model, the strong interaction between nucleons is described by QCD with quarks and gluons as fundamental degrees of freedom. 
However, the nuclear structure phenomena we are focusing on are dominated by low energies and QCD becomes non-perturbative in this regime, which so far impedes a direct derivation of the nuclear interaction from the underlying theory. 
Inspired by basic symmetries of the Hamiltonian and the meson-exchange theory proposed by Yukawa~\cite{Yukawa35}, phenomenological high-precision NN interactions, such as the Argonne V18~\cite{Wiringa1995} and CD-Bonn~\cite{Mach01} potentials, have been developed. These interactions provide an accurate description of NN systems, but sizeable discrepancies are observed in nuclear structure applications to heavier nuclei ~\cite{CaNa02,NaOr02,PhysRevC.66.044310}. This indicates the importance of many-nucleon interactions beyond the two-body level and reveal the necessity for a consistent scheme to construct the nuclear interactions.
Thus, Weinberg formulated an effective theory for the low-energy regime using nucleons and
pions as explicit degrees of freedom~\cite{Weinberg79}. The chiral EFT~\cite{Weinberg1990,Weinberg1991} uses a low-energy expansion illustrated in Fig.~\ref{fig:XEFTDiagram} in terms of $(Q/\Lambda_{\chi})^{\nu}$ that allows for a systematic improvement of the potential by an increase of the chiral order $\nu$. Here $Q$ relates to the nucleon momentum/pion mass and $\Lambda_{\chi}$ corresponds to the break down scale of the chiral expansion that is typically on the order of $1\,\text{GeV}$. 
Moreover, the chiral expansion  provides a hierarchy of NN, 3N, and many-nucleon interactions in a consistent scheme~\cite{OrRa94,VanKolck94,EpNo02,Epelbaum06}.
Since the chiral expansion is only valid at low energies it is necessary to suppress high-momentum contributions beyond a certain cutoff $\Lambda_{\text{cut}}$ by introducing a regularization. There are different possible choices for the regulator function and the cutoff, which determine the regularization scheme.  
The commonly used chiral NN interaction in nuclear structure and reaction physics is constructed by Entem and Machleidt at next-to-next-to-next-to leading order (N$^3$LO) using a cutoff $\Lambda_{\text{cut}}=500\,\text{MeV}$~\cite{Entem2003}. The low-energy constants (LECs) of this potential are fitted to the $\pi$N scattering as well as neutron-proton (np) and proton-proton (pp) data below $290\,\text{MeV}$. The accuracy of the description of NN systems is comparable to the mentioned phenomenological high-precision interactions~\cite{Entem2003}.  
This NN potential is generally supplemented by local 3N contributions at next-to next-to leading order (N$^2$LO) using a three-body cutoff $\Lambda_{\text{cut,3N}}$  of $500\mev$~\cite{Navratil2007} or $400\mev$~\cite{Roth2012}, depending on the mass region.
The 3N contributions at N$^2$LO consist of a two-pion exchange term, a one-pion exchange two-nucleon contact term and a three-nucleon contact term (see Fig.~\ref{fig:XEFTDiagram}).
The LECs in the two-pion exchange term $c_{1}$, $c_{3}$, and $c_4$ already appear for the first time in the NN force and are fitted to NN data, while the LECs $c_{D}$ and $c_{E}$ of the contact contributions appear for the first time and are fitted to the triton beta-decay half life and the $A=3$ or $A=4$-body ground-state energies~\cite{Gazit2009,Roth2014}.
This interaction is extensively studied in nuclear structure and reaction physics 
 and constitutes the starting point for the majority of the investigations in this review.    
It is important to note that the rapid developments in the construction of chiral interactions in recent years not only exploit different regularizations and fit procedures~\cite{EpGl04,EpGl05,EkBa13,EkJa15,PhysRevC.90.054323}, but extend the accessible contributions to higher chiral orders~\cite{EpKe15,BiCa15}.
The chiral interactions are currently optimized using advanced numerical techniques, showing promising results for applications to heavier nuclei beyond the p shell~\cite{EkBa13,EkJa15}.  
Moreover, the LENPIC collaboration~\cite{LENPIC} provides consistent interactions for a sequence of cutoffs constructed order by order up to N$^4$LO in combination with a prescription to determine propagated uncertainties of nuclear observables resulting from the interaction~\cite{BiCa15}.   
These developments will enhance the predictive power of \textit{ab initio} calculations and allow to determine theoretical uncertainties in the future.

%% file: SE_SRG.tex
\subsection{Similarity renormalization group method}
\label{sec:srg}
Chiral interactions are already rather soft compared to phenomenological high-precision interactions such as the Argonne V18~\cite{Wiringa1995} and CD-Bonn~\cite{Mach01} owing to the regularization that suppresses high-momentum contributions, as described in Sec.~\ref{sec:ham}.
Nevertheless, most many-body methods cannot achieve convergence in feasible model spaces due to present short-range and tensor correlation induced by the chiral interactions. 
Therefore, additional transformations, such as the unitary correlation operator method (UCOM)~\cite{Roth2010}, the $\vlowk$ renormalization group method~\cite{BoKu03,BoKu03b,BoFu07b} or the Okubo-Lee-Suzuki similarity transformation~\cite{SuLe80,Okub54} are used to soften the interactions.
The most successful transformation approach in nuclear structure physics is the SRG~\cite{Wegner1994,Bogner2007,SzPe00} that is presented in the following.
This transformation provides a model-space and nucleus independent softened interaction and allows for consistent unitary transformations of the NN and 3N components of the interaction.   

The basic concept of the SRG is the first-order differential operator equation
\begin{equation} \label{eq:srg_floweq}
  \frac{d}{ds} \hat{H}_{s} = \big{[} {\hat{\eta}_{s}},{\hat{H}_{s}}\big{]}\,,
\end{equation}
that defines the continuous unitary transformation $\hat{H}_{s}=\hat{U}^{\,\dagger}_{s} \, \hat{H} \, \hat{U}_{s}$, where the unitary operator $\hat{U}_{s}$ depends on the continuous flow-parameter $s$.
In this flow equation $\hat{H}_{s}$ denotes the SRG evolved Hamiltonian depending on the flow parameter $s$ and the anti-Hermitian dynamic generator 
\begin{equation} \label{eq:srg_gen}
\hat{\eta}_{s} = -\hat{U}^{\,\dagger}_{s} \frac{d}{ds} \hat{U}_{s} = - \hat{\eta}^{\dagger}_{s}\,.   
\end{equation}
The canonical choice for the generator (used in the majority of nuclear structure and reaction applications) is the commutator of the kinetic energy with the Hamiltonian, i.e.,
\begin{equation} \label{eq:srg_canonicgen}
  \hat{\eta}_{s} = \left(2\frac{\mu}{\hbar^2}\right)^2\; \commut{\hat T_{\rm int},\hat{H}_{s}}\,,
\end{equation}
where $\mu$ is the reduced nucleon mass and $\hat T_{\rm int}$ constitutes the trivial fix point of the flow of the Hamiltonian, such that the high- and low-momentum contributions of the interaction decouple.
For this generator choice it is reasonable to associate the flow parameter with a momentum scale, using the relation $\Lambda=s^{-(1/4)}$ as often done in the literature~\cite{Jurgenson2009,Bogner201094}.

When aiming at observables other than binding and excitation energies it is formally necessary to transform the corresponding operators $ \hat{O}_{s}=\hat{U}^{\,\dagger}_{s} \, \hat{O} \, \hat{U}_{s}$, which can be achieved by evaluating $\hat{U}_{s}$ directly or by solving the flow equation
\begin{equation}\label{eq:srg_floweqO}
\frac{d}{ds} \hat{O}_{s} = \commut{\hat{\eta}_{s},\hat{O}_{s}}\, .
\end{equation}
Because the dynamic generator contains the evolved Hamiltonian, the flow equations for the operator $\hat{O}_{s}$ and the Hamiltonian $\hat{H}_s$ need to be evolved simultaneously.
We refer to Ref.~\cite{ScQu14,Schuster2015} for recent applications and stress that there is work in progress to perform SRG transformations of observables. 

It is important to note that equation~\eqref{eq:srg_floweq} is an operator relation in  the A-body Hilbert space. Due to the repeated multiplication of the operators on the right hand side of the flow equation, irreducible many-body contributions beyond the initial particle rank of the Hamiltonian are induced. Generally, contributions beyond the three-body level cannot be considered.
This limitation causes one of the most challenging problems in context of the SRG transformation, since the unitarity is formally violated. Thus, it is necessary to confirm the invariance of physical observables under the transformation. In practice a variation of the flow-parameter $\Lambda$ is used as an diagnostic tool to access the impact of omitted many-body contributions.
Moreover, to probe the induced and initial 3N contributions individually one exploits three types of Hamiltonians. \\ 
{\bf The NN-only Hamiltonian} is obtained from an initial NN interaction performing the SRG at the two-body level and does not contain any three- or higher many-body contributions.\\
{\bf The NN+3N-ind Hamiltonian} is obtained from an initial NN interaction performing the SRG at the two- and three-body level such that the induced 3N contributions are included.\\   
{\bf The NN+3N-full or simply NN+3N Hamiltonian} is obtained from an initial NN+3N interaction performing the SRG at the two- and three-body level. This Hamiltonian contains the complete set of NN and 3N contributions.   

For practical applications the flow equation~\eqref{eq:srg_floweq} is represented in a basis and the resulting first-order coupled differential equations are solved numerically. Due to the simplicity of the evolution the SRG can be implemented in the three-body space and even beyond. 
The most efficient formulation for the SRG evolution with regard to  descriptions of finite nuclei is performed in the Jacobi HO representation~\cite{Jurgenson2009,Jurgenson2011,Roth2011} using a subsequent Talmi-Moshinsky transformation~\cite{KaKa01} to the particle representation that is utilized by the many-body approaches, see Ref.~\cite{Roth2014} for a detailed explanation. 
There are also implementations of the three-body SRG evolution performed in other basis representations, such as the partial-wave decomposed momentum-Jacobi basis~\cite{Hebe12} and the hyperspherical momentum basis~\cite{Wend13}. However, so far only the SRG in the HO basis has been used to provide reliably evolved 3N interactions and operators for nuclear structure calculations beyond the lightest nuclei.

It has been shown that the two-pion exchange part of the 3N interaction induces irreducible contributions beyond the three-body level that become sizeable in the mid-p shell~\cite{Roth2014}. As a consequence alternative formulations of the dynamic generator have been explored to avoid induced many-body contributions from the outset~\cite{Dicaire14}. In addition, it has been observed that a reduction of the 3N cutoff from $500 \mev$ to $400 \mev$ strongly suppresses the impact of induced many-body contributions~\cite{Roth2014} and allows for reliable applications beyond p- and sd-shell nuclei.

%% file: SE_NCSM_eigenstates.tex
\subsection{Square-integrable eigenstates of clusters and the compound nucleus}
\label{sec:eigenstates}

Expansions on square integrable many-body states are among the most common techniques for the description of the static properties of nuclei. 
The {\it ab initio NCSM} is one of such techniques.
Nuclei are considered as systems of $A$ non-relativistic point-like nucleons interacting through realistic inter-nucleon interactions discussed in Section~\ref{sec:ham}. All nucleons are active degrees of freedom. Translational invariance as well as angular momentum and parity of the system under consideration are conserved. 
The many-body wave function is cast into an expansion over 
a complete set of antisymmetric $A$-nucleon HO basis states containing up to  $N_{\rm max}$ 
HO excitations above the lowest possible configuration: 
\begin{equation}\label{NCSM_wav}
 \ket{\Psi^{J^\pi T}_A} = \sum_{N=0}^{N_{\rm max}}\sum_i c_{Ni}^{J^\pi T}\ket{ANiJ^\pi T}\; .
\end{equation}
Here, $N$ denotes the total number of HO excitations of all nucleons above the minimum configuration,  $J^\pi T$ are the total angular momentum, parity and isospin, and $i$ additional quantum numbers. The sum over $N$ is restricted by parity to either an even or odd sequence.
The basis is further characterized by the frequency $\Omega$ of the HO
well and may depend on either Jacobi relative~\cite{PhysRevC.61.044001} or single-particle
coordinates~\cite{Navratil2000}. In the former case, the wave function does not contain the center of mass (c.m.) motion, but antisymmetrization is complicated. In the latter case,  
antisymmetrization is trivially achieved using Slater determinants, but the c.m.\ degrees of freedom are included in the basis. The HO basis  within the $N_{\rm max}$ truncation is the only possible one that allows an exact factorization of the c.m.\ motion for the eigenstates, even when working with single-particle coordinates and Slater determinants. Calculations performed with the two alternative coordinate choices are completely equivalent. 
 
Square-integrable energy eigenstates expanded over the $N_{\rm max}\hbar\Omega$ basis, $\ket{ANiJ^\pi T}$, are obtained by diagonalizing the intrinsic Hamiltonian, 
\begin{equation}\label{NCSM_eq}
\hat{H} \ket{A \lambda J^\pi T} = E_\lambda ^{J^\pi T} \ket{A \lambda J^\pi T} \; ,
\end{equation}
with $\hat{H}$ given by Eq.~(\ref{eq:ham_Ham}) and $\lambda$ labeling eigenstates with identical $J^\pi T$. 
Convergence of the HO expansion with increasing $N_{\rm max}$ values is accelerated by the use of effective interactions derived from the underlying potential model through either Lee-Suzuki similarity transformations in the NCSM space~\cite{PhysRevC.68.034305,Navratil2000a,Navratil2000} or SRG transformations in momentum space~\cite{Bogner2007,PhysRevC.77.064003,Roth2010,Bogner201094,Jurgenson2009,Jurgenson2011} discussed in detail in Section~\ref{sec:srg}. In this latter case, the NCSM calculations are variational. Because of the renormalization of the Hamiltonian, the many-body wave function obtained in the NCSM (as well as in the NCSM/RGM and the NCSMC) are in general renormalized as well. This fact is in particular important to keep in mind when using low SRG parameters, i.e., $\Lambda\lessapprox 2$ fm$^{-1}$. Finally, we note that with the 
HO basis sizes typically used ($N_{\rm max}{\sim}10{-}14$), the NCSM $\ket{A \lambda J^\pi T}$ eigenstates lack correct asymptotic behavior for weakly-bound states and always have incorrect asymptotic behavior for resonances.

%% file: SE_binary-RGM.tex
\subsection{Binary-cluster NCSM/RGM}
\label{sec:rgm2}
A description of bound and scattering states within a unified framework can already be achieved by adopting a simplified form of $\ket{\Psi^{J^\pi T}_A}$, limited to expansions on microscopic cluster states chosen according to physical intuition and energetic arguments.
Expansions on binary-cluster states,
%
\begin{equation}
|\Phi^{J^\pi T}_{\nu r}\rangle = \Big [ \big ( \left|A-a\, \alpha_1 I_1^{\,\pi_1} T_1\right\rangle \left |a\,\alpha_2 I_2^{\,\pi_2} T_2\right\rangle\big ) ^{(s T)}
\,Y_{\ell}\left(\hat r_{A-a,a}\right)\Big ]^{(J^\pi T)}\,\frac{\delta(r-r_{A-a,a})}{rr_{A-a,a}}\,,\label{basis}
\end{equation}
are the most common, allowing to describe processes in which both entrance and exit channels are characterized by the interaction of two nuclear fragments.
 
The above translational invariant cluster basis states describe two nuclei (a target and a projectile composed of $A-a$ and $a$ nucleons, respectively), whose centers of mass are separated by the relative displacement vector $\vec r_{A-a,a}$. The translational-invariant (antisymmetric) wave functions of the two nuclei, $\left|A-a\, \alpha_1 I_1^{\,\pi_1} T_1\right\rangle$ and $\left |a\,\alpha_2 I_2^{\,\pi_2} T_2\right\rangle$, are eigenstates of the $(A-a)$- and $a$-nucleon intrinsic Hamiltonians, with angular momentum $I_i$, parity $\pi_i$, isospin $T_i$ and energy labels $\alpha_i$,  where $i=1,2$. The system is further characterized by a  $^{2s+1}\ell_J$ partial wave of relative motion, where $s$ is the channel spin resulting from the coupling of the total clusters' angular momenta, $\ell$ is the relative orbital angular momentum, and $J$ is the total angular momentum of the system. Additional quantum numbers characterizing the basis states are  parity $\pi=\pi_1\pi_2(-1)^{\ell}$  and total isospin $T$ resulting from the coupling of the clusters' isospins.   In the notation of Eq.~(\ref{basis}), all relevant quantum numbers are summarized by the index $\nu=\{A-a\,\alpha_1I_1^{\,\pi_1} T_1;\, a\, \alpha_2 I_2^{\,\pi_2} T_2;\, s\ell\}$.  

However, to be used as a continuous basis set to expand the many-body wave function, the  channel states (\ref{basis}) have to be first antisymmetrized with respect to exchanges of nucleons pertaining to different clusters, which are otherwise unaccounted for. This can be accomplished by introducing an appropriate inter-cluster antisymmetrizer, schematically
\begin{equation}
\hat{\mathcal A}_{\nu}=\sqrt{\frac{(A-a)!a!}{A!}}\left( 1+\sum_{P\neq id}(-)^pP\right)\,,
\label{antisym}
\end{equation}   
where the sum runs over all possible permutations of nucleons $P$ (different from the identical one) 
that can be carried out between two different clusters (of $A-a$ and $a$ nucleons, respectively), and $p$ is the number of interchanges characterizing them. The operator~(\ref{antisym}) is labeled by the channel index $\nu$ to signify that its form depends on the mass partition, $(A-a,a)$, of the channel state to which is applied. 

Equations (\ref{basis}) and (\ref{antisym}) lead to the RGM ansatz for the many-body wave function,
%
%
%
\begin{align}
\ket{\Psi^{J^\pi T}_A} = \sum_{\nu} \int dr\, r^2 \, \hat{\mathcal A}_{\nu} \ket{\Phi^{J^\pi T}_{\nu r}}  \frac{\gamma^{J^\pi T}_\nu(r)}{r}
\,, \label{trial}
\end{align}
where 
$\gamma^{J^\pi T}_\nu(r)$ represent continuous linear variational amplitudes that are determined by solving the RGM equations: 
%
%
%
\begin{equation}
{\sum_{\nu^\prime}\int dr^\prime r^{\prime\,2}} \left[{\mathcal H}^{J^\pi T}_{\nu\nu^\prime\,}(r,r^\prime)- E\, {\mathcal N}^{J^\pi T}_{\nu\nu^\prime\,}(r,r^\prime)\right]\frac{\gamma^{J^\pi T}_{\nu^\prime} (r^\prime)}{r^\prime} = 0  \label{RGMeq}.
\end{equation}
Here $E$ is the total energy in the c.m.\ frame, and ${\mathcal N}^{J^\pi T}_{\nu\nu^\prime}(r, r^\prime)$ and ${\mathcal H}^{J^\pi T}_{\nu\nu^\prime}(r, r^\prime)$, commonly referred to as integration kernels, are respectively the overlap (or norm) and Hamiltonian matrix elements over the antisymmetrized basis~(\ref{basis}),  i.e.  
\begin{align}
{\mathcal N}^{J^\pi T}_{\nu^\prime\nu}(r^\prime, r) = \left\langle\Phi^{J^\pi T}_{\nu^\prime r^\prime}\right|\hat{\mathcal A}_{\nu^\prime}\hat{\mathcal A}_{\nu}\left|\Phi^{J^\pi T}_{\nu r}\right\rangle\,, 
&\qquad
{\mathcal H}^{J^\pi T}_{\nu^\prime\nu}(r^\prime, r) = \left\langle\Phi^{J^\pi T}_{\nu^\prime r^\prime}\right|\hat{\mathcal A}_{\nu^\prime}\hat H\hat{\mathcal A}_{\nu}\left|\Phi^{J^\pi T}_{\nu r}\right\rangle\,.
\label{NH-kernel}
\end{align}
%
%
%
%
%
In the above equation,  $\hat H$ is the microscopic $A-$nucleon Hamiltonian of Eq.~(\ref{H}), which can be conveniently separated into the intrinsic Hamiltonians for the $(A-a)$- and $a$-nucleon systems, respectively $\hat H_{(A-a)}$ and $\hat H_{(a)}$, plus the relative motion Hamiltonian
\begin{equation}\label{Hamiltonian}
\hat H=\hat T_{\rm rel}(r)+\hat{\bar{V}}_{\rm C}(r)+\hat{\mathcal V}_{\rm rel} +\hat H_{(A-a)}+\hat H_{(a)}\,.
\end{equation}
Here, $\hat T_{\rm rel}(r)$ is the relative kinetic energy between the two clusters, $\hat{\bar{V}}_{\rm C}(r)=Z_{1\nu}Z_{2\nu}e^2/r$ ($Z_{1\nu}$ and $Z_{2\nu}$ being the charge numbers of the clusters in channel $\nu$) the average Coulomb interaction between pairs of clusters, and ${\mathcal V}_{\rm rel}$ is a localized relative (inter-cluster) potential given by:
\begin{align}
\hat{\mathcal V}_{\rm rel} &= \sum_{i=1}^{A-a}\sum_{j=A-a+1}^A \hat V^{NN}_{ij} + \sum_{i<j=1}^{A-a}\sum_{k=A-a+1}^A \hat V^{3N}_{ijk} + \sum_{i=1}^{A-a}\sum_{j<k=A-a+1}^A \hat V^{3N}_{ijk} - \hat{\bar{V}}_{\rm C}(r)\label{pot}\,.
\end{align}
Besides the nuclear components of the interactions between nucleons belonging to different clusters, it is important to notice that the overall contribution to the relative potential~(\ref{pot}) coming from the Coulomb interaction,
\begin{equation}
\label{locCoul}
\sum_{i=1}^{A-a}\sum_{j=A-a+1}^A\left(\frac{e^2(1+\tau^z_i)(1+\tau^z_j)}{4|\vec r_i-\vec r_j|} -\frac{1}{(A-a)a}\hat{\bar V}_{\rm C}(r)\right)\,,
\end{equation}
 is also localized, presenting an $r^{-2}$ behavior, as the distance $r$ between the two clusters increases.

The calculation of the many-body matrix elements of Eq.~(\ref{NH-kernel}), which contain all the nuclear structure and antisymmetrization properties of the system under consideration, represents the main task in performing RGM calculations. In the following we will review the various steps required for one of such calculations when the eigenstates of the target and the projectile are obtained within the {\em ab initio} NCSM (see Sec.~\ref{sec:eigenstates}). This is the approach known as NCSM/RGM.


\subsubsection{NCSM/RGM norm and Hamiltonian kernels}
\label{kernels}
When representing the target and projectile eigenstates by means of NCSM wave functions, it is convenient to introduce RGM cluster states in HO space (with frequency $\Omega$ identical to that used for the clusters) defined by
\begin{align}
|\Phi^{J^\pi T}_{\nu n}\rangle &= \Big [ \big ( \left|A-a\, \alpha_1 I_1^{\,\pi_1} T_1\right\rangle \left |a\,\alpha_2 I_2^{\,\pi_2} T_2\right\rangle\big ) ^{(s T)}\,Y_{\ell}\left(\hat \eta_{A-a}\right)\Big ]^{(J^\pi T)}\,R_{n\ell}(r_{A-a,a})\,.\label{ho-basis-n}
\end{align}
The coordinate-space channel states of Eq.~(\ref{basis}) can then be written as $|\Phi^{J^\pi T}_{\nu r}\rangle = \sum_n R_{n\ell}(r) |\Phi^{J^\pi T}_{\nu n}\rangle$ by making use of the closure properties of the HO radial wave functions. Following Eqs.~(\ref{antisym}) and~(\ref{Hamiltonian}), it is also useful to factorize the norm and Hamiltonian kernels into ``full-space" and ``localized" components
\begin{align}
{\mathcal N}^{J^\pi T}_{\nu^\prime\nu}(r^\prime, r)
&= \delta_{\nu^\prime\nu}\frac{\delta(r^\prime-r)}{r^\prime r} + {\mathcal N}^{\rm ex}_{\nu^\prime\nu}(r^\prime, r) \,,
\label{N-kernel-2}
\end{align}
and
\begin{align}
{\mathcal H}^{J^\pi T}_{\nu^\prime\nu}(r^\prime, r) &= \left[{\hat T}_{\rm rel}(r')+\hat{\bar{V}}_C(r')+E_{\alpha_1'}^{I_1'T_1'} +E_{\alpha_2'}^{I_2'T_2'}\right]
\mathcal{N}_{\nu'\nu}^{J^\pi T}(r', r)+\mathcal{V}^{J^\pi T}_{\nu' \nu}(r',r)\,,
\label{H-kernel-2}
\end {align}
where the exchange part of the norm, ${\mathcal N}^{\rm ex}_{\nu^\prime\nu}(r^\prime, r)$, and the potential kernel, $\mathcal{V}^{J^\pi T}_{\nu' \nu}(r',r)$, (both localized quantities) are obtained in an HO model space of size $N_{\rm max}$ consistent with those used for the two clusters as:
\begin{align}
{\mathcal N}^{\rm ex}_{\nu^\prime\nu}(r^\prime, r) &= 
\sum_{n^\prime n}R_{n^\prime\ell^\prime}(r^\prime)R_{n\ell}(r) 
\times \left\{
\begin{array}{ll}
	\left\langle\Phi^{J^\pi T}_{\nu^\prime n^\prime}\right| \sum_{P\neq id}(-)^p \hat P \left|\Phi^{J^\pi T}_{\nu n}\right\rangle& \quad {\rm if}~a^\prime=a\\
	\\
	\left\langle\Phi^{J^\pi T}_{\nu^\prime n^\prime}\right| \sqrt{\tfrac{ A! }{ (A-a^\prime)! a^\prime !}} \hat {\mathcal A}_\nu \left|\Phi^{J^\pi T}_{\nu n}\right\rangle& \quad {\rm if}~a^\prime\neq a
\end{array}
\right .
\label{Nex-kernel}
\end{align}
and 
\begin{align}
\mathcal{V}^{J^\pi T}_{\nu' \nu}(r',r) &= 
	\sum_{n^\prime n}R_{n^\prime\ell^\prime}(r^\prime)R_{n\ell}(r)\left\langle\Phi^{J^\pi T}_{\nu^\prime r^\prime}\right| 
	\sqrt{\tfrac{ A!}{ (A-a^\prime)! a^\prime !}} {\mathcal V}_{\rm rel}\hat{\mathcal A}_{\nu} 
	\left|\Phi^{J^\pi T}_{\nu r}\right\rangle\,.
	\label{V-kernel}
\end{align}
We note that in deriving the above expressions we took advantage of the commutation between antisymmetrizers~(\ref{antisym}) and $A$-nucleon Hamiltonian~(\ref{H}), $[\hat{\mathcal A}_{\nu},H]{=}0$, and used the following relationship dictated by symmetry considerations:
\begin{align}
\hat{\mathcal A}_{\nu^\prime}\hat{\mathcal A}_{\nu}|\Phi^{J^\pi T}_{\nu n}\rangle =  \sqrt{\tfrac{ A!}{ (A-a^\prime)! a^\prime !}} \hat{\mathcal A}_{\nu}|\Phi^{J^\pi T}_{\nu n}\rangle\,. 
\end{align}
Finally, while it can be easily demonstrated that the exchange part of the norm kernel is Hermitian, i.e.,
\begin{equation}
	\left\langle\Phi^{J^\pi T}_{\nu^\prime n^\prime}\right| \sqrt{\tfrac{ A! }{ (A-a^\prime)! a^\prime !}} \hat {\mathcal A}_\nu \left|\Phi^{J^\pi T}_{\nu n}\right\rangle = \left\langle\Phi^{J^\pi T}_{\nu^\prime n^\prime}\right| \hat {\mathcal A}_{\nu^\prime} \sqrt{\tfrac{ A! }{ (A-a)! a!}}  \left|\Phi^{J^\pi T}_{\nu n}\right\rangle\,,
\end{equation}
the same is not true for the Hamiltonian kernel as defined in Eq.~(\ref{H-kernel-2}), once its localized components are expanded within a finite basis. Therefore, we work with an Hermitized Hamiltonian kernel $\widetilde{\mathcal H}^{J^\pi T}_{\nu^\prime\nu}$ given by
\begin{equation}
\widetilde{\mathcal H}^{J^\pi T}_{\nu^\prime\nu}(r^\prime,r)\!=\!\left\langle\Phi^{J^\pi T}_{\nu^\prime r^\prime}\right| \tfrac 12 \left(\hat{\mathcal A}_{\nu^\prime} \hat H  \sqrt{\tfrac{ A! }{ (A-a)! a!}} + 
	\sqrt{\tfrac{ A!}{ (A-a^\prime)! a^\prime !}} \hat H \hat{\mathcal A}_{\nu} 
	 \right)\left| \Phi^{J^\pi T}_{\nu r}\right\rangle.\label{ham-herm}
\end{equation}

Being translationally-invariant  quantities, the Hamiltonian and norm kernels~(\ref{Nex-kernel},\ref{V-kernel}) can be ``naturally" derived working within a translationally invariant Jacobi-coordinate HO basis~\cite{Quaglioni2009}. However, particularly for the purpose of calculating reactions involving $p$-shell nuclei, it is computationally advantageous to use the second-quantization formalism. This can be accomplished by defining Slater-determinant (SD) channel states 
\begin{align}
|\Phi^{J^\pi T}_{\nu n}\rangle_{\rm SD}   =&    \Big [\big (\left|A-a\, \alpha_1 I_1 T_1\right\rangle_{\rm SD} 
\left |a\,\alpha_2 I_2 T_2\right\rangle\big )^{(s T)} Y_{\ell}(\hat R^{(a)}_{\rm c.m.})\Big ]^{(J^\pi T)} R_{n\ell}(R^{(a)})\,,
\label{SD-basis}
\end{align}
in which the eigenstates of the $(A-a)$-nucleon fragment are obtained in a HO SD basis as $|\Phi^{J^\pi T}_{\nu n}\rangle_{\rm SD}   = |\Phi^{J^\pi T}_{\nu n}\rangle \varphi_{00}(\vec{R}^{(A-a)})$  (while the second cluster is a NCSM Jacobi-coordinate eigenstate~\cite{Kamuntavicius1999}), with  $\vec{R}^{(A-a)} = (A-a)^{-1/2}\sum_{i = 1}^{(A-a)} \vec{r}_i$ and $\vec{R}^{(a)} = a^{-1/2}\sum_{i = A-a+1}^A \vec{r}_i$ being the vectors proportional to the center of mass coordinates of the $(A-a)$- and $a$-nucleon clusters, respectively. Indeed, it is easy to demonstrate that translationally invariant matrix elements can be extracted from those computed in the SD basis of Eq.~(\ref{SD-basis}) by inverting the following expression:
 \begin{align}
 {}_{\rm SD}\!\left\langle\Phi^{J^\pi T}_{\nu^\prime n^\prime}\right|\hat{\mathcal O}_{\rm t.i.}\left|\Phi^{J^\pi T}_{\nu n}\right\rangle\!{}_{\rm SD} = & 
\sum_{n^\prime_r \ell^\prime_r, n_r\ell_r, J_r}
 \left\langle\Phi^{J_r^{\pi_r} T}_{\nu^\prime_r n^\prime_r}\right|\hat{\mathcal O}_{\rm t.i.}\left|\Phi^{J_r^{\pi_r} T}_{\nu_r n_r}\right\rangle\nonumber\\
&  \times \sum_{NL} \hat \ell \hat \ell^\prime \hat J_r^2 (-1)^{(s+\ell-s^\prime-\ell^\prime)}
  \left\{\begin{array}{ccc}
 s &\ell_r&  J_r\\
  L& J & \ell
 \end{array}\right\}
 \left\{\begin{array}{ccc}
 s^\prime &\ell^\prime_r&  J_r\\
  L& J & \ell^\prime
 \end{array}\right\}\nonumber\\
 &\nonumber\\
& \times\langle  n_r\ell_rNL\ell | 00n\ell\ell \rangle_{\frac{a}{A-a}} 
 \;\langle  n^\prime_r\ell^\prime_rNL\ell | 00n^\prime\ell^\prime\ell^\prime \rangle_{\frac{a}{A-a}} \,.\label{Oti}
 \end{align}
Here $\hat {\mathcal O}_{\rm t.i.}$ represents any scalar, parity-conserving and translationally-invariant operator ($\hat {\mathcal O}_{\rm t.i.} = \hat{\mathcal A_\nu}$, $\hat H \hat{\mathcal A_\nu}$, etc.), $\langle  n_r\ell_rNL\ell | 00n\ell\ell \rangle_{\frac{a}{A-a}}$, $\langle  n^\prime_r\ell^\prime_rNL\ell | 00n^\prime\ell^\prime\ell^\prime \rangle_{\frac{a}{A-a}}$ are general HO brackets for two particles with mass ratio $a/(A-a)$~\cite{Trlifaj1972} and the notation $\hat \ell$ stands for $\sqrt{2\ell+1}$.

\subsubsection{Algebraic expressions for the $(A-1,1)$ mass partition} 
\label{sec:examples}
To give an example of  the algebraic expressions for the SD matrix elements on the left-hand-side of Eq.~(\ref{Oti}) here we consider the case in which the projectile in both initial and final states is a single nucleon ($a^\prime=a=1$). The antisymmetrization operator for $(A-1,1)$ mass partitions can be written as 
\begin{align}
\hat{\mathcal A}_{\nu}\equiv\hat{\mathcal A}_{(A-1,1)}=\frac{1}{\sqrt A}\left[1-\sum_{i=1}^{A-1}\hat P_{iA}\right]\,,
\end{align}
where $\hat P_{iA}$ is a permutation operator exchanging nucleon $A$ (the projectile) with the $i$-th nucleon of the target. Using the second-quantization formalism, the SD matrix elements of the exchange-part of the norm kernel~(\ref{Nex-kernel}) for a single-nucleon projectile can be related to linear combinations of
matrix elements of creation and annihilation operators between $(A-1)$-nucleon SD states as~\cite{Quaglioni2009}
\begin{align}
_{\rm SD}\langle \Phi_{\nu'\,n'}^{J^\pi T}|\sum_{P\neq id}(-)^p \hat P| \Phi_{\nu\,n}^{J^\pi T}\rangle_{\rm SD} & = -(A-1)\;_{\rm SD}\langle \Phi_{\nu'\,n'}^{J^\pi T}|\hat P_{A-1,A}| \Phi_{\nu\,n}^{J^\pi T}\rangle_{\rm SD}\nonumber\\
&= - \sum_{jj'K\tau}  \hat{s}\hat{s}'\hat{j}\hat{j}'\hat{K}\hat{\tau} (-1)^{I'_1+j'+J} (-1)^{T_1+\frac{1}{2}+T}\label{P_AAm1_SD}
\\
&\quad\times\left\{ \begin{array}{@{\!~}c@{\!~}c@{\!~}c@{\!~}} 
I_1 & \frac{1}{2} & s \\[2mm] 
\ell & J & j 
\end{array}\right\} 
\left\{ \begin{array}{@{\!~}c@{\!~}c@{\!~}c@{\!~}} I'_1 & \frac{1}{2} & s' \\ [2mm]
\ell' & J & j' \end{array}\right\} \left\{ \begin{array}{@{\!~}c@{\!~}c@{\!~}c@{\!~}} 
I_1 & K & I'_1 \\[2mm] 
j' & J & j \end{array}\right\}
\left\{ \begin{array}{@{\!~}c@{\!~}c@{\!~}c@{\!~}} 
T_1 & \tau & T'_1 \\[2mm]
\frac{1}{2} & T & \frac{1}{2} \end{array}\right\}\nonumber\\[2mm]
&\quad\times\; _{\rm SD}\langle A-1 \alpha' I'_1 T'_1 ||| (a^\dagger_{n\ell j\frac{1}{2}} \tilde{a}_{n'\ell'j'\frac{1}{2}})^{(K\tau)} ||| A-1 \alpha I_1 T_1 \rangle_{\rm SD}\,.\nonumber
\end{align}
In deriving this expression we took advantage of the symmetry properties of the $(A-1)$-nucleon target, and introduced one-body density  matrix elements of the target nucleus,
$_{\rm SD}\langle A-1 \alpha' I'_1 T'_1 ||| (a^\dagger_{n\ell j\frac{1}{2}} \tilde{a}_{n'\ell'j'\frac{1}{2}})^{(K\tau)} ||| A-1 \alpha I_1 T_1 \rangle_{\rm SD}$ (reduced both in angular momentum and isospin), where 
$\tilde{a}_{n'\ell'j'm'\frac{1}{2}m_t^\prime}=(-1)^{j'-m'+\frac{1}{2}-m_t^\prime}\;a_{n'\ell'j'-m'\frac{1}{2}-m_t^\prime}$. 

\begin{figure}[t]
\begin{center}
\includegraphics*[width=0.55\columnwidth]{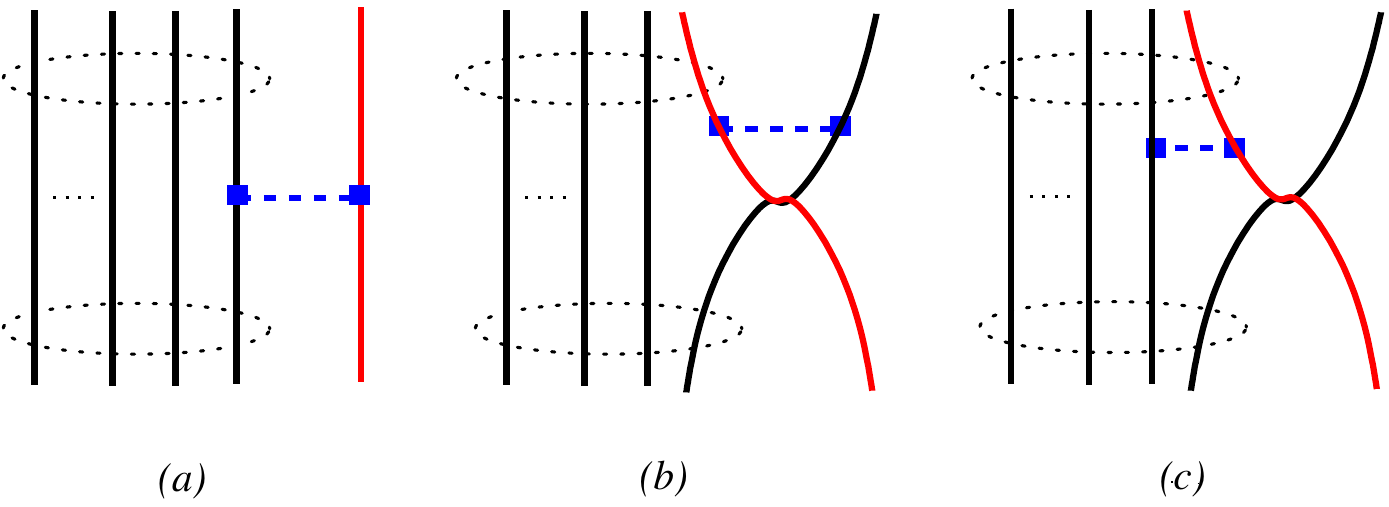}
\caption{Diagrammatic representation of ($a$ and $b$) ``direct"  and ($c$) ``exchange"  components of the potential kernel for the $(A-1,1)$ cluster basis. The first group of circled lines represents the first cluster, the bound state of $A-1$ nucleons. The separate line represents the second cluster, in the specific case  a single nucleon. Bottom and upper part of the diagrams represent initial and final states, respectively.}\label{diagram-pot}
\end{center}
\end{figure}

The derivation of the analogous matrix elements~(\ref{V-kernel}) for the NN and 3N potentials, although more involved, is straightforward. It can be demonstrated that the NN potential kernel for the same $(A-1,1)$ partition in both initial and final states ($a^\prime=a=1$) takes the form~\cite{Quaglioni2009}
\begin{align}
	_{\rm SD}\langle \Phi_{\nu'\,n'}^{J^\pi T}|\sqrt{A}~\hat{\mathcal V}_{\rm rel}^{ NN} \hat{\mathcal A}_{(A-1,1)}| \Phi_{\nu\,n}^{J^\pi T}\rangle_{\rm SD} & =  
	(A-1)\; _{\rm SD}\langle \Phi_{\nu'\,n'}^{J^\pi T}| \hat V_{A-1,A}(1-\hat P_{A-1,A})| \Phi_{\nu\,n}^{J^\pi T}\rangle_{\rm SD}\label{VNN-kernel}\\
	& \,- (A-1)(A-2)\; _{\rm SD}\langle \Phi_{\nu'\,n'}^{J^\pi T}|\hat P_{A-1,A} \hat V_{A-2,A-1}| \Phi_{\nu\,n}^{J^\pi T}\rangle_{\rm SD}\nonumber\,,
\end{align}
where in the first and second lines, respectively, on the right-hand-side of Eq.~(\ref{VNN-kernel}) one can identify a ``direct'' and an ``exchange'' term of the interaction, schematically represented by the diagrams of Fig.~\ref{diagram-pot}. 
\begin{figure}[b]
\begin{center}
\includegraphics*[width=0.55\columnwidth]{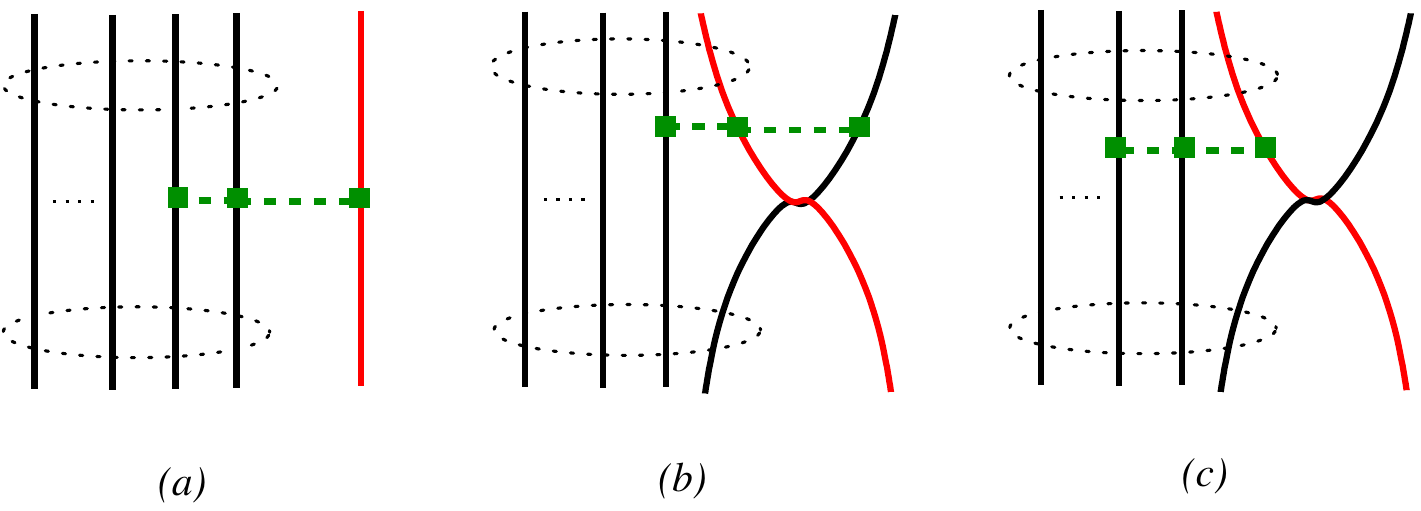}
\end{center}
\caption{Diagrammatic representation of the components of the direct ($a$ and $b$) and exchange components of the $3N$ potential kernel for the same $(A-1,1)$ partition in both initial and final states ($a^\prime=a=1$). The groups of circled lines represent the $(A-1)$-nucleon cluster. Bottom and upper part of the diagrams represent initial and final states, respectively.}\label{diagram-NNNpot}
\end{figure}
The SD matrix elements of the direct term of the potential involve operations on the projectile and one of the nucleons of the target (including a trivial exchange of the interacting nucleons shown in Fig.~\ref{diagram-pot} ($b$)) and are given by 
\begin{align}
&_{\rm SD}\langle \Phi_{\nu'\,n'}^{J^\pi T}|\hat V_{A-1,A}(1-\hat P_{A-1,A})| \Phi_{\nu\,n}^{J^\pi T}\rangle_{\rm SD}\nonumber\\
&\quad=\tfrac{1}{(A-1)}\sum_{jj'K\tau} \sum_{n_a l_a j_a}\sum_{n_b l_b j_b}\sum_{J_0 T_0}
\hat{s}\hat{s}'\hat{j}\hat{j}'\hat{K}\hat{\tau} 
\hat{J}_0^2 \hat{T}_0^2 (-1)^{I'_1+j'+J} (-1)^{T_1-\frac{1}{2}+T} 
\left\{ \begin{array}{@{\!~}c@{\!~}c@{\!~}c@{\!~}} 
I_1 & \frac{1}{2} & s \\[2mm] 
\ell & J & j 
\end{array}\right\} 
\left\{ \begin{array}{@{\!~}c@{\!~}c@{\!~}c@{\!~}} I'_1 & \frac{1}{2} & s' \\ [2mm]
\ell' & J & j' \end{array}\right\}
\nonumber\\[2mm]
&\quad\times 
\left\{ \begin{array}{@{\!~}c@{\!~}c@{\!~}c@{\!~}} 
I_1 & K & I'_1 \\[2mm] 
j' & J & j \end{array}\right\}
\left\{ \begin{array}{@{\!~}c@{\!~}c@{\!~}c@{\!~}} 
j_b & j_a & K \\[2mm]
j'  & j   & J_0 \end{array}\right\}
\left\{ \begin{array}{@{\!~}c@{\!~}c@{\!~}c@{\!~}} 
T_1 & \tau & T'_1 \\[2mm]
\frac{1}{2} & T & \frac{1}{2} \end{array}\right\}
\left\{ \begin{array}{@{\!~}c@{\!~}c@{\!~}c@{\!~}} 
\tau & \frac{1}{2} & \frac{1}{2} \\[2mm]
T_0  & \frac{1}{2} & \frac{1}{2} \end{array}\right\}
\sqrt{1+\delta_{(n_a l_a j_a),(n' \ell' j')}} \sqrt{1+\delta_{(n_b l_b j_b),(n \ell j)}}
\nonumber\\[2mm]
&\quad\times 
\langle (n_a l_a j_a \frac{1}{2}) (n' \ell' j' \frac{1}{2}) J_0 T_0 | V 
| (n \ell j \frac{1}{2}) (n_b l_b j_b \frac{1}{2}) J_0 T_0 \rangle
\nonumber\\[2mm]
&\quad\times 
\; _{\rm SD}\langle A-1 \alpha' I'_1 T'_1 ||| (a^\dagger_{n_a l_a j_a\frac{1}{2}} 
\tilde{a}_{n_b l_b j_b\frac{1}{2}})^{(K\tau)} ||| A-1 \alpha I_1 T_1 \rangle_{\rm SD} 
\label{direct-pot}
\end{align}
Different from Eqs.~(\ref{P_AAm1_SD}) and (\ref{direct-pot}), the SD matrix elements of the exchange term of the potential  involve operations on two nucleons  (projectile exchange with a first and interaction with a second nucleon) of the $(A-1)$ cluster and hence depend on two-body density matrix elements of the target nucleus (last line in the right-hand side of the equation):
\begin{align}
&_{\rm SD}\langle \Phi_{\nu'\,n'}^{J^\pi T}|\hat P_{A-1,A} \hat V_{A-2,A-1}| \Phi_{\nu\,n}^{J^\pi T}\rangle_{\rm SD}\nonumber\\
&\quad = \tfrac{1}{2(A-1)(A-2)} \sum_{jj'K\tau} \sum_{n_a l_a j_a}\sum_{n_b l_b j_b}
\sum_{n_c l_c j_c}\sum_{n_d l_d j_d}\sum_{K_a \tau_a K_{cd} \tau_{cd}} 
\hat{s}\hat{s}'\hat{j}\hat{j}'\hat{K}\hat{\tau} \hat{K}_a\hat{\tau}_a \hat{K}_{cd}\hat{\tau}_{cd}
\nonumber\\[2mm]
&\quad\times
(-1)^{I'_1+j'+J+K+j+j_a+j_c+j_d} (-1)^{T_1+\frac{1}{2}+\tau+T}  
\left\{ \begin{array}{@{\!~}c@{\!~}c@{\!~}c@{\!~}} 
I_1 & \frac{1}{2} & s \\[2mm] 
\ell & J & j 
\end{array}\right\} 
\left\{ \begin{array}{@{\!~}c@{\!~}c@{\!~}c@{\!~}} I'_1 & \frac{1}{2} & s' \\ [2mm]
\ell' & J & j' \end{array}\right\}
\left\{ \begin{array}{@{\!~}c@{\!~}c@{\!~}c@{\!~}} 
I_1 & K & I'_1 \\[2mm] 
j' & J & j \end{array}\right\}
\nonumber\\[2mm]
&\quad\times 
\left\{ \begin{array}{@{\!~}c@{\!~}c@{\!~}c@{\!~}} 
K_a & K_{cd} & K \\[2mm]
j'  & j   & j_a \end{array}\right\}
\left\{ \begin{array}{@{\!~}c@{\!~}c@{\!~}c@{\!~}} 
T_1 & \tau & T'_1 \\[2mm]
\frac{1}{2} & T & \frac{1}{2} \end{array}\right\}
\left\{ \begin{array}{@{\!~}c@{\!~}c@{\!~}c@{\!~}} 
\tau & \tau_a & \tau_{cd} \\[2mm]
\frac{1}{2} & \frac{1}{2} & \frac{1}{2} \end{array}\right\}
\sqrt{1+\delta_{(n_a l_a j_a),(n' \ell' j')}} \sqrt{1+\delta_{(n_c l_c j_c),(n_d l_d j_d)}}
\nonumber\\[2mm]
&\quad\times 
\langle (n' \ell' j' \frac{1}{2}) (n_a l_a j_a \frac{1}{2}) K_{cd} \tau_{cd} | V 
| (n_d l_d j_d \frac{1}{2}) (n_c l_c j_c \frac{1}{2}) K_{cd} \tau_{cd} \rangle
\nonumber\\[2mm]
&\quad\times 
\; _{\rm SD}\langle A-1 \alpha' I'_1 T'_1 ||| ((a^\dagger_{n \ell j\frac{1}{2}} 
a^\dagger_{n_a l_a j_a\frac{1}{2}})^{(K_a\tau_a)}(\tilde{a}_{n_c l_c j_c\frac{1}{2}}
\tilde{a}_{n_d l_d j_d\frac{1}{2}})^{(K_{cd}\tau_{cd})})^{(K\tau)} ||| A-1 \alpha I_1 T_1 \rangle_{\rm SD}
\,.
\label{exchange-pot}
\end{align}

The inclusion of the 3N force in the Hamiltonian further complicates the calculation of the $(A-1,1)$ NCSM/RGM kernels. As for the corresponding NN portion of the potential kernel, there are a direct (including a trivial exchange of the interacting nucleons) and an exchange term, described by diagrams $(a)$ and $(b)$, and diagram $(c)$ of Fig.~\ref{diagram-NNNpot}, respectively. For a summary of their expressions we refer the interested readers to Ref.~\cite{Quaglioni2012}. While the first two diagrams are similar in complexity to the NN exchange term, the third depends on three-body density matrix elements of the target nucleus.  Due to their rapidly increasing number in multi-major-shell basis spaces, storing in memory three-body density matrices is very demanding and requires the implementation of specialized computational strategies~\cite{Hupin2013}.  One of such strategies is to develop an efficient on-the-fly computation of these matrix elements by working in the $m$-scheme and exploiting the fact that the target eigenstates $|A-1\, \alpha_1 I_1^{\pi_1} M_1 T_1 M_{T_1}\rangle$ are implicitly given as expansions in HO many-body SDs within the NCSM model space. For example, the matrix elements of the operator $ \hat P_{A-1,A} \hat{V}_{A-3,A-2,A-1}$, characterizing the exchange term of the 3N force, with respect to the basis states~(\ref{SD-basis}) can be written as 
\begin{align}
\leftsub{\rm SD}{\big\langle}&\Phi^{J^\pi T}_{\nu^\prime n^\prime}\big| \op{P}_{A-1,A}  \op{V}_{A-3\,A-2\,A-1} \ket{\Phi^{J^\pi T}_{\nu n}}_{\rm SD} &\nonumber\\
& = \tfrac{1}{6(A-1)(A-2)(A-3)}  \sum_{j j^\prime} ~ \hat{s}\hat{s}^\prime\hat{j}\hat{j}^\prime (-1)^{2J+I_1+I^\prime_1+j+j^\prime} \left\{ \begin{array}{@{\!~}c@{\!~}c@{\!~}c@{\!~}} I_1 & \frac{1}{2} & s \\[2mm] \ell & J & j \end{array}\right\} \left\{ \begin{array}{@{\!~}c@{\!~}c@{\!~}c@{\!~}} I^\prime_1 & \frac{1}{2} & s^\prime \\[2mm] \ell^\prime & J & j^\prime \end{array}\right\}\nonumber\\
& \times \sum_{M_1 m_j} \sum_{M_{T_1} m_t}\sum_{M^{\prime}_1 m^{\prime}_j} \sum_{M^{\prime}_{T_1} m^{\prime}_t}     C_{I_1 M_1 j m_j}^{J M_J}
C_{T_1 M_{T_1} \frac12 m_t}^{T M_T} 
C_{I^{\prime}_1 M^{\prime}_{1} j^{\prime} m^{\prime}_j}^{J M^\prime_J} 
C_{T^{\prime}_1 M^{\prime}_{T_1} j^{\prime} m^{\prime}_t}^{T M^\prime_T} 
\notag \\
&\times
\sum_{abdef}\bra{ab\;n^{\prime}\ell^{\prime}j^{\prime}m_{j}^{\prime}\tfrac{1}{2}m^{\prime}_{t}}\op{V}^{3N}\ket{def} \; \notag \\
& \times
  \leftsub{\rm SD}{\bra{A-1\, \alpha_1^{\prime} I_1'^{\pi_1^{\prime}} M_1^{\prime} T_1^{\prime} M_{T_1}^{\prime}}} \op{a}^{\dagger}_{n\ell jm_{j}\tfrac{1}{2}m_{t}} \op{a}^{\dagger}_{b} \op{a}^{\dagger}_{a} 
\op{a}_{d} \op{a}_{e} \op{a}_{f} \ket{ A-1\, \alpha_1 I_1^{\pi_1} M_1 T_1 M_{T_1}}_{\rm SD} \, , \label{NNN-ex-express}
%
\end{align}
where we introduced  the notation $\{a,\ldots, f\}$ denoting the quantum numbers of $\ell s$-coupled HO single-particle states, i.e.\ $a  = \{n_a \ell_a j_a m_{j_a} m_{t_a}\}$. 
The two summations corresponding to the expansions in HO many-body SDs  of the eigenstates of the target (not shown explicitly in the equation) can be pulled in front of all other summations in Eq.~\eqref{NNN-ex-express}, obtaining an expression in which each term can be computed independently, i.e., ideally suited for parallel computation. In addition, the sums over HO single-particle states can of course be restricted to those combinations, which can connect the two SDs of the density matrix. Here, we can make use of the technology that was originally developed to compute $A$-body matrix elements of three-body operators during the setup of the many-body matrix in the importance-truncated NCSM \cite{Roth2007,Roth2009}. The next critical objects in Eq.~\eqref{NNN-ex-express} are the $m$-scheme matrix elements of the $3N$ interaction. The storage of these matrix elements in memory is again prohibitive if we want to proceed to large model spaces. However, we benefit from storing the matrix elements of the $3N$ interaction in the $JT$-coupled scheme developed by Roth et al.~\cite{Roth2011,Roth2014} and the corresponding efficient on-the-fly decoupling into the $m$-scheme. Finally, we note from Eq.~\eqref{NNN-ex-express} the necessity to treat the projection quantum numbers of the angular momenta and isospins of the target states explicitly, including consistent relative phases. Both can be accomplished using a single NCSM run to produce a specific eigenvector from which all other vectors with necessary projection are obtained using angular momentum raising and lowering operators.

The second option is to algebraically perform the summations over the projection quantum numbers of Eq.~\eqref{NNN-ex-express} and introduce coupled densities using the Wigner-Eckart theorem as previously done for the NN case (\ref{direct-pot}, \ref{exchange-pot}), i.e.
\begin{align}
\leftsub{\rm SD}{\langle} & \Phi^{J^\pi T}_{\nu^\prime n^\prime} |   \hat P_{A-1,A} \hat V_{A-3,A-2,A-1}  |\Phi^{J^\pi T}_{\nu n}\rangle_{\rm SD}  \nonumber\\
= &  \tfrac{1}{6(A-1)(A-2)(A-3)}  \sum_{j j^\prime} ~ \hat{s}\hat{s}^\prime\hat{j}\hat{j}^\prime (-1)^{2J+I_1+I^\prime_1+j+j^\prime} \left\{ \begin{array}{@{\!~}c@{\!~}c@{\!~}c@{\!~}} I_1 & \frac{1}{2} & s \\[2mm] \ell & J & j \end{array}\right\} \left\{ \begin{array}{@{\!~}c@{\!~}c@{\!~}c@{\!~}} I^\prime_1 & \frac{1}{2} & s^\prime \\[2mm] \ell^\prime & J & j^\prime \end{array}\right\}\nonumber\\
\times& \sum_{{\bar a} {\bar b} {\bar d} {\bar e} {\bar f}} \sum_{J_{ab} T_{ab}} \sum_{ J_0 T_0 } \sum_{J_{de} T_{de}} \sum_{ J_g T_g} {\hat J_0}{\hat T_0} {\hat J_g}{\hat T_g}{\hat K}{\hat \tau}   (-1)^{j+j'+J_{ab}+K-J_g+I_1+J + 1+T_{ab}+\tau-T_g+T_1+T}\nonumber\\
\times& \langle {\bar a}, {\bar b}  ; J_{ab} T_{ab},{n'\ell 'j'\tfrac{1}{2}} ; J_0 T_0 | \hat V^{3N}  |  {\bar d}, {\bar e}  ; J_{de} T_{de},{\bar f} ; J_0 T_0 \rangle \label{NNN-ex-express2}\\
\times&  \left\{ 
\begin{array}{ccc} 
I_1 &K&I_1' \\
 j'&J&j 
\end{array} \right\}
\left\{
\begin{array}{ccc}
j'&K&j \\
J_g&J_{ab}&J_0
\end{array}
\right\}
\left\{
\begin{array}{ccc}
T_1 &\tau&T_1'\\
\tfrac12 &T& \tfrac12
\end{array}
\right\}
\left\{
\begin{array}{ccc}
\tfrac12 &\tau& \tfrac12\\
T_g&T_{ab}&T_0
\end{array}
\right\}  \nonumber\\
\times &\leftsub{\rm SD}{\langle} A{-}1 \alpha_{1}' I_1^{\prime\,\pi_1'} T_1' ||| \Big( \Big( \big( a^{\dag}_{{\bar a}}  a^{\dag}_{{\bar b}} \big)^{J_{ab} T_{ab}} a^{\dag}_{n\ell j \tfrac{1}{2} } \Big)^{J_g T_g} \Big( \big( {\tilde a}_{{\bar d}}{\tilde a}_{{\bar e}} \big)^{J_{de}T_{de}}  {\tilde a}_{{\bar f}} \Big)^{J_0 T_0} \Big)^{K \tau} |||A{-}1 \alpha_{1} I_1^{\pi_1} T_1 \rangle_{\rm SD} ,\nonumber
\end{align}
where $\{{\bar a},\ldots,{\bar f}\}$ denote HO orbitals, i.e. ${\bar a }= \{n_{ f}, \ell_{ f}, j_{ f}\}$, and the triple vertical bars indicate that the matrix elements are reduced in both angular momentum and isospin, and $\langle {\bar a}, {\bar b}  ; J_{ab} T_{ab},{n'\ell 'j'\tfrac{1}{2}} ; J_0 T_0 | \hat V^{3N}  |  {\bar d}, {\bar e}  ; J_{de} T_{de},{\bar f} ; J_0 T_0 \rangle$ are $JT$-coupled $3N$-force matrix elements.  
Further, to avoid the storage in memory of the reduced three-body density matrix, we factorize this expression by inserting a completeness relationship over $(A-4)$-body eigenstates leading to the final expression: 
\begin{align}
\leftsub{\rm SD}{\langle} & \Phi^{J^\pi T}_{\nu^\prime n^\prime} |   \hat P_{A-1,A}\hat V_{A-3,A-2,A-1}  |\Phi^{J^\pi T}_{\nu n}\rangle_{\rm SD} \nonumber\\
& =  \tfrac{1}{6(A-1)(A-2)(A-3)} \sum_{j j^\prime} ~ \hat{s}\hat{s}^\prime\hat{j}\hat{j}^\prime (-1)^{2J+I_1+I^\prime_1+j+j^\prime} \left\{ \begin{array}{@{\!~}c@{\!~}c@{\!~}c@{\!~}} I_1 & \frac{1}{2} & s \\[2mm] \ell & J & j \end{array}\right\} \left\{ \begin{array}{@{\!~}c@{\!~}c@{\!~}c@{\!~}} I^\prime_1 & \frac{1}{2} & s^\prime \\[2mm] \ell^\prime & J & j^\prime \end{array}\right\}\nonumber\\
&\times \sum_{{\bar a} {\bar b} {\bar d} {\bar e} {\bar f}} \sum_{J_{ab} T_{ab}} \sum_{ J_0 T_0 } \sum_{J_{de} T_{de}} \sum_{ J_g T_g} \sum_{ \beta}  {\hat J_0}{\hat T_0} {\hat J_g}{\hat T_g}   ~ \left\{ \begin{array}{ccc} I_{\beta} &J_g &I_1' \\ J_0&J_{ab}& j' \\ I_1&j&J \end{array} \right\}   \left\{ \begin{array}{ccc} T_{\beta} &T_g&T_1' \\ T_0&T_{ab}& \tfrac12  \\ T_1&\tfrac12&T \end{array} \right\} \nonumber\\
&\times \langle {\bar a}, {\bar b}  ; J_{ab} T_{ab},n'\ell' j' \tfrac{1}{2} ; J_0 T_0 |  \hat V^{3N}  |  {\bar d}, {\bar e}  ; J_{de} T_{de},{\bar f} ; J_0 T_0 \rangle\nonumber\\
&\times\leftsub{\rm SD}{\langle} A-1\,\alpha_{1}' I_1^{\prime\,\pi_1'} T_1' ||| \Big( \big(a^{\dag}_{{\bar a}}  a^{\dag}_{{\bar b}} \big)^{J_{ab} T_{ab}}a^{\dag}_{n\ell j \tfrac{1}{2} }\Big)^{J_g T_g} ||| A-4 \,\alpha_{\beta} I_{\beta}^{\pi_\beta} T_{\beta} \rangle_{\rm SD} \nonumber \\
&\times \leftsub{\rm SD}{\langle} A-1\,\alpha_{1} I_1^{\pi_1} T_1 ||| \Big( \big( {a}^{\dag}_{{\bar d}} {a}^{\dag}_{{\bar e}} \big)^{J_{de}T_{de}}  {a}^{\dag}_{{\bar f}} \Big)^{J_0T_0}  ||| A-4\,\alpha_{\beta}I_{\beta}^{\pi_\beta} T_{\beta} \rangle_{\rm SD} \; . \label{NNN-ex-express3}
\end{align}
Compared to Eq.~(\ref{NNN-ex-express2}), there is an additional summation over the index $\beta$ labeling the eigenstates $| A-4 \,\alpha_{\beta},I_{\beta}^{\pi_\beta},T_{\beta} \rangle$ of the $(A-4)$-body system. In this second approach, we first calculate and store in memory the reduced matrix elements of the tensor operator
$\Big( \big( {a}^{\dag}_{{\bar d}} {a}^{\dag}_{{\bar e}} \big)^{J_{de}T_{de}}  {a}^{\dag}_{{\bar f}} \Big)^{J_0T_0}$
and compute the factorized three-body density of Eq.~(\ref{NNN-ex-express3}) on the fly. This strategy reduces the computational burden and computer memory required to perform the calculation. We work directly with the $JT$-coupled $3N$ matrix elements exploiting their symmetries and using the appropriate Racah algebra if necessary. The main limitation of this approach is the factorization of the reduced density which is feasible only for light systems where a complete set of $(A-4)$-body eigenvectors can be obtained, i.e., the four- and five-nucleon systems for the specific case of nucleon-nucleus collisions. For such systems, however, it is still a more efficient approach when many excited states of the target are included in the calculation as discussed in  Sec.~\ref{sec:s-shell} in the case of $n$-$^4$He scattering calculations with seven \elem{He}{4} eigenstates. In terms of numerics we have achieved a load-balanced parallel implementation by using a non-blocking master-slave algorithm. Finally, following the application of the Wigner-Eckart theorem, the isospin breaking terms of the nuclear Hamiltonian in the potential kernels are not treated exactly in this approach. Rather, they are approximated by isospin averaging~\cite{Hupin2013}. However, it should be noted that no isospin-symmetry breaking terms are typically included in the chiral 3N interaction. 
With the exception of the treatment of isospin symmetry, the two implementations described in this section are formally equivalent. By storing in memory the reduced densities, the latter is more efficient for reactions with different projectiles while the former is ideally suited for addressing heavier targets as in the case of the $^9$Be study of Sec.~\ref{sec:Be9}.

More in general, 
the complexity of the integration kernels rapidly increases with projectile mass, number of projectile/target states, and number of channels included, making the NCSM/RGM a computationally intensive approach. In this section we have presented a review of the algebraic expressions for the SD matrix elements entering the norm and Hamiltonian kernels for equal $(A-1,1)$ mass partitions in both initial and final states. The explicit form of the inter-cluster antisymmetrizer for the case in which the projectile is a deuterium nucleus ($a=2$), together with algebraic expressions for the SD matrix elements of $\hat {\mathcal O}_{\rm t.i.} = \hat{\mathcal A}_\nu$ and ${\mathcal V}_{\rm rel}^{NN}\hat{\mathcal A}_\nu$ for equal mass partitions in initial and final states can be found in Ref.~\cite{Navratil2011}. For reactions involving a deuterium-nucleus entrance and nucleon-nucleus exit channels [{\em e.g.}, $^3$H$(d,n)^4$He] or vice versa, and, more in general, whenever both nucleon-nucleus and deuterium-nucleus channel basis states are used in the RGM model space, one has to address the additional contributions coming from the off-diagonal matrix elements between the two mass partitions: $(A-1,1)$ and $(A-2,2)$. A summary of their expressions in the case of a two-body Hamiltonian can be found in Ref.~\cite{Quaglioni2012}.
Finally, the NCSM/RGM formalism can be generalized to the description of collisions with heavier projectiles, such as $^3$H/$^3$He-nucleus scattering~\cite{Dohet-Eraly:2015fsa} and, in principle, $\alpha$-nucleus scattering.

\subsubsection{Orthogonalization of the RGM equations}
\label{orthog}
An important point to notice, is that Eq.~(\ref{RGMeq}) does not represent a system of multichannel Schr\"odinger equations, and $\gamma^{J^\pi T}_\nu(r)$ do not represent Schr\"odinger wave functions. This feature, which is highlighted by the presence of the norm kernel ${\mathcal N}^{J^\pi T}_{\nu^\prime\nu}(r^\prime, r)$ and is caused by the short-range non-orthogonality induced by the non-identical permutations in the inter-cluster anti-symmetrizers~(\ref{antisym}),  can be removed by working with orthonormalized binary-cluster states
\begin{equation}\label{eq:formalism_70}
  \sum_{\nu'}\int dr' {r'}^2 \; \mathcal{N}_{\nu \nu'}^{-\frac{1}{2}}(r,r')
                            \; \hat{\mathcal{A}}_{\nu'} \ket{\Phi_{\nu' r'}^{J^\pi T}} \; ,
\end{equation}
and applying the inverse-square root of the norm kernel, $\mathcal{N}_{\nu \nu'}^{-\frac{1}{2}}(r,r')$, to both left and right-hand sides of the square brackets in Eq.~(\ref{RGMeq}). Here, we review how this can be done in practice.

Following Eq.~(\ref{Nex-kernel}), the norm kernel in $r$-space representation can be written as a convolution of a localized kernel
\begin{align}
	\mathcal{N}^{J^\pi T}_{\nu n \nu' n'} = \delta_{\nu\nu'}\delta_{n n'} - 
	\left\{
\begin{array}{ll}
	\left\langle\Phi^{J^\pi T}_{\nu^\prime n^\prime}\right| \sum_{P\neq id}(-)^p P \left|\Phi^{J^\pi T}_{\nu n}\right\rangle& \quad {\rm if}~a^\prime=a\\
	\\
	\left\langle\Phi^{J^\pi T}_{\nu^\prime n^\prime}\right| \sqrt{\tfrac{ A! }{ (A-a^\prime)! a^\prime !}} \hat {\mathcal A}_\nu \left|\Phi^{J^\pi T}_{\nu n}\right\rangle& \quad {\rm if}~a^\prime\neq a
\end{array}
\right .
\label{N-mod}
\end{align}
plus a correction owing to the finite size of the model space, i.e.
\begin{align}
\label{Northo}
\mathcal{N}^{J^\pi T}_{\nu \nu'}(r,r')   = \delta_{\nu\nu'}\left[ 
            \frac{\delta(r-r')}{rr'} 
           -\sum_{nn'} R_{n\ell}(r)\delta_{nn'}R_{n'\ell'}(r')
      \right]  +\sum_{nn'} R_{n\ell}(r) \mathcal{N}^{J^\pi T}_{\nu n \nu' n'} R_{n'\ell'}(r') \;.
\end{align}
Square and inverse-square roots 
of the norm kernel can then be defined in an analogous way as:
\begin{align}
\mathcal{N}_{\nu \nu'}^{\pm \frac{1}{2}}(r,r')
 = \delta_{\nu\nu'}\left[ 
                  \frac{\delta(r-r')}{rr'} 
                 -\sum_{nn' } R_{n\ell}(r) \delta_{nn'}R_{n'\ell'}(r')
           \right] +\sum_{nn' } R_{n\ell}(r) \mathcal{N}^{\pm \frac{1}{2}}_{\nu n \nu' n'} R_{n'\ell'}(r')\,,
\end{align}
where the matrix elements $\mathcal{N}^{\pm \frac12}_{\nu n \nu' n'}$, can be obtained from the eigenvalues and eigenstates of the matrix~(\ref{N-mod}) by using the spectral theorem, and it can be easily demonstrated that, e.g., the convolution of $\mathcal{N}^{\frac12}_{\nu\mu}(r,y)$ with $\mathcal{N}^{- \frac12}_{\mu \nu'}(y,r')$ yields the full-space identity, $\delta_{\nu\nu'}\delta(r-r')/r r'$. Similarly, the Hermitized Hamiltonian kernel  within the orthonormal basis of Eq.~(\ref{eq:formalism_70}) follows from applying the inverse-square root of the norm from the left and right-hand sides of Eq.~(\ref{ham-herm}), i.e.
\begin{align}
\label{eq:formalism_90}
\overline{\mathcal{H}}^{J^\pi T}_{\nu \nu'}(r,r')  =  \sum_{\mu\mu'}\int\!\! \int \!\! dy dy' {y}^2 {y'}^2 
                                    \mathcal{N}_{\nu \mu}^{-\frac{1}{2}}(r,y)
                                    \widetilde{\mathcal{H}}^{J^\pi T}_{\mu \mu'}(y,y')
                                    \mathcal{N}_{\mu' \nu'}^{-\frac{1}{2}}(y',r')\,,
\end{align}
and the orthogonalized RGM equations take the form of a set of non-local coupled channel Schr\"odinger equations: 
\begin{align}
{\sum_{\nu^\prime}\int dr^\prime\, r^{\prime\,2}} \, \overline{\mathcal{H}}^{J^\pi T}_{\nu\nu^\prime\,}(r,r^\prime)\frac{\chi^{J^\pi T}_{\nu^\prime} (r^\prime)}{r^\prime} = E\,\frac{\chi^{J^\pi T}_{\nu} (r)}{r} \,. \label{RGMorteq} 
\end{align}
Here, the Schr\"odinger wave functions of the relative motion $\chi^{J^\pi T}_{\nu} (r)$ are the new unknowns, related to the original functions $\gamma^{J^\pi T}_{\nu} (r)$ by
\begin{align}
\frac{\chi^{J^\pi T}_\nu(r)}{r} = \sum_{\nu^\prime}\int dr^\prime\, r^{\prime\,2} {\mathcal N}^{\frac12}_{\nu\nu^\prime}(r,r^\prime)\,\frac{\gamma^{J^\pi T}_{\nu^\prime}(r^\prime)}{r^\prime}\,,
\end{align}
For more details on the NCSM/RGM kernels we refer the interested reader to Ref.~\cite{Quaglioni2009}.

%% file: SE_ternary-RGM.tex
\subsection{Three-cluster NCSM/RGM}
\label{sec:rgm3}

\begin{figure}[t]
\centering
\epsfig{file=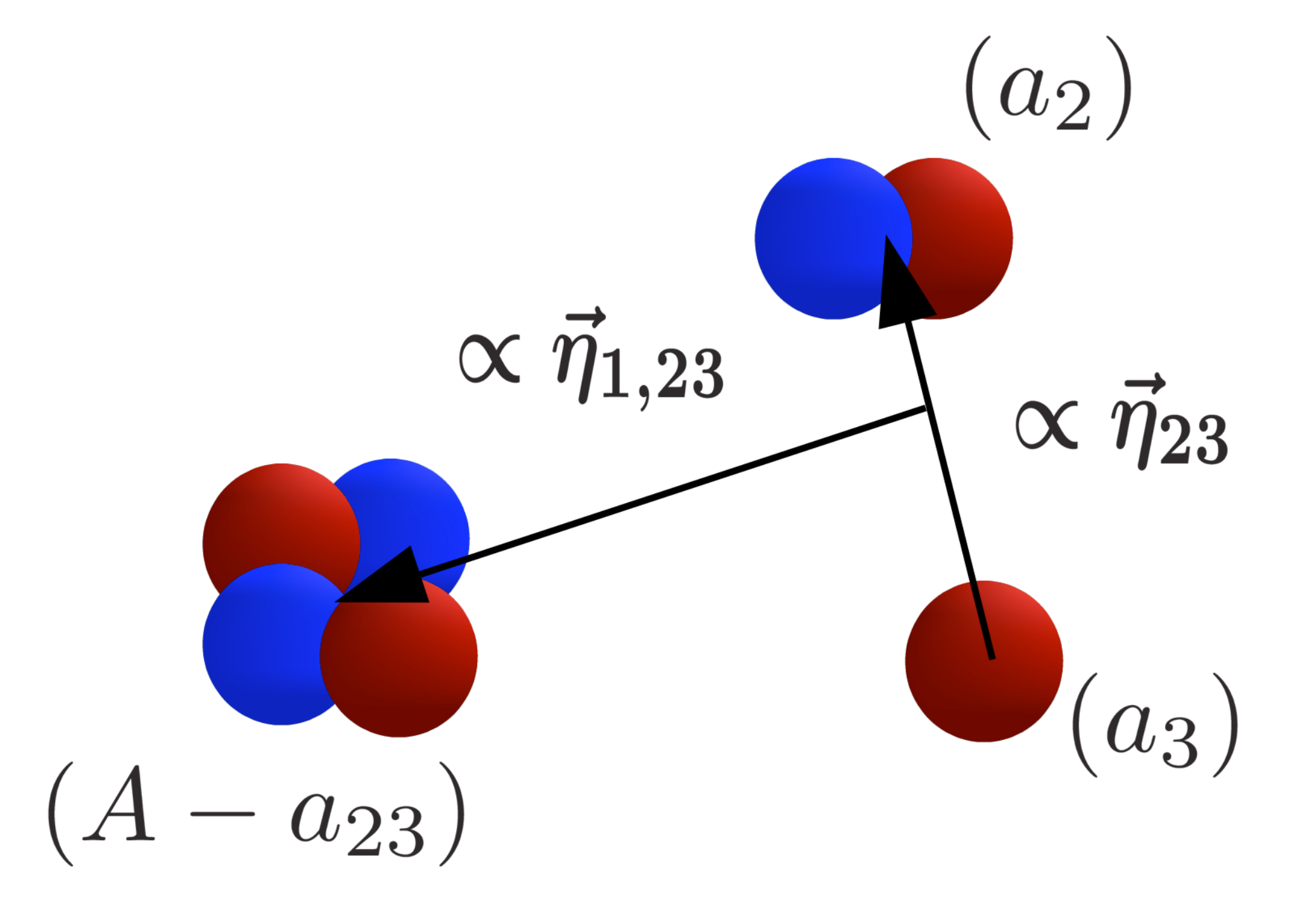,width=60mm}
\caption{(Color online) Jacobi coordinates for three cluster configurations, 
 $\vec\eta_{1,23}$ (proportional to the vector
between the c.m.\ of the first cluster and that of the
residual two fragments) and $\vec\eta_{23}$ (proportional to the vector between the c.m.\ of
clusters 2 and 3). In the
figure, a case with three clusters of four, two and one nucleons are shown, however the formalism is
general and could be used to describe any three cluster configuration.}
\label{FigCoor}
\end{figure}

When considering ternary clusters, the NCSM/RGM emerges in the same way as
for binary clusters. However, in this case, the relative motion behavior
of the wave function must be described in terms of two different
coordinates that characterize the relative position among the clusters.
Therefore, we can represent a system of $A$ nucleons 
arranged into three clusters respectively of mass number $A-a_{23}$, $a_2$, 
and $a_3$ ($a_{23}=a_2+a_3< A$), by the many-body wave function
\begin{align}
        |\Psi^{J^\pi T}\rangle  = \sum_{\nu} \iint dx \, dy \, x^2\, y^2 \, G_{\nu}^{J^\pi T}(x,y) \, \hat {\mathcal A}_\nu\, |\Phi^{J^\pi T}_{\nu x y} \rangle \,,
        \label{eq:trialwf}
\end{align}
where $G_{\nu}^{J^\pi T}(x,y)$ are continuous variational amplitudes of the integration variables $x$ and $y$, $\hat {\mathcal A}_\nu$ is an appropriate intercluster antisymmetrizer introduced to guarantee the exact preservation of the Pauli exclusion principle, and

\begin{align}
         |\Phi^{J^\pi T}_{\nu x y} \rangle  =&  
        \Big[\Big(|A-a_{23}~\alpha_1I_1^{\pi_1}T_1\rangle 
        \left (|a_2\, \alpha_2 I_2^{\pi_2} T_2\rangle |a_3\, \alpha_3 I_3^{\pi_3}T_3\rangle \right)^{(s_{23}T_{23})}\Big)^{(ST)}
 \nonumber \\
        &\left(Y_{\ell_x}(\hat{\eta}_{23})Y_{\ell_y}(\hat{\eta}_{1,23})\right)^{(L)}\Big]^{(J^{\pi}T)} 
         \times \frac{\delta(x-\eta_{23})}{x\eta_{23}} \frac{\delta(y-\eta_{1,23})}{y\eta_{1,23}}\,,
        \label{eq:3bchannel}    
\end{align}
are three-body cluster channels of total angular momentum $J$, parity $\pi$ and isospin $T$.  Here, $|A-a_{23}~\alpha_1I_1^{\pi_1}T_1\rangle$, $|a_2\, \alpha_2 I_2^{\pi_2} T_2\rangle$ and $|a_3\, \alpha_3 I_3^{\pi_3} T_3\rangle$ denote the microscopic (antisymmetric) wave functions of the three nuclear fragments, 
which are labelled by the spin-parity, isospin and energy quantum numbers $I_i^{\pi_i}$, $T_i$, and $\alpha_i$, respectively, with $i=1,2,3$. Additional quantum numbers characterizing the basis states (\ref{eq:3bchannel}) are the spins $\vec s_{23}=\vec I_2 + \vec I_3$ and $\vec S = \vec I_1+ \vec s_{23}$, the orbital angular momenta $\ell_x$, $\ell_y$ and $\vec L = \vec\ell_x+\vec\ell_y$, and the isospin $\vec T_{23}=\vec T_2+\vec T_3$. In our notation, all these quantum numbers are grouped under the cumulative index $\nu = \{A-a_{23}\, \alpha_1I_1^{\pi_1}T_1; a_2\, \alpha_2 I_2^{\pi_2} T_2; a_3\, \alpha_3 I_3^{\pi_3}T_3; s_{23} \,T_{23}\, S \,\ell_x \,\ell_y \, L\}$. Besides the translationally invariant coordinates 
(see e.g.\ Ref.~\cite{Quaglioni2009} Sec.\ II.C) used to describe the internal dynamics of clusters 1, 2 and 3, respectively, in Eq.~(\ref{eq:3bchannel}) we have introduced the Jacobi coordinates $\vec\eta_{1,23}$ and $\vec\eta_{23}$ where
\begin{align}
        \vec\eta_{1,23} & = \eta_{1,23} \hat{\eta}_{1,23}  \label{eq:etay}
         = \sqrt{\tfrac{a_{23}}{A(A-a_{23})}}  \sum_{i=1}^{A-a_{23}} \vec{r}_i - \sqrt{\tfrac{A-a_{23}}{A\,a_{23}}} \sum_{j=A-a_{23}+1}^A \vec{r}_j 
\end{align}
is the relative vector proportional to the displacement between the center of mass (c.m.) of the first cluster and that of the residual two fragments, and
\begin{align}
        \vec\eta_{23} & = \eta_{23} \hat{\eta}_{23} \label{eq:etax} 
         =\sqrt{\tfrac{a_3}{a_{23}\,a_2}}  \sum_{i=A-a_{23}+1}^{A-a_3} \vec{r}_i - \sqrt{\tfrac{a_2}{a_{23}\,a_3}} \sum_{j=A-a_3+1}^A \vec{r}_j      
\end{align}
is the relative coordinate proportional to the distance between the centers of mass of cluster
2 and 3 (See figure \ref{FigCoor}). Here, $\vec{r}_i$ denotes the position vector of the $i$-th nucleon.

When using the expansion~(\ref{eq:trialwf}), 
the many-body problem can be described through a set of coupled 
integral-differential equations that arise from projecting the 
$A$-body Schr\"odinger equation onto the cluster basis states 
$\hat {\mathcal A}_\nu\, |\Phi^{J^\pi T}_{\nu x y} \rangle$, 
\begin{align}
\label{eq:3beq1} 
        \sum_\nu \iint \!\!dx \, dy \, x^2 y^2 \Big [ {\mathcal H}^{J^\pi T}_{\nu^\prime\nu}(x^\prime,y^\prime,x,y) 
         - E \, {\mathcal N}^{J^\pi T}_{\nu^\prime\nu}(x^\prime,y^\prime,x,y) \Big] G_{\nu}^{J^\pi T}(x,y) = 0
\end{align}
where $G_{\nu}^{J^\pi T}(x,y)$ are the unknown continuum amplitudes and $E$
is the energy of the system in the c.m. frame. The integration kernels
are defined now as
\begin{align}
        {\mathcal H}^{J^\pi T}_{\nu^\prime\nu}(x^\prime,y^\prime,x,y) & = 
                \left\langle\Phi^{J^\pi T}_{\nu^\prime x^\prime y^\prime} \right| \hat {\mathcal A}_{\nu^\prime} \hat H \hat {\mathcal A}_\nu \left | \Phi^{J^\pi T}_{\nu x y} \right\rangle\,, \label{eq:Hkernel} \\
        {\mathcal N}^{J^\pi T}_{\nu^\prime\nu}(x^\prime,y^\prime,x,y) & = 
                \left\langle\Phi^{J^\pi T}_{\nu^\prime x^\prime y^\prime} \right| \hat{\mathcal A}_{\nu^\prime}  \hat {\mathcal A}_\nu \left | \Phi^{J^\pi T}_{\nu x y} \right\rangle   \label{eq:Nkernel}
\end{align}
where $\hat H$ is the intrinsic $A$-body Hamiltonian.

The system of multi-channel equations (\ref{eq:3beq1}) can be Hermitized and 
orthogonalized obtaining the analogous of Eq. (\ref{RGMorteq})
for binary cluster systems, i.e.:
 
\begin{align}
\label{3B_ortho}
 \sum_\nu \iint \!\!dx \, dy \, x^2 y^2 \Big [ \bar{\mathcal H}^{J^\pi T}_{\nu^\prime\nu}
(x^\prime,y^\prime,x,y)  - E \,\delta_{\nu\nu'}\frac{\delta(x'-x)}{x'x}\frac{\delta(y'-y)}{y'y} \Big
] \chi_{\nu}^{J^\pi T}(x,y) = 0\,.
\end{align} 
In this case, the relative motion wave functions $\chi_{\nu}^{J^\pi T}(x,y)$ depend on two relative coordinates.
In order to solve these equations, it is convenient to perform a
transformation to the hyperspherical hamornics (HH) basis. 
This basis has the great advantage that its elements are eigenfunctions of the hyper-angular part
of the relative kinetic 
operator when written in hyperspherical coordinates \cite{delaripelle83}. The hyperspherical coordinates 
(hyperradius $\rho$ and hyperangle $\alpha$) can 
be defined in terms of the relative coordinates of Eqs. (\ref{eq:etay}) and (\ref{eq:etax}) as:
\begin{align}
        \eta_{23} = \rho_\eta \sin\alpha_\eta\,, & &x=\rho\sin\alpha \,, \\
        \eta_{1,23} = \rho_\eta \cos\alpha_\eta\,, & &y=\rho\cos\alpha \,.
\end{align}
The continuous amplitudes $\chi_{\nu}^{J^\pi T}(x,y)$ can then be expanded in HH functions as

\begin{equation}
\chi_{\nu}^{J^{\pi}T}(\rho,\alpha)=\frac{1}{\rho^{5/2}}\sum_K u^{J^{\pi}T}_{K\nu}(\rho)
\phi_K^{\ell_x,\ell_y}(\alpha)\,
\label{expansionHH}
\end{equation}
where the basis elements are 

\begin{equation}
\phi_K^{\ell_x,\ell_y}(\alpha)=N_K^{\ell_x\ell_y}(\sin\alpha)^{\ell_x} (\cos \alpha)^{\ell_y}
P_n^{\ell_x+\frac{1}{2},\ell_y+\frac{1}{2}}(\cos 2\alpha), 
\label{eq:phi}
\end{equation}
here $P_n^{\alpha,\beta}(\xi)$ are Jacobi polynomials, and $N_K^{\ell_x\ell_y}$ normalization constants.

When projecting Eqs. (\ref{3B_ortho}) over the HH basis and integrating on the hyperangle $\alpha$,
this expansion allows to reduce those equations to a set of nonlocal integral-differential equations 
in the hyperradial coordinate:

\begin{align}
        \sum_{K\nu}\int d\rho \rho^5 \bar{\cal H}_{\nu'\nu}^{K'K}(\rho',\rho) \frac{u^{J^{\pi}T}_{K\nu}(\rho)}{\rho^{5/2}} 
        = E \frac{u^{J^{\pi}T}_{K^\prime\nu^\prime}(\rho^\prime)}{\rho^{\prime\,5/2}}\,,
        \label{RGMrho}
\end{align}
which is the three-cluster analogous of Eq. (\ref{RGMorteq}).

For details on the method adopted to solve the above set of equations, we
refer the interested reader to Sec. \ref{r_matrix}. 

At the moment, the method has been implemented exclusively for systems in which two of the clusters are single
nucleons. In this case, the calculation of the integration kernels (\ref{eq:Hkernel}) 
is performed, up to a great extent,
through the same expressions as when studying a binary system with a two-nucleon projectile, which  
can be found in Ref. \cite{Navratil2011}. However, it is important to note that 
in the case of the two-body projectile formalism
the interaction between those nucleons has been already taken into account
 through NCSM eigenstates of the projectile.  
Therefore, when considering such nucleons as different clusters, 
one has to additionally account for the intrinsic two-nucleon
hamiltonian $\hat H (x)$. In particular,
the specific expression for the kernel
produced by the interaction between the nucleons $\hat V (x)$ 
can be found in Eq. (39) of Ref. \cite{Quaglioni:2013kma}. 
In the absence of Coulomb interaction between the two nucleons, 
such term is localized in the $x$ coordinate, but not in the $y$ variables
where a Dirac's $\delta$ appear. 
For computational purposes, this $\delta$ is approximated by
an extended-size expansion in HO
 radial wave functions that goes well beyond the adopted HO model space ($N_{\rm ext}>>N_{\rm max}$). 
The convergence of the results with respect to the newly introduced parameter $N_{\rm ext}$ is 
discussed in Sec. \ref{3B_conv} in order
to determine the effect of such approximation.       

%% file: SE_NCSMC.tex
\subsection{NCSMC}
\label{sec:ncsmc}
The unified description of structure and reaction properties of an $A$-nucleon system is most efficiently obtained working within a generalized model space spanned by fully antisymmetric $A$-body basis states including both square-integrable wave functions, $\ket{A \lambda J^\pi T}$, 
and continuous RGM binary-cluster (and/or multi-cluster, depending on the particle-emission channels characterizing the nucleus in consideration) channel states, $\hat{\mathcal{A}}_\nu\ket{\Phi_{\nu r}^{J^\pi T}}$, 
of angular momentum $J$, parity $\pi$ and isospin $T$:
\begin{align}
\label{NCSMC_wav}
\ket{\Psi^{J^\pi T}_A} & =  \sum_\lambda c^{J^\pi T}_\lambda \ket{A \lambda J^\pi T}  + \sum_{\nu} \int dr \, r^2 \;
                               \frac{\gamma_{\nu}^{J^\pi T}(r)}{r}
                               \;\hat{\mathcal{A}}_\nu\ket{\Phi_{\nu r}^{J^\pi T}}.
\end{align}
%
When the compound, target and projectile wave functions are eigenstates of their respective intrinsic Hamiltonians computed within the NCSM [see Eqs.~\eqref{NCSM_wav} and \eqref{NCSM_eq} of Sec.~\ref{sec:eigenstates} and Sec.~\ref{sec:rgm2}], this approach is known as no-core shell model with continuum (NCSMC)  and the 
unknown discrete,  $c^{J^\pi T}_\lambda$, and continuous, $\gamma_\nu^{J^\pi T}(r)$ 
linear variational amplitudes can be simultaneously obtained by solving the set of coupled equations,
\begin{align}
&\left(
\begin{array}{cc}
        {\mathbb E} & \bar{h}\\  
	   \bar{h}  &\overline{\mathcal{H}} 
\end{array}
\right)
\left(
\begin{array}{c}
	c \\ \chi
\end{array}
\right)  =   E \left(
\begin{array}{cc}
        {\mathbb I}&\bar{g} \\  
        \bar{g}  &{\mathcal I} 
\end{array}
\right) \left(
\begin{array}{c}
c\\
\chi
\end{array}
\right)\,,
\label{eq:NCSMC-eq}
\end{align}
%
where $ \chi^{J^\pi T}(r)$ are the relative motion wave functions in the NCSM/RGM sector when working within the orthogonalized cluster channel states of Eq.~\eqref{eq:formalism_70}. The two by two block-matrices on the left- and right-hand side of Eq.~\eqref{eq:NCSMC-eq} represent, respectively, the NCSMC Hamiltonian and norm (or overlap) 
kernels.  The upper diagonal blocks are given by the Hamiltonian (overlap) matrix elements over the square-integrable part of the basis. In particular, as the basis states are NCSM eigenstates of the $A$-nucleon Hamiltonian, these are trivially given by the diagonal matrix ${\mathbb E}_{\lambda\lambda^\prime} = E_\lambda \delta_{\lambda\lambda^\prime}$ of the eigenergies (the identity matrix ${\mathbb I}_{\lambda\lambda^\prime}= \delta_{\lambda\lambda^\prime}$). Similarly, those over the 
continuous portion of the basis appear in the lower diagonal blocks and are given by the orthogonalized, Hermitized Hamiltonian kernel of Eq.~\eqref{eq:formalism_90} and 
${\mathcal I}_{\nu\nu^\prime}(r,r^\prime) = \delta_{\nu\nu^\prime}\delta(r-r^\prime)/r r^\prime$.
The off-diagonal blocks contain the couplings between the two sectors of the basis, with 
\begin{align}
\bar{g}_{\lambda \nu}(r)= \sum_{\nu^\prime}\int dr^\prime\, r^{\prime\,2} \bra{A\, \lambda J^\pi T} {\mathcal A}_{\nu^\prime}\ket{\Phi^{J^\pi T}_{\nu^\prime r^\prime}} \, {\mathcal N}^{-1/2}_{\nu^\prime\nu}(r^\prime,r)
\label{g-bar}
\end{align}
the cluster form factor, and the coupling form factor analogously given by 
\begin{align}
\bar{h}_{\lambda \nu}(r)= \sum_{\nu^\prime}\int dr^\prime\, r^{\prime\,2} \bra{A\, \lambda J^\pi T} \hat H {\mathcal A}_{\nu^\prime}\ket{\Phi^{J^\pi T}_{\nu^\prime r^\prime}} \, {\mathcal N}^{-1/2}_{\nu^\prime\nu}(r^\prime,r)\,.
\end{align}
As for the RGM kernels in Secs.~\ref{kernels} and \ref{sec:examples} these form factors can be computed using the second-quantization formalism. Their algebraic expressions can be found in Refs.~\cite{Baroni2013,Baroni2013a}.
Similar to the NSCM/RGM in Sec.~\ref{sec:rgm3}, the NCSMC formalism presented here can also be generalized for the description of three-cluster dynamics. A detailed presentation of such formalism will be given in Ref.~\cite{Romero-Redondo2016}.

\subsubsection{Orthogonalization of the NCSMC equations}
\label{orthog-ncsmc}
The NCSMC equations can be orthogonalized in an analogous way to that presented in Sec.~\ref{orthog} for the binary-cluster NCSM/RGM. To define the square and inverse square roots of the NCSMC norm in $r$-space representation, we first rewrite the two-by-two matrix on the left-hand-side of Eq.~(\ref{eq:NCSMC-eq})
 as the convolution of a NCSMC model-space norm kernel,
\begin{align}
N^{\tilde\lambda\tilde\lambda^\prime}_{\tilde\nu n\,\tilde\nu^\prime n^\prime }  \equiv
\left(
     \begin{array}{cc}
        \delta_{\tilde\lambda \tilde\lambda^\prime} & \bar{g}_{\tilde\lambda \tilde\nu^\prime n^\prime} \\[2mm]
        \bar{g}_{\tilde\lambda^\prime \tilde\nu n} & \delta_{\tilde\nu \tilde\nu^\prime}\delta_{nn'}
     \end{array}
 \right)\,,
\end{align}
plus a correction owing to the finite size of  the HO model-space, i.e.:
\begin{align}\label{eq:formalism_170}
             N^{\lambda\lambda'}_{\nu r \nu' r'} 
  &\equiv 
\left(
\begin{array}{cc}
        {\mathbb I}_{\lambda\lambda^\prime}&\bar{g}_{\lambda\nu^\prime}(r^\prime) \\  
        \bar{g}_{\lambda^\prime\nu}(r)  &{\mathcal I}_{\nu\nu^\prime}(r,r^\prime) 
\end{array}
\right)
\nonumber\\
& =
  \left(
     \begin{array}{ccc}
        0 & &0 \\
        0  & & \delta_{\nu\nu'}\frac{\delta(r-r')}{rr'} - \delta_{\nu\nu'} R_{n\ell}(r) \delta_{nn'} R_{n'\ell'}(r')
     \end{array}
  \right) 
  \nonumber\\
 & +
  \left(
     \begin{array}{cc}
        \delta_{\lambda \tilde\lambda} & 0 \\
        0  & R_{\nu r \tilde\nu n}
     \end{array}
  \right)
  N^{\tilde\lambda\tilde\lambda^\prime}_{\tilde\nu n\,\tilde\nu^\prime n^\prime } 
   \left(
     \begin{array}{cc}
        \delta_{\tilde\lambda^\prime \lambda^\prime} & 0 \\
        0  & R_{\nu^\prime r^\prime \tilde\nu^\prime n^\prime}
     \end{array}
  \right)\,,
\end{align}
Here, $R_{\nu r \tilde\nu n} = R_{n\ell}(r)\delta_{\nu\tilde\nu}$, the model-space cluster form factor is related to the $r$-space one through $\bar{g}_{\lambda \nu}(r)=\sum_n R_{nl}(r)\bar{g}_{\lambda \nu n}$, and the sum over the repeating indexes $\tilde\lambda, \tilde\nu, n, \tilde\lambda^\prime, \tilde\nu^\prime$, and $n^\prime$ is implied.  
The square and inverse square roots of $N$ can then be defined as:
\begin{align}\label{eq:formalism_180}
             (N^{\pm \frac{1}{2}})^{\lambda\lambda'}_{\nu r \nu' r'}
& \equiv
\left(
\begin{array}{cc}
        (N^{\pm \frac{1}{2}})^{\lambda\lambda^\prime}&(N^{\pm \frac{1}{2}})^{\lambda}_{\nu^\prime}(r^\prime) \\  
        (N^{\pm \frac{1}{2}})^{\lambda^\prime}_{\nu}(r)  &(N^{\pm \frac{1}{2}})_{\nu \nu^\prime}(r,r^\prime) 
\end{array}
\right)
\nonumber\\
  & =  
  \left(
     \begin{array}{ccc}
        0 &  &0 \\
        0  &  & \delta_{\nu\nu'}\frac{\delta(r-r')}{rr'} - \delta_{\nu\nu'} R_{n\ell}(r) \delta_{nn'} R_{n'\ell'}(r')
     \end{array}
  \right)
   \nonumber\\ 
   &+ \left(
     \begin{array}{cc}
        \delta_{\lambda\tilde\lambda} & 0 \\
        0  & R_{\nu r \tilde\nu n}
     \end{array}
  \right) 
(N^{\pm\frac12})^{\tilde\lambda\tilde\lambda^\prime}_{\tilde\nu n\, \tilde\nu^\prime n^\prime}
 \left(
     \begin{array}{cc}
             \delta_{\tilde\lambda^\prime \lambda^\prime} & 0 \\
                     0  & R_{\nu^\prime r^\prime \tilde\nu^\prime n^\prime}
     \end{array}
  \right).
\end{align}
These expressions can be easily generalized to the case in which expansion (\ref{NCSMC_wav}) contains RGM components of the three-cluster type.
In general, inserting the identity $N^{-\frac{1}{2}}N^{+\frac{1}{2}}$ in both left- and right-hand sides of Eq.~(\ref{eq:NCSMC-eq}), and multiplying
by $N^{-\frac{1}{2}}$ from the left, one finally obtains,
\begin{equation}\label{eq:formalism_200}
  \overline{H}
  \left(
     \begin{array}{c}
          \bar{c} \\
		  \bar\chi
     \end{array}
  \right)
  = 
  E
  \left(
     \begin{array}{c}
          \bar{c} \\ 
		  \bar\chi
     \end{array}
  \right)\,,
\end{equation}
where the orthogonalized NCSMC Hamiltonian $\overline{H}$ is given by,
\begin{equation}\label{eq:formalism_210}
  \overline{H} =
  N^{-\frac{1}{2}} 
\left(
\begin{array}{cc}
        {\mathbb E} & \bar{h}\\  
	   \bar{h}  &\overline{\mathcal{H}} 
\end{array}
\right)
  N^{-\frac{1}{2}},
\end{equation}
and the orthogonal wave functions by
\begin{equation}\label{eq:formalism_220}
  \left(
     \begin{array}{c}
          \bar{c} \\
		  \bar\chi
     \end{array}
  \right)
  = 
  N^{+\frac{1}{2}} 
  \left(
     \begin{array}{c}
          c \\ 
		  \chi
     \end{array}
  \right).
\end{equation}
Finally, starting from Eq.~(\ref{NCSMC_wav}) the orthogonalized NCSMC wave function takes then the general form:
%
\begin{align}\label{NCSMC_wave_orth}
\ket{\Psi^{J^\pi T}_A} 
= & \sum_\lambda \ket{A \lambda J^\pi T} \left[\sum_{\lambda^\prime} (N^{-\frac{1}{2}})^{\lambda\lambda'} \bar{c}_{\lambda^\prime}
+\sum_{\nu^\prime} \int  dr^\prime\, r^{\prime\,2} \; (N^{-\frac{1}{2}})^{\lambda}_{\nu^\prime }(r^\prime)\; \bar\chi_{\nu^\prime}(r^\prime)\right]
\nonumber \\
&+ \sum_{\nu\nu^\prime} \int d r\, r^2  \int d r^\prime\, r^{\prime\,2}\;  \hat{\mathcal{A}}_\nu\ket{\Phi_{\nu r}^{J^\pi T}} \;
\mathcal{N}_{\nu \nu'}^{-\frac{1}{2}}(r,r') 
 \\
&\times 
\left[\sum_{\lambda^\prime} (N^{-\frac{1}{2}})^{\lambda'}_{\nu'}( r')\; \bar{c}_{\lambda'} 
+ \sum_{\nu''}  \int  d r'' r^{\prime\prime\, 2} (N^{-\frac{1}{2}})_{\nu'  \nu'' }(r',r'')\; \bar\chi_{\nu''}(r'')\right].\nonumber
\end{align}
%

%% file: SE_R-matrix.tex
\subsection{Microscopic R-matrix approach}
\label{r_matrix}

Within the RGM, solving the integro-differential Eqs. (\ref{RGMorteq}) or (\ref{3B_ortho}),
for binary or ternary clusters respectively,  
 provides the continuum 
coefficients of the cluster expansion, i.e., the relative motion wave functions.       
In order to solve these equations, we use the coupled-channel $R$-matrix method on a 
Lagrange mesh \cite{Hesse1998,Hesse2002}. 
The formalism for binary and ternary clusters is completely analogous and
therefore, here we present details only for the binary cluster case. Details of
the generalization for ternary clusters can be found on \cite{Quaglioni:2013kma}.

The configuration space is first divided in two regions delimited by a 
matching radius $r=a$.  
In the internal region, the complete internuclear interaction is considered
 and the radial wave function is expanded
on a Lagrange basis.

In the external region, only the Coulomb interaction 
is assumed to be relevant. Therefore, in this region the wave function 
is approximated by its known asymptotic form, which is proportional to the 
Whittaker functions $W_{\ell}(\eta_{\nu},\kappa_{\nu}r)$ for bound states,

\begin{equation}
u^{J^{\pi}T}_{\nu,ext}(r)=C^{J^{\pi}T}_{\nu}W_{\ell}(\eta_{\nu},\kappa_{\nu}r)  
\end{equation}
with $C^{J^{\pi}T}$ being the asymptotic normalization constant.
It can be written in terms of incoming and outgoing functions $H^{\pm}(\eta_{\nu},\kappa_{\nu}r)$ 
and the Scattering matrix $S^{J^{\pi}T}_{\nu i}$ when studying 
continuum states:

\begin{equation}
\label{cont_asym}
u^{J^{\pi}T}_{\nu,ext}(r)=\frac{i}{2}v_{\nu}^{-\frac{1}{2}}\left[\delta_{\nu i}
H^-_{\ell}(\eta_{\nu},\kappa_{\nu}r)-
S^{J^{\pi}T}_{\nu i}H^+_{\ell}(\eta_{\nu},\kappa_{\nu}r)\right].  
\end{equation}
with $v_{\nu}$ the speed, $\kappa_{\nu}$ the wave number, and $\eta_{\nu}$ the 
Sommerfield parameter of the final state. Labels $i$ and $\nu$ refer to the initial and final state,
respectively.  
The $R$-matrix formalism is expressed in terms of the Bloch-Schr\"odinger equations:
\begin{eqnarray}
\nonumber 
&\sum_{\nu}
\int dr r^2 
\Bigg(\bar{\cal H}_{\nu'\nu}(r',r) 
+\hat{L}_{\nu}(r)  
-E\frac{\delta(r-r')}{r^2}\delta_{\nu'\nu}\Bigg) 
 \frac{u_{\nu,int}^{J^{\pi}T}(r)}{r}  \\ 
&=\sum_{\nu}
\int dr r^2 \hat{L}_{\nu}(r)\frac{u_{\nu,ext}^{J^{\pi}T}(r)}{r}
\label{bloch-schr}
\end{eqnarray}
where the Bloch operator, which has the dual function of restoring the Hermiticity of the Hamiltonian
in the internal region and enforcing a continuous derivative at the matching radius \cite{Descouvemont2010}, 
is defined as 
\begin{equation}
\label{bloch:binary}
\hat{L}_{\nu}(r)=\frac{\hbar}{2\mu_{\nu}}\delta(r-a)\left(\frac{d}{dr}-\frac{B_{\nu}}{r} \right) 
\end{equation}
here the constants $B_{\nu}$ are arbitrary and therefore can be
chosen to facilitate the solution of the equations (\ref{bloch-schr}).  
When calculating a bound state 
the constants $B_{\nu}$ are chosen as the logarithmic derivative of $u^{J^{\pi}T}_{\nu,ext}(r)$
evaluated in the matching radius. This election cancels the right hand side of Eq. (\ref{bloch-schr})  
and gives rise to an eigenvalue problem that can be solved iteratively starting from $B_{\nu}=0$ 
(convergence is typically reached within a few iterations). When calculating a continuum
state the constants $B_{\nu}$ are chosen to be zero, and the scattering matrix is obtained
through the calculation of the $R$-matrix. The details of this procedure can be found in \cite{Hesse1998}.      

The solution of Eq. (\ref{bloch-schr}), is conveniently achieved on a Lagrange mesh by 
projecting the equations on a set of N Lagrange functions.
Due to its properties, the Lagrange basis provides 
straightforward expressions for the matrix elements of the relative kinetic operator, the
Bloch operator and the non-local Hamiltonian kernel. 
The basis is defined as a set of $N$ functions $f_n(x)$ 
(see \cite{Hesse1998} and references therein), given by
\begin{equation}
f_n(x)=(-1)^na^{-1/2}\sqrt{\frac{1-x_n}{x_n}}\frac{xP_N(2x/a-1)}{x-ax_n}\,,
\end{equation}
where $P_N(x)$ are Legendre polynomials, and $x_n$ satisfy
\begin{equation}
P_N(2x_n-1)=0\,.
\end{equation}
The Lagrange mesh associated with this basis consists of $N$ points $ax_n$ 
on the interval $[0,a]$ and satisfies the Lagrange condition 
\begin{equation}
f_{n'}(ax_n)=\frac{1}{\sqrt{a\lambda_n}}\delta_{nn'},
\label{LagCond}
\end{equation}
where the coefficients $\lambda_n$ are the weights corresponding to a
Gauss-Legendre quadrature approximation for the $[0,1]$ interval, i.e.
\begin{equation}
\int_0^1g(x)dx \sim \sum_{n=1}^N \lambda_ng(x_n)\,.
\end{equation}
Using the Lagrange conditions of Eq.~(\ref{LagCond}), it is straightforward to see that
within the Gauss approximation the Lagrange functions are orthogonal, i.e.
\begin{equation}
\int_0^a f_n(x) f_{n'}(x)dx\sim \delta_{nn'}.
\end{equation}

\subsubsection{Using the R-matrix method within the NCSMC.}

When working within the NCSMC formalism, the wave function is described by a short/mid range 
contribution from the discrete NCSM basis and a long range contribution
coming from the cluster basis. Due to the fact that at the matching radius 
only the long range behavior is present,  
the only relevant component of the wave function comes from the 
cluster basis states. Therefore, it is possible to perform the matching between internal and
external regions using only this contribution.

This is accomplished by defining a matrix Bloch surface operator in which only the right-bottom block
is non-zero
\begin{equation} 
\hat{L}_{\nu}= \begin{pmatrix}  
0 & 0\\ 
0 & \hat{L}_{\nu}(r)
\end{pmatrix}
\end{equation}     
where the operator 
  $\hat{L}_{\nu}(r)$ is given by Eq. (\ref{bloch:binary}) 
for binary clusters. The Bloch-Schr\"odinger 
equations can be written as

\begin{equation}
(\bar{H}+\hat{L}-E)\begin{pmatrix} \bar{c}\\ \bar{\chi}\end{pmatrix}=\hat{L}
\begin{pmatrix} \bar{c}\\ \bar{\chi}\end{pmatrix}   
\end{equation}
and their solution is obtained in an analogous way as within the NCSM/RGM approach.      

%% file: SE_convergence.tex
\subsection{Convergence properties}
\label{sec:cv}

In this section we give an overview of the convergence properties of the NCSM/RGM (for both binary and ternary scattering processes) and NCSMC (for binary reactions) approaches with respect to all relevant parameters characterizing their model spaces. In particular, as an example in Sec.~\ref{sec:binconv} we present an analysis performed on the $N$+$^4$He and $d$+$^4$He scattering phase shifts obtained with the SRG-evolved chiral $NN$+$3N$-full Hamiltonian with resolution scale $\Lambda{=}2.0$ fm$^{-1}$ by solving the NCSM/RGM equations~(\ref{eq:formalism_90}) and NCSMC equations~\eqref{eq:formalism_200} with scattering boundary conditions by means of the microscopic $R$-matrix method on Lagrange mesh outlined in Sec.~\ref{r_matrix}.  Because of the additional expansion on HH basis states, the treatment of three-cluster dynamics introduces a new set of parameters controlling the behavior of the results. For this reason, as an example in Sec.~\ref{3B_conv} we also show results for the bound and scattering states of the $^4$He+$n$+$n$ system obtained by solving the NCSM/RGM equations~\eqref{3B_ortho} with bound and scattering boundary conditions starting from the SRG N$^3$LO $NN$-only interaction evolved to $\Lambda=1.5$ fm$^{-1}$.

\subsubsection{Binary clusters.}
\label{sec:binconv}
For binary reactions, a channel radius of $a=18\,\rm fm$ is typically large enough for the clusters to interact only through the Coulomb force and about $N_s=36$ mesh points (roughly 2 mesh points per fm) are usually sufficient to obtain convergence with respect to the Lagrange expansion. In this section we will present results obtained with these Lagrange parameters and concentrate on the remaining convergence properties of the scattering calculation.  
\begin{figure}[t]
\includegraphics[width=75mm]{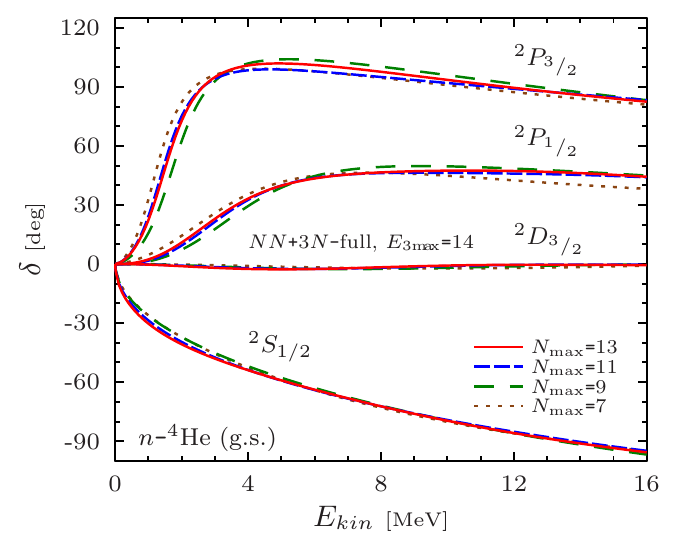}
\
\includegraphics[width=75mm]{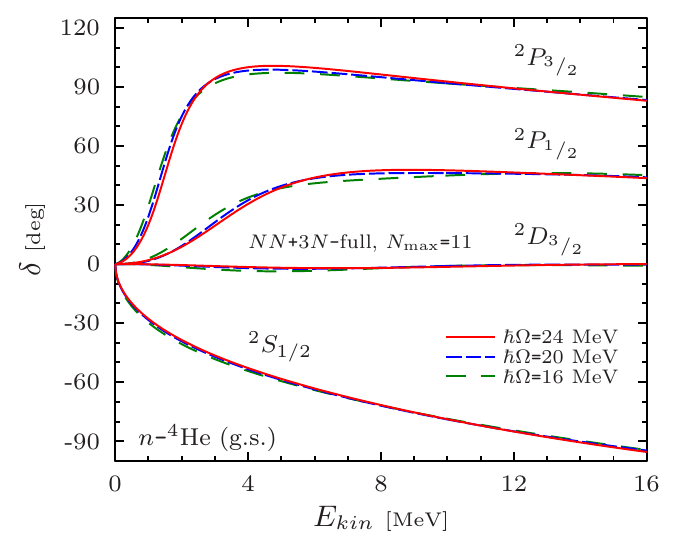}
\caption{Left panel: Convergence of the $n$-$^4$He $^1S_{1/2}$, $^2P_{1/2}$, $^2P_{3/2}$, and $^2D_{3/2}$ phase shifts with respect to the model-space size $N_{\rm max}$ at $\hbar\Omega{=}20$ MeV. Brown dotted lines, green long-dashed lines, blue dashed lines and red solid lines correspond to $N_{\text{max}} = 7,9,11$ and 13, respectively. Right panel: Dependence of the $n$-$^4$He phase shifts on the HO frequency. Green long-dashed lines, blue dashed lines and red solid lines correspond to $\hbar\Omega= 16, 20$ and 24, respectively. The model space is truncated at $N_{\rm max}=11$. In both panels all curves were obtained including the g.s.\ of $^4$He and employing the SRG-evolved chiral $NN$+$3N$-full interaction with $\Lambda=2.0$ fm$^{-1}$.}
\label{fig:Nmax-convergence}
\end{figure}

In both NCSM/RGM and NCSMC methods, the NCSM eigenstates of the target and projectile (and, for the second, also of the compound nucleus) are essential building blocks. Naturally the properties of the these eigenstates propagate to the NCSM/RGM and NCSMC calculations and we find the HO frequency $\Omega$ and the maximum number of major HO shell $N_{\rm max}$ (of the HO basis used to expand the NCSM wave functions of the clusters and the localized parts of the NCSM/RGM kernels) to be parameters regulating the behavior of the results. Therefore, we begin our overview by presenting in Fig.~\ref{fig:Nmax-convergence} the dependence of NCSM/RGM calculations of the $n$-$^4$He phase shifts with respect to $N_{\rm max}$ and $\hbar\Omega$. The left panel presents single channel calculations for $N_{\rm max}{=}7,9,11$, and 13 carried out using $n$-$^4$He channel states with the $^4$He in its ground state (g.s.). The phase shifts for the first four partial waves exhibit a good convergence behavior. With the exception of the $^2P_{3/2}$ resonance, where we can observe a difference of less than 5 deg in the energy region $4\le E_{kin}\le 10$ MeV, the $ N_{\rm max}{=}11$ and $ N_{\rm max}{=}13$  phase shifts are on top of each other. An analogous behavior is obtained when using the $NN$+$3N$-induced interaction and  the N$^3$LO $NN$-only interaction evolved to the same $\Lambda$ value~\cite{Navratil2010}. In the latter case, working only with a two-body potential, we were able to obtain results for $N_{\rm max}$ values as high as 17, but no substantial differences were found with respect to those obtained for $N_{\max}{=}13$. Therefore, given the large scale of these computations, we study the sensitivity with respect to the HO frequency within an $N_{\rm max}=11$ model space. This is shown in the right panel of Fig.~\ref{fig:Nmax-convergence}, where we compare $\hbar\Omega{=}16$, $20$ and $24$ MeV results. We find essentially no dependence on the HO frequency when comparing the $\hbar\Omega=20$ and $24$ MeV, i.e., the phase shifts are in good agreement: the $^2 S_{1/2}$ and $^2 D_{3/2}$ phase shifts are on top of each other while the $^2 P_{1/2}$ and $^2 P_{3/2}$ phase shifts show very small deviations around the resonance positions. At the same time, we note that using the lower frequency of $\hbar\Omega=16\,\text{MeV}$ is problematic due to the finite size of the HO model space used during the SRG transformation. This has to be cured using a frequency conversion technique \cite{Roth2014}. 

\begin{figure}[b]
\begin{minipage}{0.48\columnwidth}
\includegraphics[width=75mm]{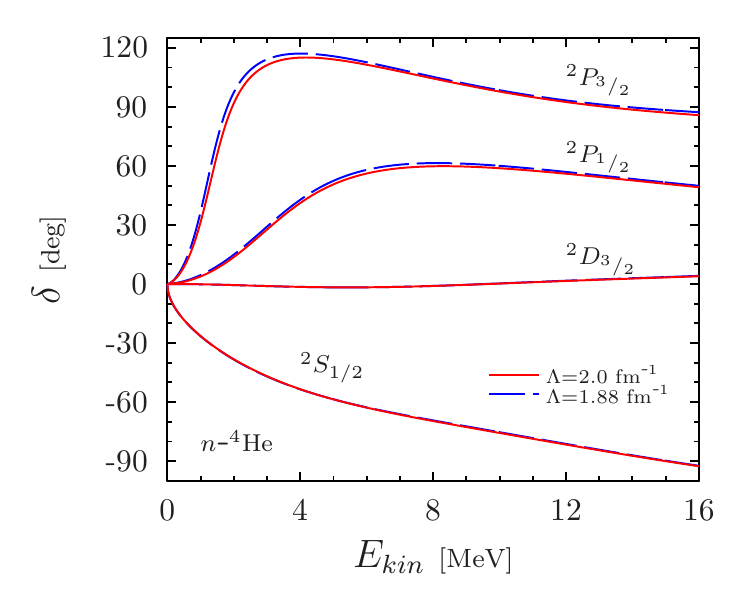}
\end{minipage}
\
\begin{minipage}{0.51\columnwidth}
\caption{Sensitivity of the $n$-$^4$He $^1S_{1/2}$, $^2P_{1/2}$, $^2P_{3/2}$, and $^2D_{3/2}$ phase shifts with respect to SRG scale $\Lambda$ using the SRG-evolved chiral $NN$+$3N$-full interaction and a model-space size $N_{\rm max}{=}13$ major HO shells at $\hbar\Omega{=}20$ MeV. Blue dashed lines and red solid lines correspond to $\Lambda = 1.88$ and $2.0$ fm$^{-1}$, respectively. Additionally, the first six excited state of $^4$He are included in the calculation.}
\label{fig:srg-convergence}
\end{minipage}
\end{figure}
As explained previously in Sec.~\ref{sec:srg}, employing an effective interaction obtained from the SRG method requires to account for induced effects, a priori of an $A$-body nature. This is impractical to achieve for larger $A$ values. Therefore the SRG scale $\Lambda$ became a parameter of the model and, as long as the reminding higher-order induced effects remain negligible we can, in the context of an {\em ab initio} framework, claim that renormalized results in a finite model space are almost unitary equivalent to their bare counterpart. Thus it is essential to study our reaction observables with respect to $\Lambda$. While the proper way to address this is to show that renormalized results are consistent with those of $\Lambda=\infty$, because of the finiteness of the model space together with the sensitivity of reaction observables to g.s. energies of the reactant nuclei, it can only be done within a restricted range of values. In figure~\ref{fig:srg-convergence}, we show a sensitivity study of the $n$-$^4$He phase shifts with respect to $\Lambda$. Only two values are displayed due to the computational difficulty of the calculation. We can see that the effects of the missing four- and five-body SRG induced interactions are small, of the order or smaller than the effects of the $N_{\rm max}$ and $\hbar \Omega$ parameters. In the corresponding study of Ref.~\cite{Hupin2013}, it was shown that the $3N$-induced interaction was essential to correct for the SRG scale dependence of the $NN$-only phase shifts for the same values of $\Lambda$.

\begin{figure}[t]
\includegraphics[width=76mm]{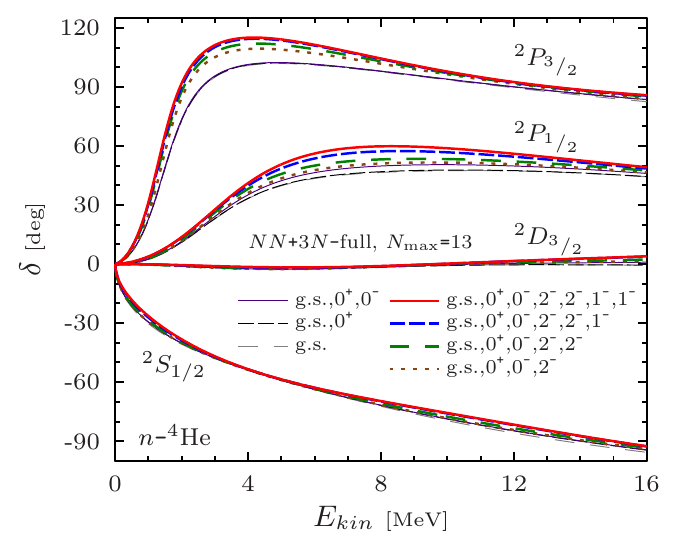}
\
\includegraphics[width=74mm]{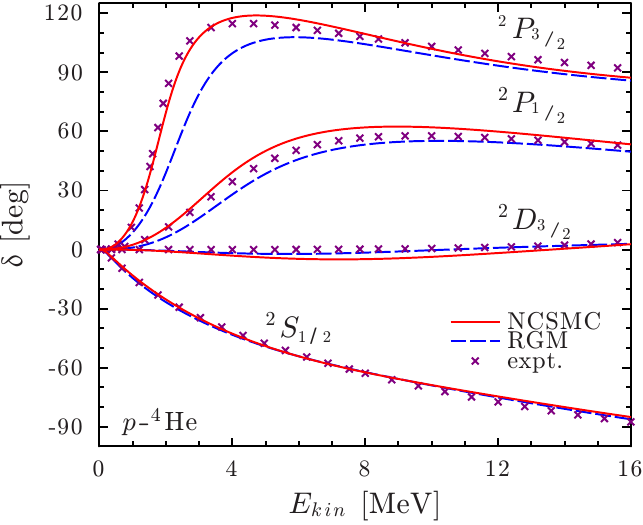}
\caption{Left panel: Dependence of the $n$-$^4$He phase shifts 
on the considered target eigenstates. Results with only the g.s.\ of $^4$He (thin gray long-dashed lines) are compared to those obtained by including in addition up to the 0$^+$0 (thin black dashed lines), 0$^-$0 (thin violet lines), 2$^-$0 (thick brown dotted lines), 2$^-$1 (thick green long-dashed lines), 1$^-$1 (thick blue dashed lines), and 1$^-$0 (thick red lines) excited states of \elem{He}{4}, respectively.  Right panel: Comparison between the phase shifts obtained using the NCSM/RGM (dashed blue lines) and NCSMC (continuous red lines) methods for the $^4$He($p$,$p$)$^4$He reaction. The $R$-matrix analysis of data from Ref.~\cite{Hale} is shown as guidance (purple crosses). In both calculations, the first five excited states of $^4$He are included and the chiral $NN$+$3N$ interaction SRG-evolved to a typical scale of $2$ fm$^{-1}$ is used. Additionally, in the NCSMC, all the influential NCSM eigenstates of $^5$Li are accounted for. In both panels the model space is truncated at $N_{\rm max}{=}13$. Other parameters are identical to those of the left panel of Fig.~\ref{fig:Nmax-convergence}.}
\label{fig:NCSMC_RGM}
\end{figure}
\begin{figure}[t]
\begin{minipage}{0.45\columnwidth}
\includegraphics[width=\columnwidth]{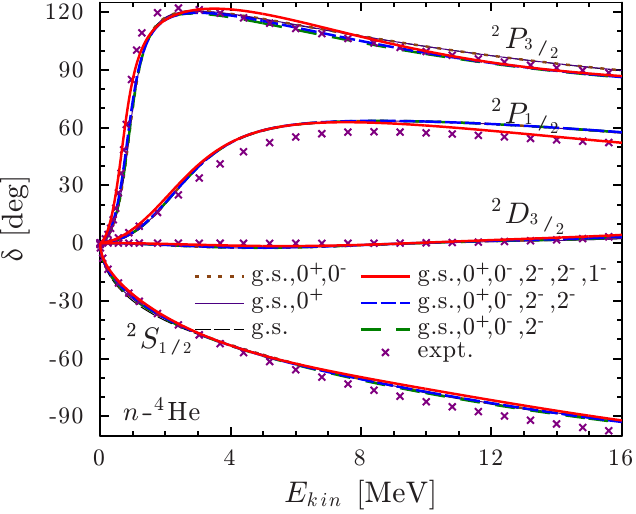}
\end{minipage}
\begin{minipage}{0.55\columnwidth}
\centering
\caption{Convergence of the $^4$He($n$,$n$)$^4$He phase shifts obtained with the NCSMC with respect to the number of excited states of $^4$He included in the calculation from g.s. (black dashed line) to the first five excited states (continuous red line). The $R$-matrix analysis of data from Ref.~\cite{Hale} is shown as guidance (purple crosses). Additional parameters of the computed phase shifts are identical to those of Fig~\ref{fig:NCSMC_RGM}.}\label{fig:NCSMC_nalpha_polarization}
\end{minipage}
\end{figure}

With the behavior with respect to the parameters $\Omega$, $N_{\rm max}$ and $\Lambda$ of our many-body space discussed above, we can now focus on the convergence properties of the phase shifts with respect to the number and types of NCSM/RGM cluster states. In the case of the $n$-$^4$He scattering phase shifts, this means investigating the dependence on the number of target eigenstates included in the calculation. This is shown in the left panel of Fig.~\ref{fig:NCSMC_RGM}, where we employ our largest model space of $N_{\rm max}{=}13$. As can be seen in the figure, the target excitations are crucial in particular for the resonant phase shifts, where they lead to an enhancement. Specifically, the resonance of the $^{2}P_{3/2}$ wave is strongly influenced by the inclusion of the $J^\pi T{=}2^-0$ state, while the $^{2}P_{1/2}$ phase shift is slowly enhanced near its resonance with the addition of the odd-parity states, among which the $1^-$ states have the strongest effect. On the other hand, the $^{2}S_{1/2}$ wave is Pauli blocked and mostly insensitive to these polarizations effects. However, despite the inclusion of up to the first seven eigenstates of the $^4$He, overall the convergence is clearly slow.  This is further demonstrated by the right panel of Fig.~\ref{fig:NCSMC_RGM}, comparing NCSM/RGM calculations (dashed blue lines) to the corresponding results obtained within the NCSMC (continuous red lines) by including the same number of target eigenstates. Here, the figure shows the phase shifts for a proton elastically colliding on a $^4$He target. We can see that the coupling to the discrete $^5$Li NCSM eigenstates is instrumental to reproduce the experimental phase shifts obtained from an $R$-matrix analysis of the data, in particular the position and width of the resonances. 
More in detail, the inclusion of the eigenstates of the compound nucleus tends to influence the resonances around the energy ($E_{kin}$) corresponding to their eigenvalues $E_\lambda$.
The more efficient simultaneous description of both short and medium-to-long range correlations, and hence faster convergence, obtained within the NCSMC approach ~\cite{Baroni2013} by augmenting the NCSM/RGM basis with NCSM $A$-body eigenstates of the compound (here $^5$He) system is further demonstrated by the plot of Fig.~\ref{fig:NCSMC_nalpha_polarization}. The convergence of the $n-^4$He scattering phase shits with respect to the number of target excitations included in the calculations is now excellent.
\begin{figure}[b]
\begin{minipage}{0.48\columnwidth}
\includegraphics[width=\columnwidth]{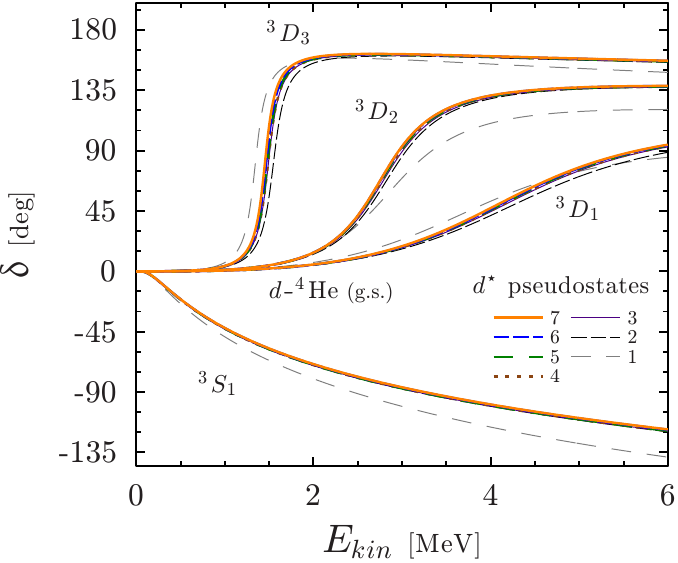}
\end{minipage}
\begin{minipage}{0.51\columnwidth}
\centering
\caption{Computed $S$- and $D$-wave $^4$He($d$,$d$)$^4$He phase shifts obtained with the NCSMC using fifteen $^6$Li eigenstates and up to seven $^2$H in each of three most influential channels ($^3S_1{-}{}^3D_1$, $^3D_2$ and $^3D_3{-}{}^3G_3$). Only the two-body part of the SRG-evolved chiral $NN$ interaction is used. The HO quanta maximum is $N_{\rm max}=8$ while other parameters are identical to those of Fig~\ref{fig:NCSMC_RGM}. (Figure adapted from Ref.~\cite{Hupin2015})}\label{fig:NCSMC_dalpha_CV}
\end{minipage}
\end{figure}

Similarly to this case, the distortion that arises from the projectile polarization can play a role in scattering calculations. This is illustrated in Fig.~\ref{fig:NCSMC_dalpha_CV} where the phase shifts of a deuterium impinging on a $^4$He target computed within the NCSMC approach are compared starting from the case where only the $^2$H g.s. is included in the model space, to the case in which up to seven deuterium pseudo states in each of the influential channels ($^3S_1{-}{}^3D_1$, $^3D_2$ and $^3D_3{-}{}^3G_3$) are considered. Stable results are found with as little as three deuteron states per channel. 
This is a strong reduction of the $d^\star$ influence with respect to the NCSM/RGM study of Ref.~\cite{Navratil2011}, lacking the coupling of square-integrable $^6$Li eigenstates.  
Still, although the convergence in term of pseudo states is fast, their effect cannot be entirely disregarded. This is due to the fact that the deuterium is only bound by $2.224$ MeV resulting in a very low breakup threshold for this reaction system. In turn, it means that the system transitions from a two- to a three-body nature. Thus the discretization of the deuterium continuum consists in an approximation above this threshold, where the appropriate three-body asymptotic should be accounted for. Part of these effects are absorbed in the coupling to the discrete $^6$Li eigenstates, but a true modeling of the system above $2.224$ MeV of excitation energy should include a ternary cluster basis.

\subsubsection{Ternary clusters.}
\label{3B_conv}
For ternary clusters, we present convergence properties when
performing calculations with the NCSM/RGM basis alone. (The implementation
of the NCSMC within ternary clusters and its convergence behavior will be presented 
elsewhere \cite{Romero-Redondo2016}.)   In this case, the asymptotic condition is reached for larger values of the matching (hyper) radius $a$. This can be understood from the fact that for a three-cluster motion, a large hyper radius can result from configurations in which two of the clusters are still relatively close to each other. In practice the value of $a$ and the number of radial mesh points $N_s$ needed to achieve convergence have to be carefully  chosen for each partial wave and tend to grow together with the value of $N_{\rm max}$ and $N_{\rm ext}$ (the parameter introduced to approximately describe the effect of the interaction between the two single-nucleon clusters, see Sec.~\ref{sec:rgm3}). Typically, $a$ ranges between $30$ and $45$ fm and one needs on the order of $60\le N_s\le 130$ radial mesh points to achieve convergence.   

\begin{figure*}[t]
    \begin{minipage}[c]{8.6cm}\hspace*{-6mm}
      \includegraphics[width=8.0cm,clip=,draft=false]{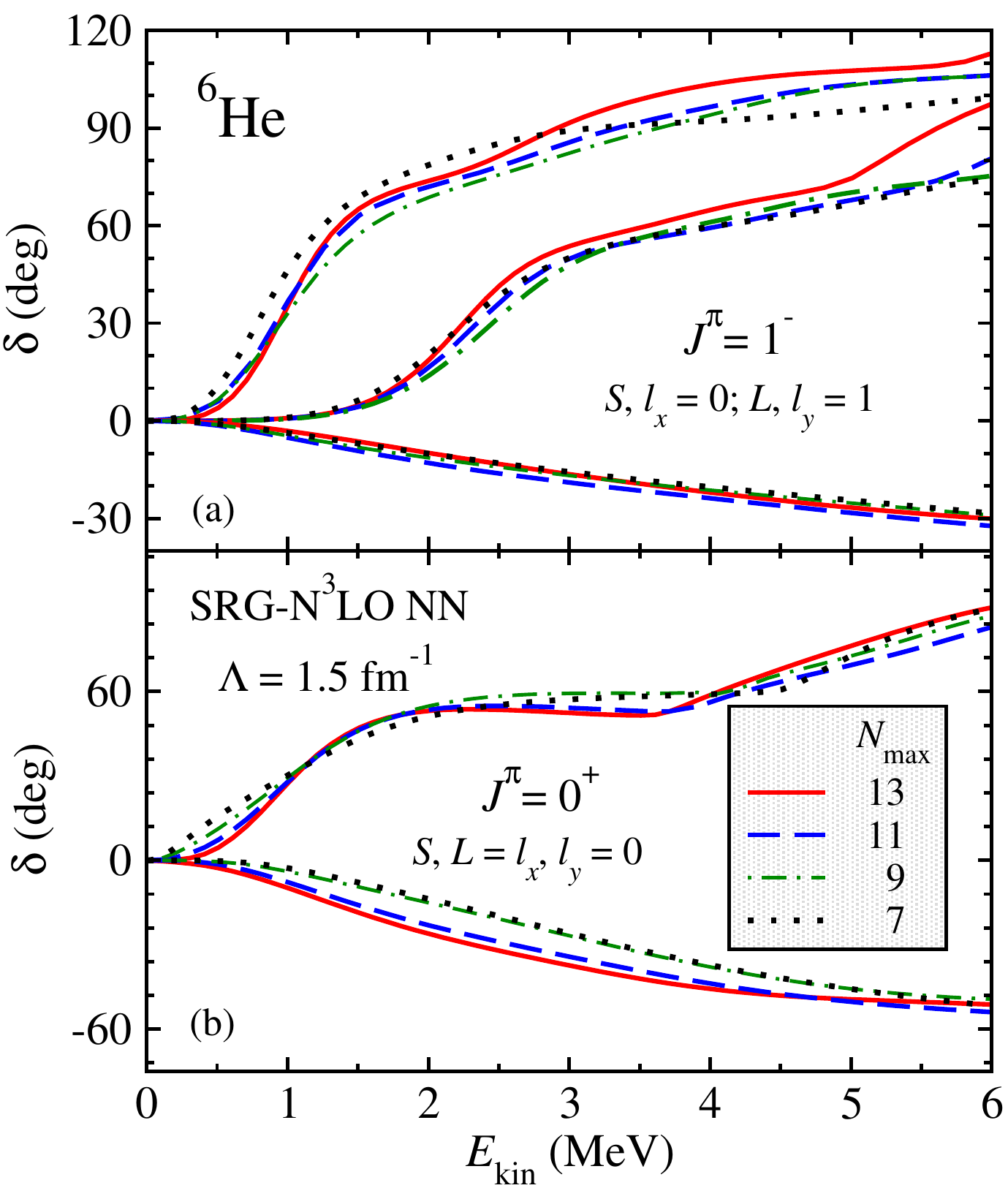}
    \end{minipage}
    \begin{minipage}[c]{8.6cm}\hspace*{-6mm}
      \includegraphics[width=8.0cm,clip=,draft=false]{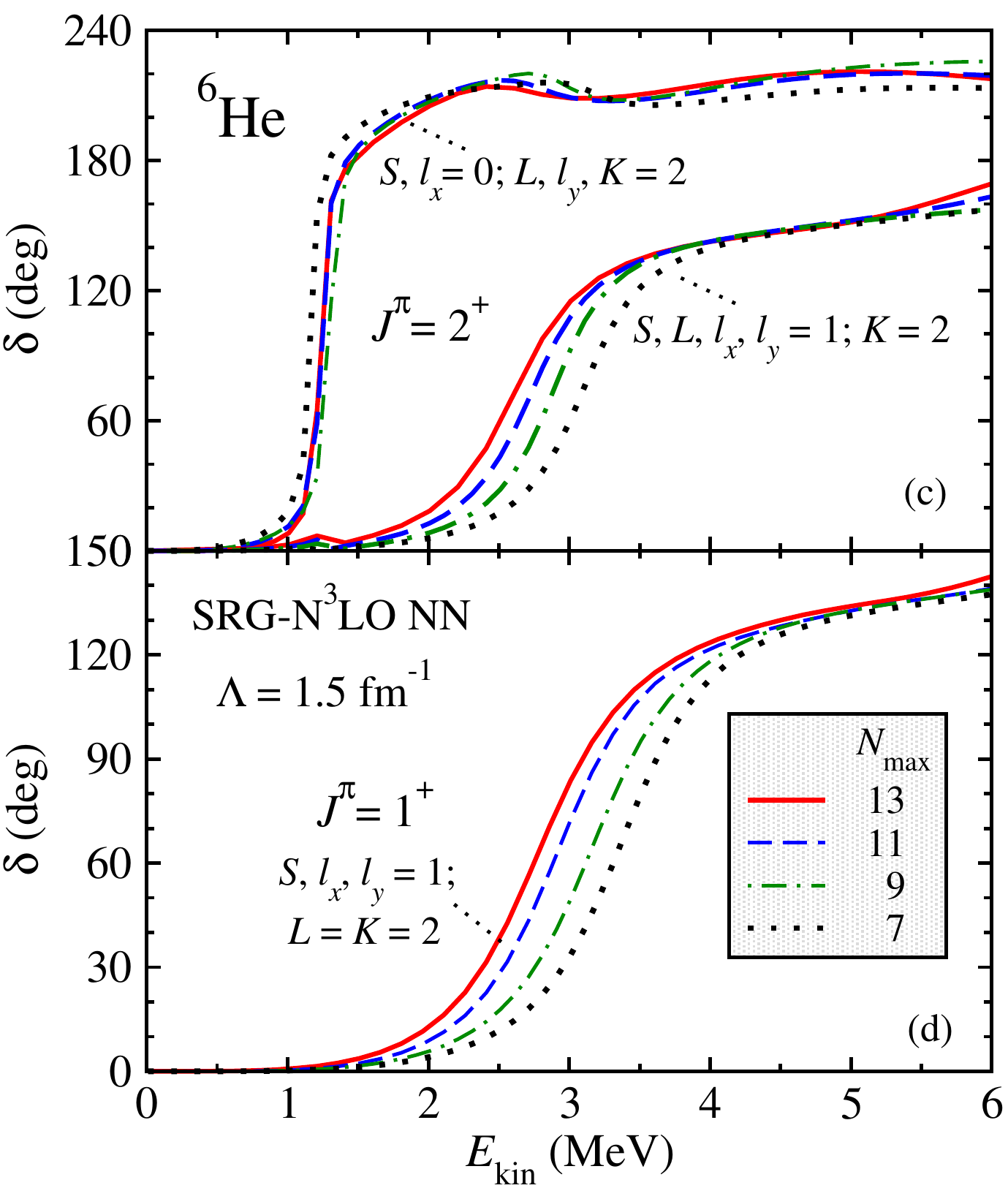}
    \end{minipage}
\caption{(Color online) Convergence behavior of calculated $^4{\rm He}+n+n$ (a) $J^\pi = 1^-$ and (b) $0^+$ eigenphase shifts at $K_{\rm max}=19$ and $28$, respectively, and (c) $2^+$ and (d) $1^+$ diagonal phase shifts at $K_{\rm max}=20$ with respect to the size $N_{\rm max}$ of the NCSM/RGM model space. For these calculations we used a matching radius of $a=30$ fm, $N=60$ Lagrange mesh points, and an extended HO model space of $N_{\rm ext}=70$.}
\label{fig:3b_ps_conv}
\end{figure*}

The convergence pattern with respect to the size of the HO
model space is similar to the one seen in the binary cluster case.
Although, when studying continuum states, 
the computational
challenge of these calculations prevents us from obtaining accurate 
quantitative results, the degree of convergence is sufficient to provide a
very good qualitative description of the continuum. As an example, in 
Fig. \ref{fig:3b_ps_conv}, we show convergence of the $^4{\rm He}+n+n$
eigenphase shifts for $J^\pi = 1^-, 0^+, 2^+$ and $1^+$ channels 
 with respect to the size of the model space $N_{\rm max}$. Here, it can be seen
that, despite the lack of complete convergence, the presence of resonances
is well determined as they clearly appear even at low $N_{\rm max}$. This is easily
seen in the right panel of the figure where two $2^+$ resonances and one $1^+$ are shown.  
An approximate position and width of those resonances can also be extracted.       

It is important to additionally study the convergence of the
results with respect to the parameters that appear exclusively when performing a three-cluster
calculation: the maximum hypermomentum $K_{\rm max}$ included in the hyperspherical
expansion (\ref{expansionHH}) and the extended model space used for the description
of the potential kernel ($N_{\rm ext}$). 
Further, one also has to consider the convergence with respect to the integration in the hyperangles, which we perform numerically using a Chebyshev-Gauss quadrature and is usually well under control using $N_\alpha=40$ angular mesh points. 

\begin{figure}[t]
\includegraphics[width=75mm]{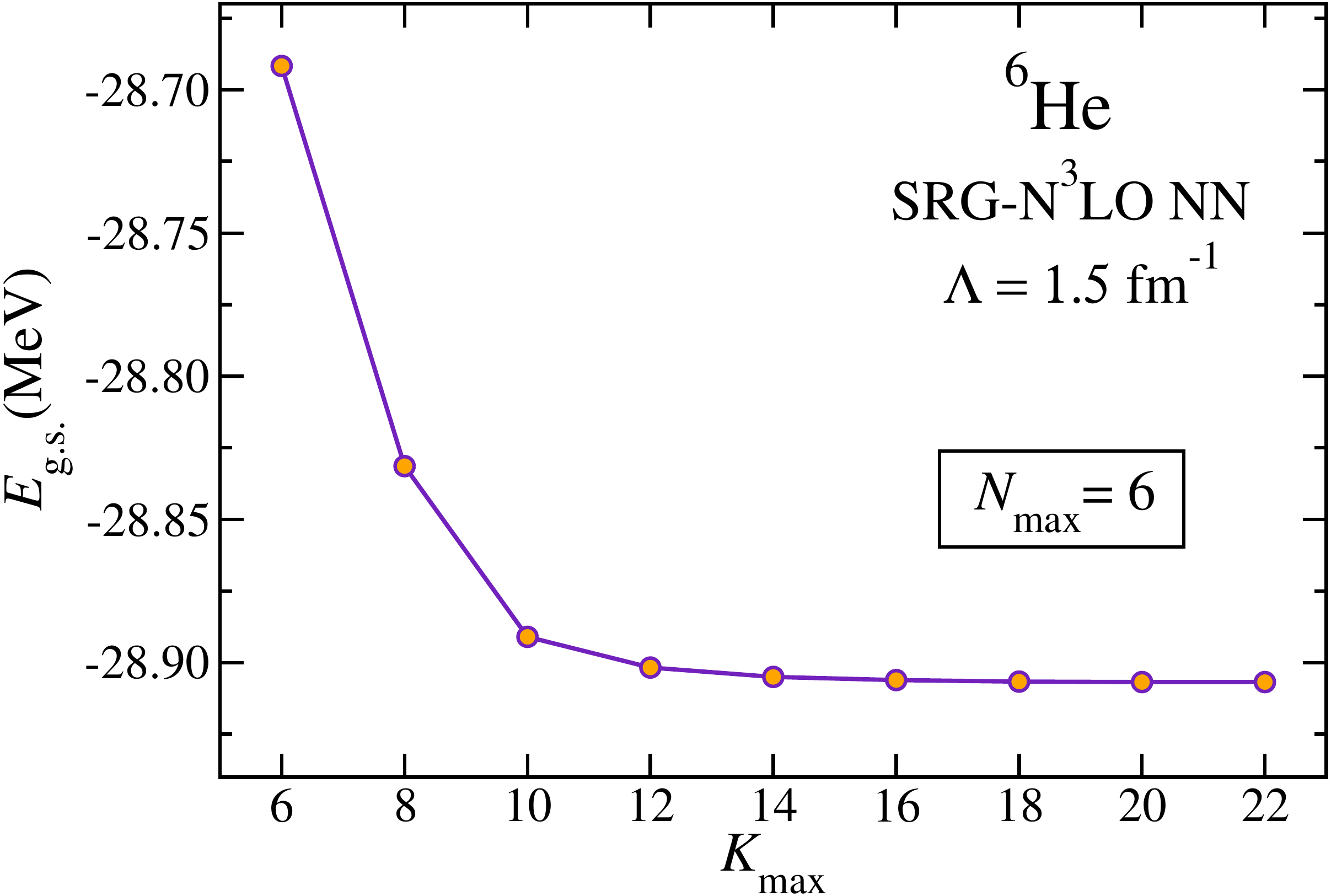}
\
\includegraphics[width=75mm]{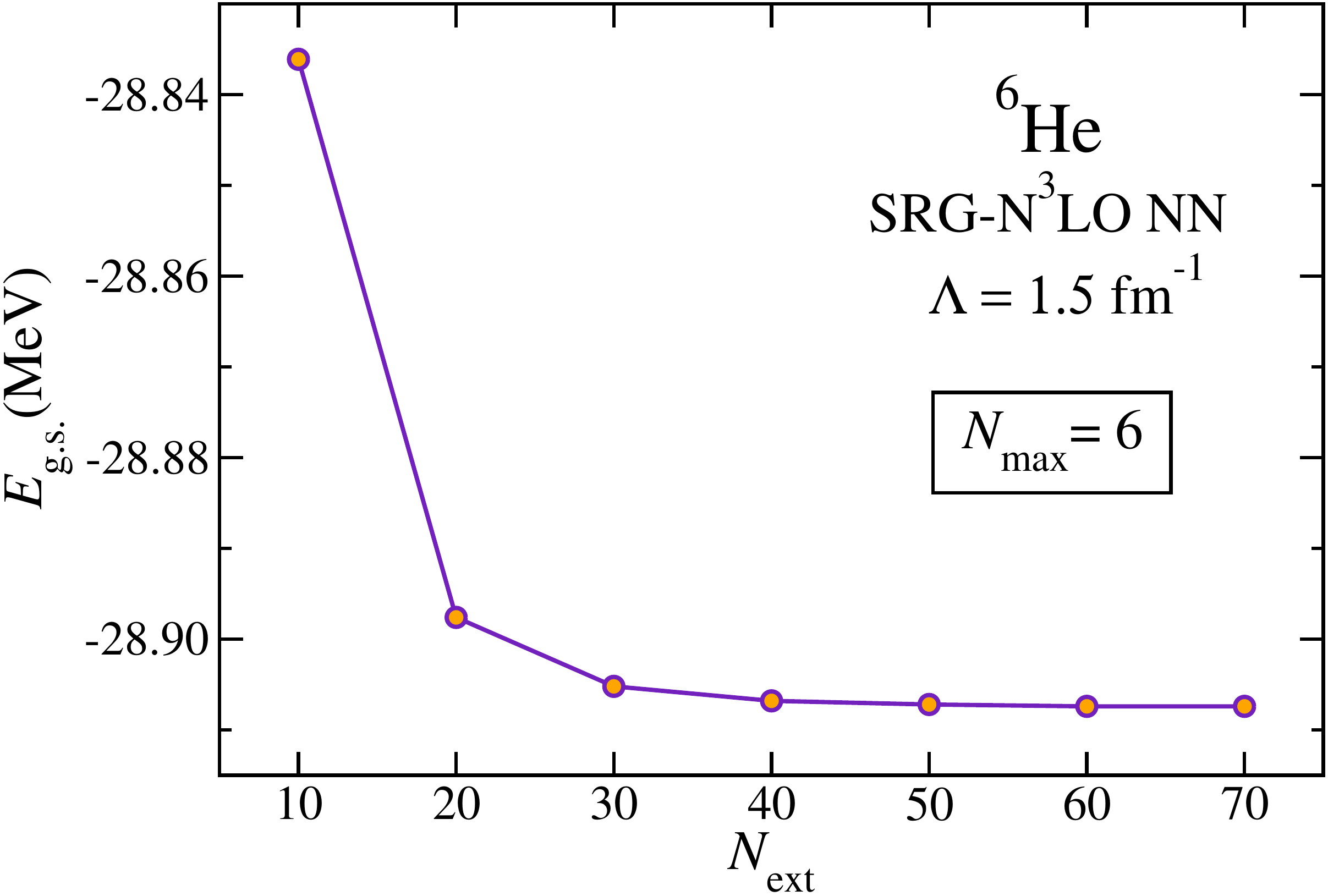}
\caption{(Color online) Dependence of the NCSM/RGM calculated $^6$He g.s.\ energy 
at $N_{\rm max}=6$ within a $^4$He+$n$+$n$ basis using a matching radius of $a=20$
on (left panel) the maximum value of the hypermomentum $K_{\rm max}$ used in the 
HH expansion ($N_{\rm ext}=40$) and (right panel) the size of the extended HO model space 
$N_{\rm ext}$ (results for $K_{\rm max}=20$).}
\label{fig:conv_gs_3B}
\end{figure}

In figure \ref{fig:conv_gs_3B}, we show the convergence pattern of the
ground state energy of $^6$He respect to $K_{\rm max}$ and $N_{\rm ext}$
for an $N_{\rm max}=6$ calculation. In both cases very stable
results are reached. 
It is found that for higher values of $N_{\rm max}$ convergence patterns are
very similar, however convergence is reached at slighter higher values of the
correponding parameters.   
      
\begin{figure}[t]
\centering
\includegraphics[width=8.0cm,clip=,draft=false]{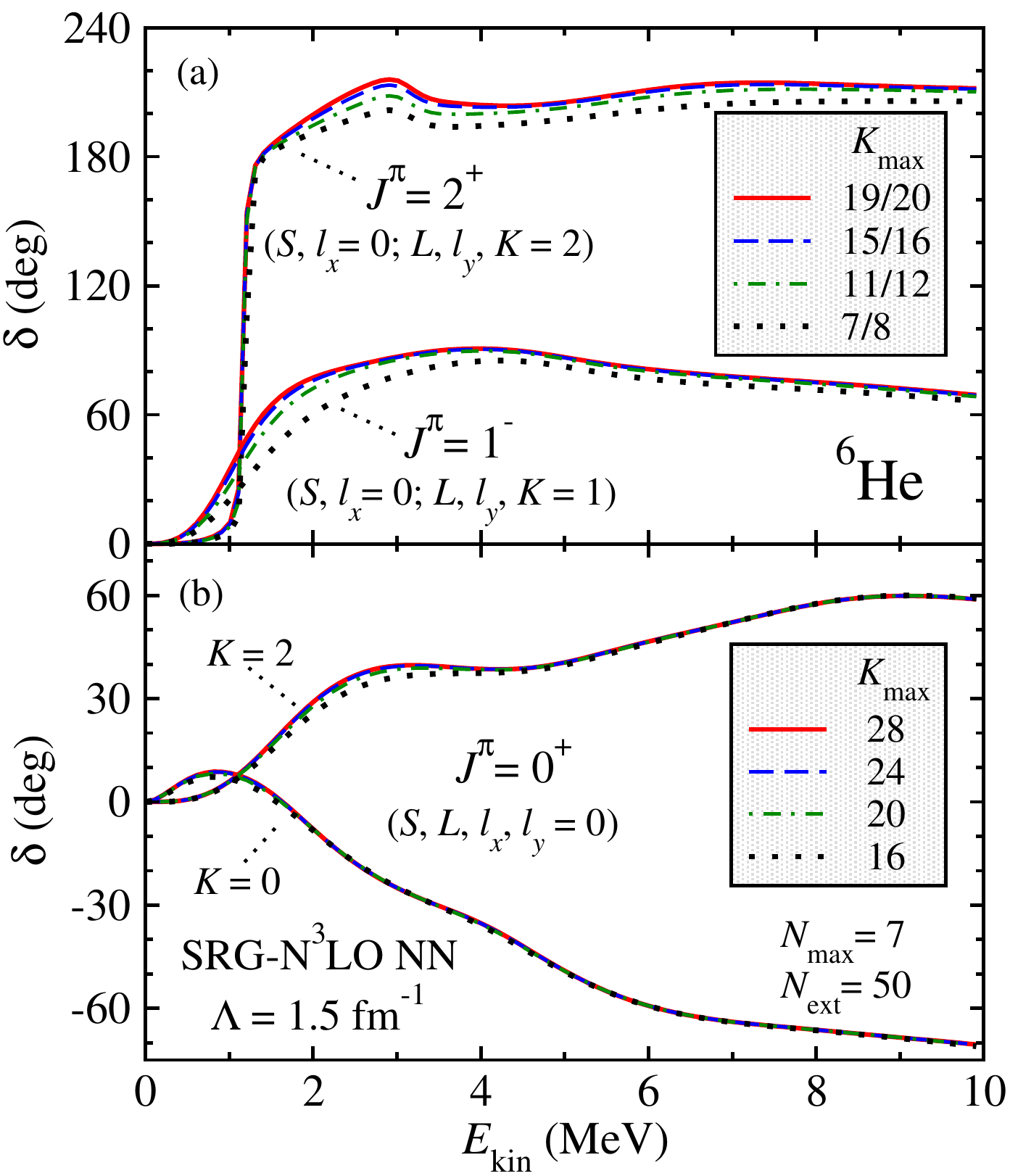}
\caption{(Color online) Convergence behavior respect to the maximum hypermomentum used in the hyperspherical expansion (\ref{expansionHH})   
of the calculated $^4{\rm He}+n+n$ (a) $J^\pi = 2^+, 1^-$ and (b) $0^+$   
diagonal phase shifts at $N_{\mbox{max}}=7$ and $N_{\mbox{ext}}=50$.}   
\label{fig:conv_kmax_ps}
\end{figure}

When studying continuum states, the values needed in  
order to reach convergence depend  
greatly on the particular channel that is being considered. For example, 
in Fig.~\ref{fig:conv_kmax_ps}, convergence with respect to the maximum hypermomentun 
$K_{\rm max}$ used in the expansion (\ref{expansionHH}) when calculating
phase shifts is shown. For this parameter the convergence pattern is very smooth, 
however, the particular value needed for $K_{\rm max}$ in order to reach convergence
 can be as low a 19 for the $1^-$ channel or
as high as 24 for the $0^+$ case in an $N_{\rm max}=7$ calculation.

%% file: SE_transitions.tex
\subsection{Electric dipole transitions in the NCSMC}
\label{sec:E1}
The {\em ab initio} NCSMC approach, introduced in Sec.~\ref{sec:ncsmc}, provides an efficient 
simultaneous description of bound and scattering states associated
with a microscopic Hamiltonian. It can
thus be naturally applied to the description of radiative-capture
reactions, which involve both scattering (in the initial channels)
and bound states (in the final channels). While the main components of the formalism have been introduced in Sec.~\ref{sec:ncsmc}, here we provide the algebraic expressions for the matrix elements -- between NCSMC basis states -- of the electric dipole operator 
\begin{align}
\label{E1op}
	\vec{E1} & = e \sum_{i=1}^{A} \frac{1+\tau^{(3)}_i}{2} \left( \vec{r}_i - \vec{R}^{(A)}_{\rm c.m.} \right) \,,
\end{align} 
(usually the dominant electromagnetic multipole at low excitation energies, when the long wavelength limit applies) required to compute radiative-capture cross sections. Here $e$ is the electric charge, $\vec{r}_i$ and $\tau_i$ are the position vector and isospin of the i$th$ nucleon, and
\begin{align}
	\vec{R}^{(A)}_{\rm c.m.} = \frac{1}{A} \sum_{i=1}^A \vec{r}_i
\end{align}
the center of mass (c.m.) coordinate of the $A$-nucleon system.
Working within a binary cluster basis, it is convenient to re-write Eq.~(\ref{E1op}) in terms of three components: $i)$ an operator acting exclusively on the first $A-a$ nucleons (pertaining to the first cluster or target); $ii)$ an operator acting exclusively on the last $a$ nucleons (belonging to the second cluster or projectile); and, finally, $iii)$ an operator acting on the relative motion wave function between target and projectile: 
\begin{align}
	\label{E1op2}
	\vec{E1} & = e \sum_{i=1}^{A-a} \frac{1+\tau^{(3)}_i}{2} \left( \vec{r}_i - \vec{R}^{(A-a)}_{\rm c.m.} \right) \nonumber\\
	& + e \sum_{j=A-a+1}^{A} \frac{1+\tau^{(3)}_j}{2} \left( \vec{r}_i - \vec{R}^{(a)}_{\rm c.m.} \right) \nonumber\\
	&+ e \frac{Z_{(A-a)} a - Z_{(a)} (A-a)}{A} \vec{r}_{A-a,a}\,.
\end{align}
Here, $\vec{R}^{(A-a)}_{\rm c.m.}$ and $\vec{R}^{(a)}_{\rm c.m.}$ are the c.m.\ coordinates the $(A-a)$- and $a$-nucleon systems,  respectively, and $\vec{r}_{A-a,a}=\vec{R}^{(A-a)}_{\rm c.m.}-\vec{R}^{(a)}_{\rm c.m.}$ is the relative displacement vector between the two clusters, while $Z_{(A-a)}$ and $Z_{(a)}$ represent respectively the charge numbers of the target and of the projectile.
It can be easily demonstrated that Eqs.~(\ref{E1op}) and (\ref{E1op2}) are exactly equivalent. 

Noting that the dipole operator can be expanded in terms of spherical basis vectors $\{\hat{e}_\mu, \mu =0,\pm1\}$ as $\vec E1 = \sqrt{\tfrac{4\pi}{3}}\sum_\mu {\mathcal M^{E}_{1\mu}} \hat e_\mu$, with components 
\begin{equation}
	{\mathcal M^{E}_{1\mu}} =  e \sum_{j=1}^{A} \frac{1+\tau^{(3)}_j}{2} \big | \vec{r}_j - \vec{R}^{(A)}_{\rm c.m.}\big |\, Y_{1\mu}(\widehat{r_j -R^{(A)}_{\rm c.m.}})\,, 
\end{equation}
it is convenient to introduce the reduced matrix elements between two bound states of an $A$-body nucleus with spin $J_i$, parity $\pi_i$, isospin $T_i$, energy $E_i$ in the initial state and $J_f, \pi_f, T_f, E_f$ in the final state: 
\begin{align}
	{\mathcal B}^{E1}_{fi} \equiv&  \Big \langle \Psi^{J_f^{\pi_f}T_f}_{A} (E_f) \Big |\Big| {\mathcal M^{E}_{1}} \Big| \Big |  \Psi^{J_i^{\pi_i}T_i}_{A} (E_i) \Big\rangle \label{ME1}\\[2mm]
		           =&  \frac{\sqrt{2J_f+1}}{C_{J_iM_i 1 \mu}^{J_f M_f}} 
		           \Big \langle \Psi^{f M_f}_{A} (E_f) \Big | e \sum_{j=1}^{A} \frac{1+\tau^{(3)}_j}{2} \big | \vec{r}_j - \vec{R}^{(A)}_{\rm c.m.}\big |\, Y_{1\mu}(\widehat{r_j -R^{(A)}_{\rm c.m.}})  \Big | \Psi^{i M_i}_{A} (E_i) \Big\rangle\,.\nonumber
\end{align}
In the second line of Eq.~(\ref{ME1}) we have introduced 
the short notation $f(i)$ for the group of quantum numbers $\{J_{f(i)}^{\pi_{f(i)}} T_{f(i)}\}$ that will be used throughout the rest of this section. 
In the NCSMC formalism the matrix element of Eq.~(\ref{ME1}) is given by the sum of four components, specifically, the  reduced matrix element in the NCSM sector of the wave function, the ``coupling" reduced matrix elements between NCSM and NCSM/RGM (and {\em vice versa}) basis states, and the reduced matrix element in the NCSM/RGM sector:
\begin{align}
	{\mathcal B}^{E1}_{fi} = & \phantom{+}\sum_{\lambda \lambda^\prime} c^{*f}_{\lambda^\prime}   \matrELred{A \lambda^\prime J_f^{\pi_f} T_f}{{\mathcal M}^E_1}{A \lambda J_i^{\pi_i} T_i} \, c^i_{\lambda}  \nonumber\\
	                        & + \sum_{\lambda^\prime\nu} \int dr r^{2}   \, c^{*f}_{\lambda^\prime} \matrELred{A \lambda^\prime J_f^{\pi_f} T_f}{{\mathcal M}^E_1\hat{\mathcal A}_{\nu} } {\Phi^i_{\nu r}} \, \frac{\gamma^i_{\nu}(r)}{r}   \nonumber\\
	                        & + \sum_{\lambda\nu^\prime} \int dr^\prime r^{\prime\,2}  \frac{\gamma^{*f}_{\nu^\prime}(r^\prime)}{r^\prime} \matrELred{\Phi^f_{\nu^\prime r^\prime}}{\hat{\mathcal A}_{\nu^\prime} {\mathcal M}^E_1} {A \lambda J_i^{\pi_i} T_i} \, c^i_{\lambda}  \nonumber\\		
	                        & +  \sum_{\nu\nu^\prime} \int dr^\prime r^{\prime\,2} \int dr r^{2}  \frac{\gamma^{*f}_{\nu^\prime}(r^\prime)}{r^\prime} \matrELred{\Phi^f_{\nu^\prime r^\prime}}{\hat{\mathcal A}_{\nu^\prime} {\mathcal M}^E_1 \hat{\mathcal A}_{\nu}} {\Phi^i_{\nu r}} \, \frac{\gamma^i_{\nu}(r)}{r}	\,.	
\end{align}   
The algebraic expression for the reduced matrix elements in the NCSM sector  $\matrELred{A \lambda^\prime J_f^{\pi_f} T_f}{{\mathcal M}^E_1}{A \lambda J_i^{\pi_i} T_i}$ can be easily obtained working  in the single-particle SD harmonic oscillator basis. 
In the following, we consider the reduced matrix elements in the NCSM/RGM sector. First, we notice that the inter-cluster antisymmetrizer commutes with the $A$-nucleon $\vec{E1}$ dipole operator of Eq.~(\ref{E1op}) and
\begin{align}
	\label{NCSMRGM-E1}
	\matrELred{\Phi^f_{\nu^\prime r^\prime}}{\hat{\mathcal A}_{\nu^\prime} {\mathcal M}^E_1 \hat{\mathcal A}_{\nu}} {\Phi^i_{\nu r}}
\! = \!\tfrac{1}{2}\!\left(\sqrt{\tfrac{ A! }{ (A-a)! a !}}\matrELred{\Phi^f_{\nu^\prime r^\prime}}{\hat{\mathcal A}_{\nu^\prime} 
	{\mathcal M}^E_1 } {\Phi^i_{\nu r}} \!+\! \sqrt{\tfrac{ A! }{ (A-a^\prime)! a^\prime !}}\matrELred{\Phi^f_{\nu^\prime r^\prime}}{ {\mathcal M}^E_1
	\hat{\mathcal A}_{\nu} } {\Phi^i_{\nu r}}\right)\!. 
\end{align}
Second, using the $\vec{E1}$ operator in the form of Eq.~(\ref{E1op2}) we can rewrite, e.g., the first matrix element in the right-hand side of Eq.~(\ref{NCSMRGM-E1})  as:
\begin{align}
	\label{NCSMRGM-E1-2}
	\matrELred{\Phi^f_{\nu^\prime r^\prime}}{\hat{\mathcal A}_{\nu^\prime}
	{\mathcal M}^E_1 } {\Phi^i_{\nu r}} 
	& = \phantom{+}\matrELred{\Phi^f_{\nu^\prime r^\prime}}{\hat{\mathcal A}_{\nu^\prime}
	\, e \sum_{i=1}^{A-a} \frac{1+\tau^{(3)}_i}{2} \left| \vec{r}_i - \vec{R}^{(A-a)}_{\rm c.m.} \right| Y_1(\widehat{\vec{r}_i - \vec{R}^{(A-a)}_{\rm c.m.}}) } {\Phi^i_{\nu r}} \nonumber\\
	& \phantom{=} +  \matrELred{\Phi^f_{\nu^\prime r^\prime}}{\hat{\mathcal A}_{\nu^\prime}
	\, e \sum_{j=A-a+1}^{A} \frac{1+\tau^{(3)}_j}{2} \left| \vec{r}_j - \vec{R}^{(a)}_{\rm c.m.} \right| Y_1(\widehat{\vec{r}_j - \vec{R}^{(a)}_{\rm c.m.}}) } {\Phi^i_{\nu r}} \nonumber\\
	& \phantom{=} + e \frac{Z_{(A-a)} a - Z_{(a)} (A-a)}{A} \matrELred{\Phi^f_{\nu^\prime r^\prime}}{\hat{\mathcal A}_{\nu^\prime}
	\, r_{A-a,a} Y_1(\hat{r}_{A-a,a})} {\Phi^i_{\nu r}}
\end{align}
Given the long-range nature of the electric dipole operator and the fact that the effect of the exchange part of the antisymmetrization operator is short-ranged, if there are no allowed E1 transitions between the target (projectile) eigenstate in the initial state and that in the final state (e.g., only positive-parity eigenstates of the target/projectile are included in the model space), the first two terms on the right hand side of Eq.~(\ref{NCSMRGM-E1-2}) are expected to be negligible and one obtains:  
\begin{align}
	\label{NCSMRGM-E1-3}
	\matrELred{\Phi^f_{\nu^\prime r^\prime}}{\hat{\mathcal A}_{\nu^\prime}
	\,  {\mathcal M}^E_1 } {\Phi^i_{\nu r}} 
	& \simeq  e \frac{Z_{(A-a)} a - Z_{(a)} (A-a)}{A} \matrELred{\Phi^f_{\nu^\prime r^\prime}}{\hat{\mathcal A}_{\nu^\prime}
	\,  r_{A-a,a} Y_1(\hat r_{A-a,a}) } {\Phi^i_{\nu r}} \nonumber\\ 
	& = e \frac{Z_{(A-a)} a - Z_{(a)} (A-a)}{A}  \sqrt{(2J_i+1)(2J_f+1)(2\ell+1)} \,(-)^{s+J_f} \,\delta_{T_iT_f}\nonumber \\
	& \phantom{=} \times \sum_{\tilde\ell} \sqrt{2\tilde\ell+1} 
	\left(
	\begin{array}{ccc}
		1 & \ell & \tilde\ell \\
		0 & 0 & 0
	\end{array}
	\right)
	\left\{
	\begin{array}{ccc}
		\ell & s & J_i \\[0.5mm]
		J_f & 1 & \tilde\ell
	\end{array}
	\right\}
	{\mathcal N}^f_{\nu^\prime \tilde\nu} (r^\prime, r) \, r \,,
\end{align}
where all quantum numbers in the index $\tilde\nu$ are identical to those in the index $\nu$ except for the angular momentum $\ell$, which is replaced by $\tilde\ell$.

The ``coupling" E1 reduced matrix elements between NCSM and NCSM/RGM components of the basis can be derived  making similar considerations:
\begin{align}
	\label{coupling-E1-1}
	&\matrELred{A \lambda^\prime J_f^{\pi_f} T_f}{ {\mathcal M}^E_1 \hat{\mathcal A}_{\nu}} {\Phi^i_{\nu r}}  \nonumber\\[2mm]
	&\qquad\qquad = \matrELred{A \lambda^\prime J_f^{\pi_f} T_f}{\hat{\mathcal A}_{\nu} {\mathcal M}^E_1 } {\Phi^i_{\nu r}}	 \nonumber\\
	&\qquad\qquad = \matrELred{A \lambda^\prime J_f^{\pi_f} T_f}{ \hat{\mathcal A}_{\nu}  \, e \sum_{i=1}^{A-a} \frac{1+\tau^{(3)}_i}{2} \left| \vec{r}_i - \vec{R}^{(A-a)}_{\rm c.m.} \right| Y_1(\widehat{\vec{r}_i - \vec{R}^{(A-a)}_{\rm c.m.}})} {\Phi^i_{\nu r}} \nonumber\\
	& \qquad\qquad\phantom{=} +   \matrELred{A \lambda^\prime J_f^{\pi_f} T_f}{ \hat{\mathcal A}_{\nu}  \,e \sum_{j=A-a+1}^{A} \frac{1+\tau^{(3)}_j}{2} \left| \vec{r}_j - \vec{R}^{(a)}_{\rm c.m.} \right| Y_1(\widehat{\vec{r}_j - \vec{R}^{(a)}_{\rm c.m.}}) } {\Phi^i_{\nu r}} \nonumber\\
	& \qquad\qquad\phantom{=} +  e \frac{Z_{(A-a)} a - Z_{(a)} (A-a)}{A} \matrELred{A \lambda^\prime J_f^{\pi_f} T_f}{ \hat{\mathcal A}_{\nu}  \,r_{A-a,a} Y_1(\hat r_{A-a,a})} {\Phi^i_{\nu r}}\,.
\end{align}
Once again, the first two terms in the right-hand side of Eq.~(\ref{coupling-E1-1}) are expected to be negligible provided there are no E1 transitions between the $(A-a)$-nucleon ($a$-nucleon) eigenstates included in the model space. This can be easily understood by inserting -- to the left of the antisymmetrization operator -- an approximate closure relationship with respect to the binary cluster basis. In such a case, one can make the approximation:
\begin{align}
	\matrELred{A \lambda^\prime J_f^{\pi_f} T_f}{ {\mathcal M}^E_1 \hat{\mathcal A}_{\nu}} {\Phi^i_{\nu r}}  
	& \simeq e \frac{Z_{(A-a)} a - Z_{(a)} (A-a)}{A} \matrELred{A \lambda^\prime J_f^{\pi_f} T_f}{ \hat{\mathcal A}_{\nu}  \,r_{A-a,a} Y_1(\hat r_{A-a,a}) } {\Phi^i_{\nu r}}  \label{coupling-E1-2}\\ 
	& = e \frac{Z_{(A-a)} a - Z_{(a)} (A-a)}{A}  \sqrt{(2J_i+1)(2J_f+1)(2\ell+1)}  \,\delta_{T_iT_f}\nonumber\\\
	&\times \,(-)^{s+J_f}\sum_{\tilde\ell} \sqrt{2\tilde\ell+1} 
	\left(
	\begin{array}{ccc}
		1 & \ell & \tilde\ell \\
		0 & 0 & 0
	\end{array}
	\right)
	\left\{
	\begin{array}{ccc}
		\ell & s & J_i \\[0.5mm]
		J_f & 1 & \tilde\ell
	\end{array}
	\right\}
	g^f_{\lambda^\prime \nu}(r) \, r \,,\nonumber
\end{align}
where $g^f_{\lambda^\prime \nu}(r) = \braket{A \lambda^\prime J_f^{\pi_f} T_f}{\hat{\mathcal{A}}_{\nu} \Phi_{\nu r}^{J_f^{\pi_f} T_f}}$ is the non-orthogonalized cluster form factor, related to the form factor of Eq.~(\ref{g-bar}) by the relationship 
\begin{align}
	g_{\lambda \nu}(r) &= \sum_{\nu^\prime}\int dr^\prime r^{\prime\, 2} \bar g_{\lambda \nu^\prime}(r^\prime) {\mathcal N}^{\frac12}_{\nu^\prime \nu} (r^\prime,r) \,,
\end{align}
and $\tilde\nu$ has the same meaning as in Eq.~(\ref{NCSMRGM-E1-3}).

To summarize, in the following we provide the algebraic expressions for the reduced matrix elements of the $ {\mathcal M}^E_1$ operator within the fully orthogonalized NCSMC basis, that is:
\begin{align}
\bar {\mathcal B}^{E1}_{fi} & =
\left( \bar c^f  \quad \bar \chi^f \right) \, {N^f}^{-\tfrac{1}{2}}\,
\left(
\begin{array}{cc}
A^{fi} & B^{fi} \\[2mm]
\bar B^{fi} & C^{fi}
\end{array}
\right)
\, {N^{i}}^{-\tfrac{1}{2}}\,
\left(
\begin{array}{c}
\bar c^i\\[2mm]
\bar \chi^i
\end{array}
\right)\,,
\end{align}
where $\bar{c}^{i(f)}$ and $\bar{\chi}^{i(f)}$ are, respectively, the NCSM and NCSM/RGM components of the orthogonal NCSMC basis function of Eq.~(\ref{NCSMC_wave_orth}), ${N^{f(i)}}^{-\tfrac{1}{2}}$ the inverse-square roots of the NCSMC norm kernel, defined in Eq.~(\ref{eq:formalism_180}), and $A^{fi}, B^{fi}, \bar B^{fi}$ and $C^{fi}$ four matrices given by: 
\begin{align}
A^{fi}_{\lambda^\prime \lambda} & = \matrELred{A \lambda^\prime J_f^{\pi_f} T_f}{ {\mathcal M}^E_1}{A \lambda J_i^{\pi_i} T_i} \,, \\[2mm]
B^{fi}_{\lambda^\prime \mu y}	
& = \sum_{\tilde\mu} \int d\tilde y \, \tilde y^2 \, \bar{g}^f_{\lambda^\prime \tilde\mu} (\tilde y)\, E^{fi}_{\tilde\mu \mu} (\tilde y, y)\\[2mm]
\bar B^{fi}_{\mu^\prime y^\prime \lambda }	
& =  \sum_{\tilde\mu^\prime} \int d\tilde y^\prime \, \tilde y^{\prime\,2} \, \bar E^{fi}_{\mu^\prime \tilde\mu^\prime} (y^\prime , \tilde y^\prime) 
	\, \bar{g}^i_{\lambda \tilde\mu^\prime} (\tilde y^\prime) \\[2mm]
C^{fi}_{\mu^\prime y^\prime \mu y} & = \tfrac{1}{2} \left[ E^{fi}_{\mu^\prime \mu}(y^\prime,y) + \bar E^{fi}_{\mu^\prime \mu}(y^\prime,y) \right] \,.
\end{align}
%
Here, we have introduced the orthogonalized E1 integration kernel
\begin{align}
E^{fi}_{\mu^\prime \mu} (y^\prime, y) & \simeq  \delta_{T_iT_f} \,\sqrt{(2J_i+1)(2J_f+1)} \, \sum_{\nu\tilde\ell} e \frac{Z_{(A-a)} a - Z_{(a)} (A-a)}{A}\nonumber\\
	& \phantom{=} \times  (-)^{s+J_f}\sqrt{(2\ell+1)(2\tilde\ell+1)} 
	\left(
	\begin{array}{ccc}
		1 & \ell & \tilde\ell \\
		0 & 0 & 0
	\end{array}
	\right)
	\left\{
	\begin{array}{ccc}
		\ell & s & J_i \\[0.5mm]
		J_f & 1 & \tilde\ell
	\end{array}
	\right\}  \nonumber\\
	& \phantom{=} \times  \int dr \, r^2 \, {{\mathcal N}^f}^{\tfrac{1}{2}}_{\mu^\prime\tilde\nu}(y^\prime,r) \, r \, {{\mathcal N}^i}^{-\tfrac{1}{2}}_{\nu \mu}(r, y)\,,
\end{align}
%
%
and its Hermitian conjugate
\begin{align}
\bar E^{fi}_{\mu^\prime \mu} (y^\prime , y) & = (-)^{J_f-J_i} \, E^{if}_{\mu \mu^\prime } (y, y^\prime )  \nonumber\\[2mm]
& \simeq \delta_{T_iT_f}\, \sqrt{(2J_i+1)(2J_f+1)} \, \sum_{\nu^\prime\tilde\ell^\prime} e \frac{Z_{(A-a^\prime)} a^\prime - Z_{(a^\prime)} (A-a^\prime)}{A}    \nonumber\\
	& \phantom{=} \times  (-)^{s^\prime+J_f} \sqrt{(2\ell^\prime+1)(2\tilde\ell^\prime+1)} 
	\left(
	\begin{array}{ccc}
		1 & \ell^\prime & \tilde\ell^\prime \\
		0 & 0 & 0
	\end{array}
	\right)
	\left\{
	\begin{array}{ccc}
		\ell^\prime & s^\prime & J_f \\[0.5mm]
		J_i & 1 & \tilde\ell^\prime
	\end{array}
	\right\}  \nonumber\\
	& \phantom{=} \times
	\int dr^\prime r^{\prime\,2} {{\mathcal N}^f}^{-\tfrac{1}{2}}_{\mu^\prime\nu^\prime}(y^\prime,r^\prime) \, r^\prime {{\mathcal N}^i}^{\tfrac{1}{2}}_{\tilde\nu^\prime \mu}(r^\prime, y)\,,
	\end{align} 
where $a^\prime$ is the mass number of the projectile in the final state, the quantum numbers in the index $\tilde\nu^\prime$ are identical to those in the index $\nu^\prime$ except for $\ell^\prime$, which is replaced by $\tilde\ell^\prime$, and  the  phase $(-1)^{J_f-J_i}$ is a result of the symmetry properties of the reduced matrix elements of the $ {\mathcal M}^E_1$ operator under Hermitian conjugation:
\begin{align}
	\matrELred{n^\prime J_f}{ {\mathcal M}^E_1}{n \, J_i} = (-1)^{J_f-J_i} \matrELred{n\, J_i}{ {\mathcal M}^E_1}{n^\prime J_f}^* \,. 
\end{align}

The formalism presented in this section for the matrix elements of the electric dipole operator can be extended to any electromagnetic multipole of interest, although in general -- different form the $\vec E1$ case -- the partition of the operator into terms acting exclusively on the target, projectile and relative motion wave functions [see Eq.~\eqref{E1op2}] will only be approximate. For a general multipole operator the assumption of zero internal transitions between target's (projectile's) eigenstates used to arrive at Eqs.~\eqref{NCSMRGM-E1-3} and \eqref{coupling-E1-2} may also no longer be valid. In those instances, the matrix elements of the target's and projectile's portion of the operators can be estimated by inserting an approximate closure relationship with respect to the binary cluster basis.

%% file: SE_3-intro-section3.tex
\section{Structure and reaction observables of light nuclei with chiral two- and three-nucleon forces} 
\label{sec:s-p-shell-res}

Light nuclei exhibit strong clustering effects mostly due to the tightly bound alpha particle, i.e., the $^4$He nucleus. 
One of the consequences is the presence of low-lying  thresholds resulting in few (or even no) bound states and many resonances at low excitation energies. 
A realistic description of light nuclei must take into account the presence of these thresholds and include a continuum description of the inter-cluster motion.
Further, it has been known for a long time that significant effects due to the 3N forces are manifest in the structure of $p$-shell nuclei. Consequently, for a realistic description of light nuclei the 3N interactions should also be included. As described in the previous section, we have now developed a capability to include both these aspects in the NCSMC formalism. In this section, we present results for nucleon scattering on $^4$He and properties of $A=5$ nuclei, for a simultaneous description of the structure of $^6$Li and the deuteron scattering on $^4$He, and for the structure of $^9$Be with a focus on continuum and 3N-force effects.

%% file: SE_scatt-s-shell.tex
\subsection{Nucleon and deuterium scattering on $s$-shell targets} \label{sec:s-shell}

A few-nucleon projectile impinging on an $s$-shell target at low energy is among the most suited systems to build the foundations of a theory designed to tackle simultaneously bound and resonant states. Even in such a restricted mass region, nuclei exhibit a sharp transition from a unit of MeV of binding energy per nucleons for the deuteron ($d$), to $7$ MeV per nucleon for the $\alpha$ particle, thus indicating a leap from dilute systems to the dense and tightly bound $\alpha$ particle. As a direct consequence the low-energy continuum of $^5$He ($^5$Li) consists of a single open channel per total angular momentum $J$ and orbital angular momentum $\ell$, up to an energy of $17.638$ MeV ($18.35$ MeV) where the $^3$H($d$,$n$)$^4$He ($^3$He($d$,$p$)$^4$He) channel become energetically opened. These simple features have turned this scattering system into the tool of choice to test {\it ab initio} structure augmented by a reaction framework. Despite the conceptual simplicity of these two systems, it is only recently that an {\it ab initio} description of the elastic collision of a neutron and a $\alpha$ particle has been performed by Nollett {\it et al.}~\cite{Nollett2007}. In the work of Nollett {\it et al.}, the realistic two-nucleon AV18 and two different (UIX and IL2) three-nucleon force models are employed. The findings of Nollett {\it et al.} showed that an accurate three-nucleon interaction is crucial to reproduce the low-energy phase shifts, and spurred global interest on the subject. In the meantime, the development of chiral Effective Field Theory\cite{Weinberg1990} had reached its maturity allowing nuclear physicists to unravel nuclear properties starting from the fundamental theory of QCD~\cite{Entem2003}. These interactions are non-local by nature and therefore difficult to implement in the Green's Function Monte Carlo method used by Nollett {\it et al.}, but can be applied in the context of the NCSM/RGM, which is able to address non-local interactions~\cite{Quaglioni2008,Quaglioni2009,Hupin2013}. Nevertheless, using the GFMC framework, a recent work of Lynn {\it et al.}~\cite{Lynn:2015jua} employed the local part of the chiral N$^2$LO two-nucleon interaction to describe this reaction system.
\begin{figure}[h]
\begin{minipage}{0.45\columnwidth}
\includegraphics[width=\columnwidth]{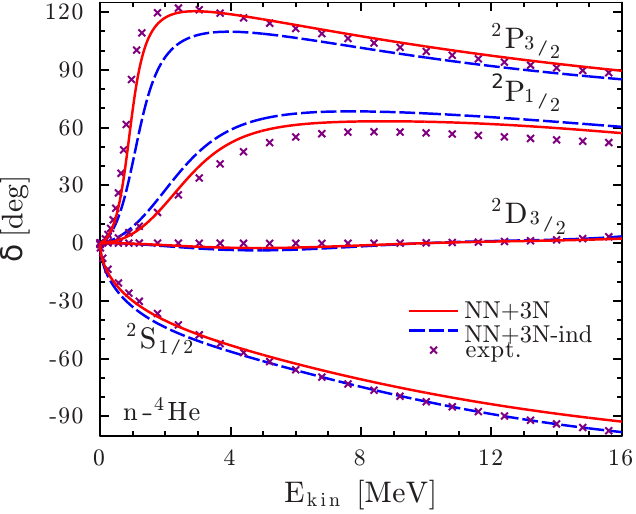}
\end{minipage}
\begin{minipage}{0.54\columnwidth}
\centering
\caption{Computed $^4$He($n$,$n$)$^4$He phase shifts obtained with the NCSMC using the SRG-evolved N$^3$LO NN interaction augmented with the three-nucleon SRG-induced (NN, blue dashed lines) and total NN+3N (continuous red lines) Hamiltonian. The $R$-matrix analysis of data from Ref.~\cite{Hale} is shown as guidance (purple crosses). The results are computed in a HO model of $N_{\rm max}=13$ with a HO frequency of $\hbar\Omega=20$ MeV, all the influential eigenstates of the compound nuclei ($^5$He) and only the g.s. of the target nuclei are included. The SRG-resolution scale is $\Lambda=2$ fm$^{-1}$, which, for this system, approximates well a unitary representation of the initial interaction~\cite{Hupin2013}.}\label{fig:NCSMC_nalpha1}
\end{minipage}
\end{figure}
Figure~\ref{fig:NCSMC_nalpha1} shows the first application of the NCSMC to this reaction system. The calculated phase shifts are obtained with the SRG-evolved N$^3$LO two-nucleon chiral interaction supplemented by the induced three-nucleon forces (blue dashed lines) and the complete NN+3N chiral interaction (continuous red lines). At the present HO frequency ($\hbar\Omega=20$ MeV) and SRG-resolution scale ($\Lambda=2$ fm$^{-1}$), the former and the latter are representations of the initial chiral NN and NN+3N interaction, respectively, as the SRG unitarity is broken only mildly in this case (see Fig.~\ref{fig:srg-convergence} as well as Figs. 6 and 7 in Ref.~\cite{Hupin2013} and the related discussion). The observed disagreement between the two-nucleon force model results and experiment, particularly regarding the relative position  of the $P_{\frac32}$ to $P_{\frac12}$ centroids, corroborates the conclusions of Nollett {\it et al}. Accordingly, the inclusion of the chiral three-nucleon force is necessary to the reproduction of the observed splitting between the $P$-waves and given that the spin-orbit interaction is responsible of the fine tuning of the relative position of the $\frac32^-$ and $\frac12^-$ resonances, we conclude that the chiral three-nucleon force brings an important part of the nuclear spin physics. Here we stress that it would not have been possible to draw such conclusions working within the many-body model space of the NCSM/RGM, where it is difficult to account for the short-range many-body correlations. This is exemplified by Figs.~\ref{fig:NCSMC_RGM} and \ref{fig:NCSMC_nalpha_polarization} of Sect.~\ref{sec:cv}. 

Therefore, the advent of the NCSMC treating on the same footing bound and resonant states permits us to reach convergence
with respect to the parameters of the HO model space henceforth revealing insights into the details of the nuclear
interaction.
\begin{figure}[h]
\begin{minipage}{0.48\columnwidth}
\includegraphics[width=\columnwidth]{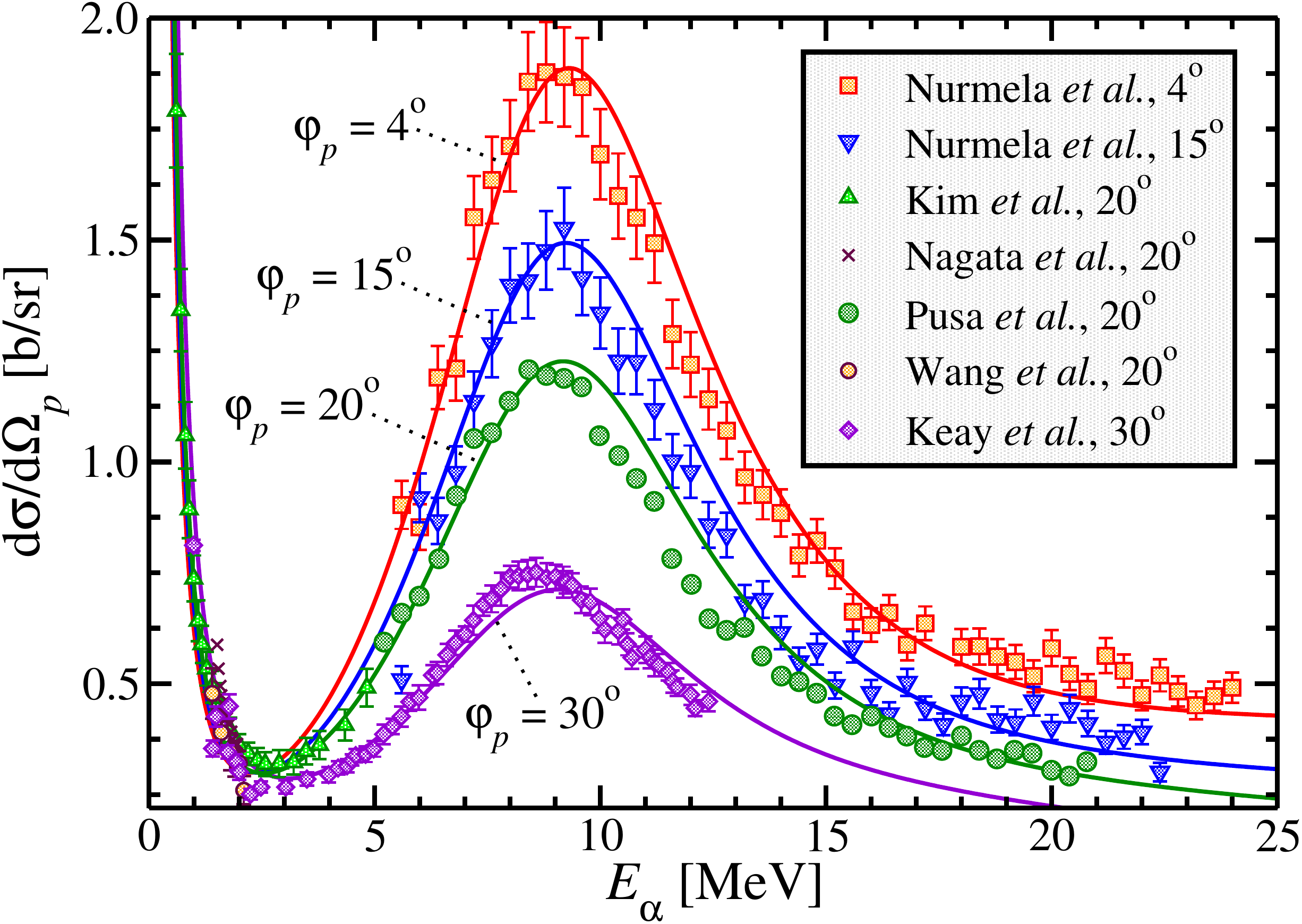}
\end{minipage}
\begin{minipage}{0.51\columnwidth}
\centering
\caption{The computed $^1$H($\alpha$,$p$)$^4$He angular differential cross-section at proton recoil angle of $\varphi_p=4^\circ,~16^\circ,~20^\circ$ and $30^\circ$ as a function of the proton incident energy is plotted versus the data (symbols) of Refs.~\cite{Nurmela1997,Kim2001,Nagata1985,Pusa2004,Wang1986,Keay2003}. Parameters of the computed angular distributions are identical to those of fig~\ref{fig:NCSMC_nalpha1}. (Figure adapted from Ref.~\cite{Hupin2014})}\label{fig:NCSMC_palpha}
\end{minipage}
\end{figure}
A typical example of the precision that can be attained is displayed in Fig.~\ref{fig:NCSMC_palpha}. There, the 
$^1$H($\alpha$,p)$^4$He angular differential cross-section calculated with the NCSMC is compared to data of 
Refs.~\cite{Nurmela1997,Kim2001,Nagata1985,Pusa2004,Wang1986,Keay2003} for a set of proton recoil angles ($\varphi$).  
The range in energy of the impinging $\alpha$ particle covers the $\frac32^-$ and $\frac12^-$ resonances where the cross 
section deviates the most from the Rutherford limit (limit of structureless charged particle scattering). Numerous 
experiments have been performed in this region to understand the nuclear enhancement and thus obtain precise 
cross-section needed for Ion-Beam Analysis. For instance, the cross section shown here are essential for the 
determination of the concentration and depth of $^4$He impurity in superconductors used in fusion energy research. Here, 
we benefit from the wealth of data and use it to probe the precision of the (SRG-evolved NN+3N) nuclear force model. For almost all angles, 
the theoretical results are within the error bars of data, and only at small angles we see disagreements that could be 
related to the remaining inaccuracies of the interaction to reproduce the centroid positions. Despite this, the present 
NCSMC formalism and the state-of-the-art chiral interaction has reached the stage where it can be used as a predictive 
tool in particular in light nuclei where the convergence and the SRG sensitivity is more-or-less under control~\cite{Hupin2014}. Furthermore, it is essential to stress that a consistent framework for bound and resonant states is constitutive to the agreement of the cross section with data. For instance, Fig.~\ref{fig:NCSMC_RGM} in Sect.~\ref{sec:cv} sheds the light on the effects of the coupling to the NCSM $^5$Li eigenstates. 
\begin{figure}[h]
\centering
\begin{minipage}{0.40\columnwidth}
\includegraphics[width=\columnwidth]{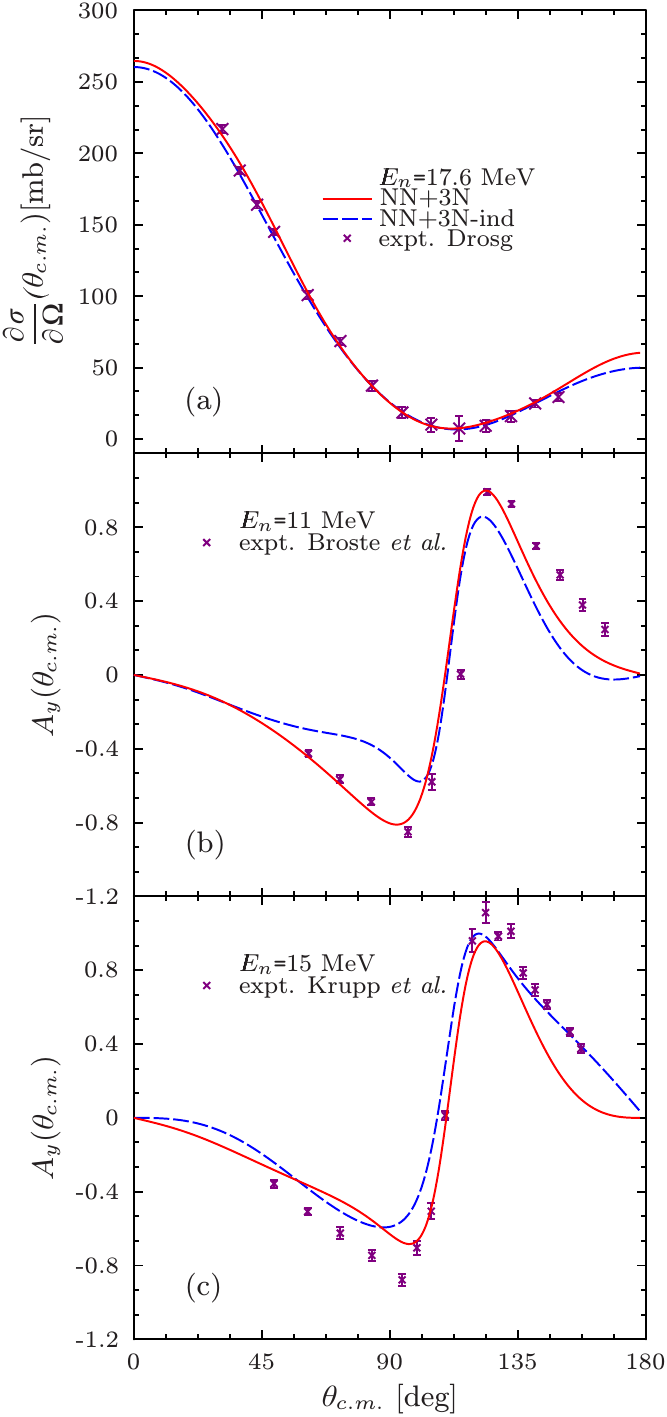}
\end{minipage}
\begin{minipage}{0.54\columnwidth}
\caption{Comparison between angular differential cross-section and polarization observable for $^4$He($n$,$n$)$^4$He computed with 
SRG-evolved N$^3$LO NN interaction augmented with the three-nucleon SRG-induced (NN, blue dashed lines) and total 
NN+3N (red lines) Hamiltonian, and data (purple crosses) from Refs.~\cite{Drosg1978,Broste1972,Krupp1984}. From top 
to bottom, respectively, the angular differential cross-section at neutron incident energy of $17.6$ MeV, the 
polarization observable at $11$ and $15$ MeV are shown. Parameters of the computed cross section are identical to those 
of fig~\ref{fig:NCSMC_nalpha1}.}\label{fig:NCSMC_nalpha2}
\end{minipage}
\end{figure}

In Fig.~\ref{fig:NCSMC_nalpha2} the $^4$He($n$,$n$)$^4$He angular differential cross section and a polarization 
observable ($A_y$) of Refs.~\cite{Drosg1978,Broste1972,Krupp1984} are compared to NCSMC results using the chiral N$^3$LO NN and its 3N induced interaction (dashed blue lines) and the full NN+3N (continuous red lines), which includes the chiral N$^2$LO 3N interaction. 
In panel (a), the agreement between the NCSMC results and experiment is nearly within the error bars of data and the effect of the initial three-nucleon force is almost indistinguishable. In this respect, panels (b) and (c) give more insight as the $A_y$ analyzing power is a more sensitive probe of the spin-orbit force. In particular we can see that the differences between the chiral two- and two-plus-three-nucleon force models are mostly concentrated in the spin physics, which is better reproduced in the latter model. The remaining disagreement with experiment, in particular in panel (b), hints that there is still room for improvement of our understanding of the nuclear interaction. Additionally, we can see that as the incident neutron energy increased from panels (b) to (c), the disagreement between the NCSMC NN+3N and experiment widens. This is related to the missing distortion effects induced by the closed but neighboring $^3$H($d$,$n$)$^4$He channel.
\begin{figure}[h]
\begin{minipage}{0.45\columnwidth}
\includegraphics[width=\columnwidth]{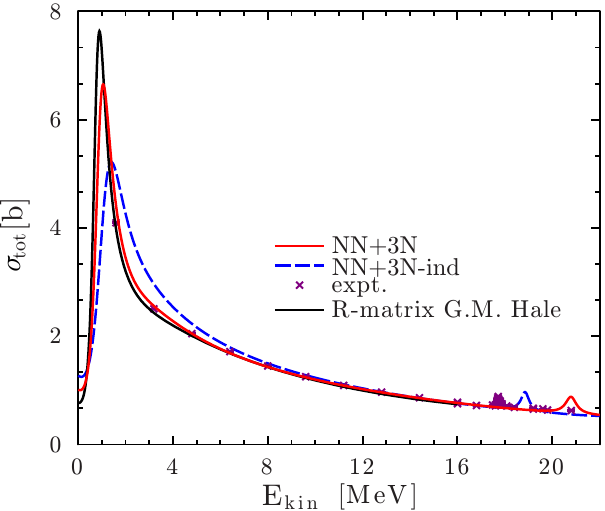}
\end{minipage}
\begin{minipage}{0.54\columnwidth}
\centering
\caption{Comparison between the computed $^4$He($n$,$n$)$^4$He cross-section obtained with the NCSMC using the SRG-evolved N$^3$LO NN interaction augmented with the three-nucleon SRG-induced (NN, blue dashed lines), the total NN+3N (continuous red lines) Hamiltonian, $R$-matrix analysis from ENDF (black line) and data (purple crosses). Parameters of the computed cross section are identical to those of fig~\ref{fig:NCSMC_nalpha1}.}\label{fig:NCSMC_xsect}
\end{minipage}
\end{figure}
This is more readily visible in Fig.~\ref{fig:NCSMC_xsect} where the total cross-section of the $^5$He continuum is plotted up to $22$ MeV of excitation energy. The calculated cross-sections correspond to the phase shifts and angular distributions shown in Fig.~\ref{fig:NCSMC_nalpha1} and \ref{fig:NCSMC_nalpha2} with the same color coding. The calculations are compared to data (purple crosses) and $R$-matrix fit (black line) from the ENDF library. Once again, the difference between the two nuclear force models is apparent around the positions of the two low-lying resonances while, at higher energies, the cross sections are indistinguishable and differences can only probe using more sensitive observables such as the $A_y$ shown in Fig.~\ref{fig:NCSMC_nalpha2}. We can see here that the enhancement of the cross-section due to $d+^3$H fusion is already present in the NCSMC calculation however at the wrong energy due to the lack of the appropriate $^3$H-$d$ cluster states in the basis, which encode the correct reaction threshold as well as escape width.

Along the path towards the treatment of $^3$H($d$,$n$)$^4$He transfer channel (or, in the present case, $d+^3$H fusion) 
within the NCSMC, the formalism needs to be extended to treat two-nucleon projectile. 
In the NCSM/RGM framework, reaction channels are treated as an essential building block of the theory and thus need to be implemented incrementally as they become energetically relevant to the reaction mechanism (see for instance Sects.~\ref{sec:rgm2} and \ref{sec:rgm3}). In addition to this technical aspect, among the possible two-nucleon projectile only the deuteron is bound henceforth it can be observed in the final state of a nuclear collision, though with its binding energy of $2.224$ MeV, the deuteron is likely to break into a neutron and a proton. Consequently, the problem takes on a three-body nature and we have the additional challenge of accounting for the breakup channel. However, below the breakup threshold we can incorporate the distortion effects through the inclusion of pseudo-states of the deuterium ($E>0$) such that the continuum is discretized. Above the breakup threshold, we make an approximation using this scheme, which should be tested with the implementation of the three-body cluster technique described in Sect.~\ref{sec:rgm3}. This approach is feasible and we have demonstrated in Sect.~\ref{sec:cv} that the combination of the NCSMC and the discretization of the deuteron continuum results in a satisfactory convergence pattern and represents a reliable approximation even above the breakup threshold (see Fig.~\ref{fig:NCSMC_dalpha_CV}). This step of extending the binary-cluster to heavier projectile mass is necessary to address reaction systems where the compound nuclei breaks apart ejecting, for instance, a deuteron.  
\begin{figure}[h]
\centering
\begin{minipage}{0.45\columnwidth}
\includegraphics[width=1\columnwidth]{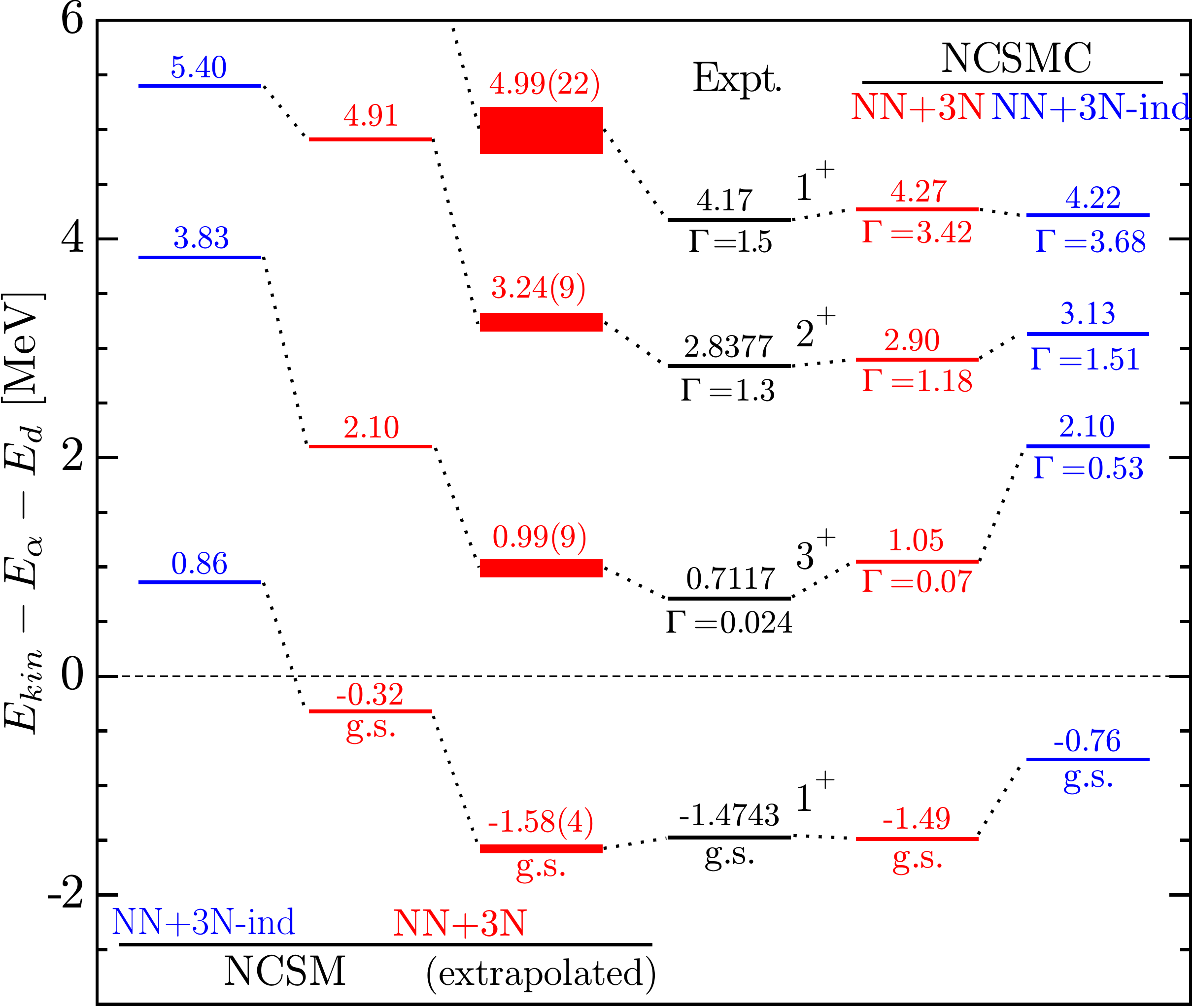}
\end{minipage}
\begin{minipage}{0.54\columnwidth}
\caption{Comparison between the positive parity low-lying states of $^6$Li and their width $\Gamma$ computed with (red lines) and without (blue lines) the initial chiral 3N force, and augmented (right-hand side) or not (left-hand side) with the coupling between NCSM/RGM and NCSM $^6$Li wave functions within the NCSMC. The NCSM extrapolated spectra (red thick lines) towards $N_{\rm max}\to \infty$ is obtained using an exponential form. All parameters are identical to those of Fig~\ref{fig:NCSMC_nalpha1} but the $N_{\rm max}$ is limited, for computational reasons, to $11$ major HO shell. (Figure adapted from Ref.~\cite{Hupin2015})}\label{fig:Li6_spectrum}
\end{minipage}
\end{figure}

As a summary of the power of the NCSMC using chiral NN+3N forces, we show in Fig.~\ref{fig:Li6_spectrum}, the low-lying spectrum of $^6$Li comparing the NCSM, experiment (black) and NCSMC on the left, middle and right, respectively. All the influential NCSM eigenvalues of $^6$Li and pseudo states of the deuteron are included. The chiral NN interaction (blue) is compared to the chiral NN+3N (red) nuclear interaction. Both are softened via the SRG method to a resolution scale of $\Lambda=2$ fm$^{-1}$ providing negligible four-body induced effects for the energy of the lowest lying states in $^6$Li~\cite{Jurgenson2011,Roth2011}. We see that the three-nucleon force is essential to the reproduction of the g.s. energy of $^6$Li yielding $-32.01$ MeV compared to the experimental $-31.994$ MeV and, at the same time, to account for the correct spacing between the $3^+$ and $2^+$ excited states, but slightly overestimates the energy of the $3^+$ resonance by $350$ keV. On the other hand, the comparison at a given $N_{\rm max}$, i.e., without taking into account the Nmax extrapolation of the NCSM result (left versus right spectra), shows that the NCSMC is able to grasp long-range correlations far above the  $N_{\rm max}$ truncation of traditional NCSM. Nevertheless, once the eigenenergies of the NCSM are extrapolated to $N_{\rm max}\to \infty$ (in the present case assuming an exponential form) thus accounting for the finiteness of the HO model space, all but the high-energy resonant states  are reproduced. In fact due to their resonant nature, no bound-state techniques, like the NCSM, is able to fully describe them. 
\begin{figure}[h]
\centering
\begin{minipage}{0.53\columnwidth}
\includegraphics[width=\columnwidth]{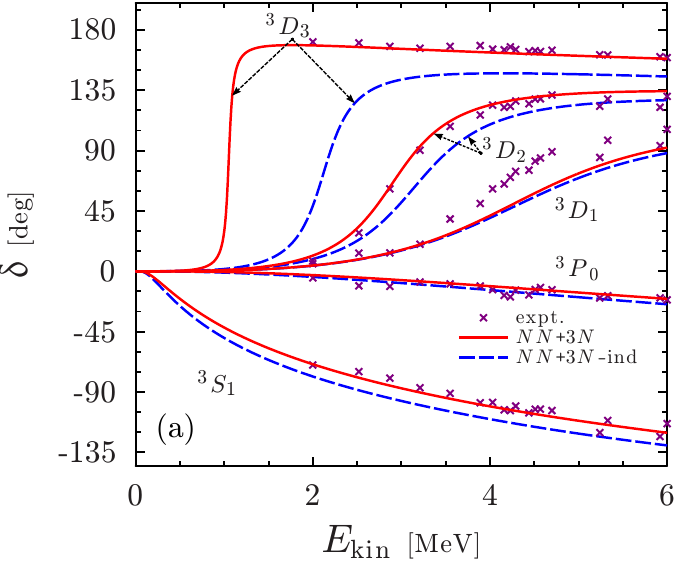}
\end{minipage}
\begin{minipage}{0.46\columnwidth}
\includegraphics[width=\columnwidth]{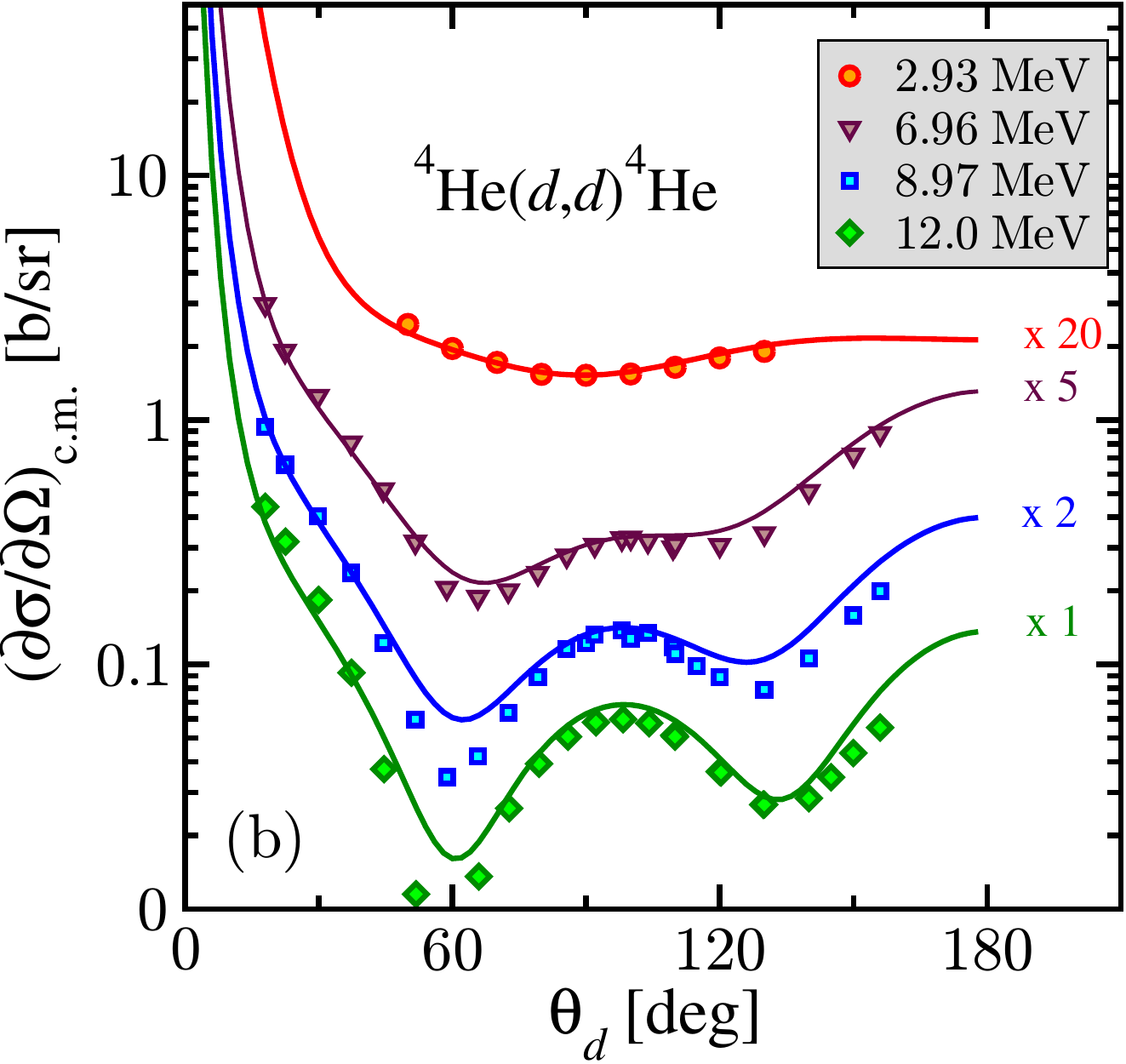}
\vspace*{-8pt}
\end{minipage}
\caption{(a) $^4$He($d$,$d$)$^4$He phase shifts computed with the SRG-evolved N$^3$LO NN interaction augmented with the three-nucleon SRG-induced (blue dashed lines) and total NN+3N (red lines) Hamiltonian, compared to $R$-matrix fits of data (purple crosses) from Refs.~\cite{Gruebler1975,Jenny1983}. In panel (b), the calculated c.m. angular distributions (lines) is compared to the measured one (symbols) at $E_d=2.93$, $6.96$, $8.97$~\cite{Jett1971} and $12$ MeV~\cite{Senhouse1964}. The cross sections are scaled by an appropriate factor to fit in the figure. Other parameters are identical to those of Fig~\ref{fig:Li6_spectrum}. (Figure adapted from Ref.~\cite{Hupin2015})}\label{fig:NCSMC_dalpha}
\end{figure}
Accordingly, we can describe the reaction observables using the same wave functions that yield Fig.~\ref{fig:Li6_spectrum} starting from the elastic $^4$He($d$,$d$)$^4$He reaction. The corresponding $S$-, $^3P_0$- and $D$-wave computed phase-shifts are shown in panel (a) of Fig.~\ref{fig:NCSMC_dalpha}. We use the same color coding as in Fig.~\ref{fig:Li6_spectrum} for the NN+3N-ind and NN+3N nuclear force models. We find again that the chiral 3N force affects essentially the splitting between the $^3D_3$- and $^3D_2$-partial waves, which corresponds to the main difference between the two spectra at the right-hand side of Fig.~\ref{fig:Li6_spectrum}. Thus, owing to a fairly good reproduction of the g.s. and low-lying spectrum of $^6$Li with the chiral NN+3N interaction model, we compare directly differential cross-section to data of Refs.~\cite{Jett1971,Senhouse1964} in panel (b). At the available experimental energies of $E_d=2.93$, $6.96$, $8.97$ and $12$ MeV, the computed angular distribution reproduces the bulk and resonant structure. However it would fail around the $3^+$ resonance due to the remaining $350$ keV discrepancy between NN+3N calculation and experiment. In addition to the phase shifts shown in panel (a), the negative-parity partial-waves are mandatory to the computation of the cross-section while the partial-wave expansion runs up to $J=6$.

Moreover, only in the NCSMC case do the wave functions present the correct asymptotic of the $^6$Li ground state, which is essential for the extraction of the asymptotic normalization constant yielding a $D$- to $S$-state ratio of $-0.027$ in agreement with a determination from $^6$Li-$^4$He elastic scattering~\cite{George1999} and the value previously obtained by Nollett{\it et al.} using variational Monte Carlo~\cite{Nollett2001}. As shown in Refs.~\cite{Navratil2011,Hupin2015} and exemplified by Fig.~\ref{fig:Li6_spectrum}, the combination of NCSMC and a realistic NN+3N force model is essential to the reproduction of the cross-section of such a reaction system. This is particularly true for an accurate description of the low-energy regime that is essential for the determination of astrophysical S-factor or, in the present case, $^2$H($\alpha$,$\gamma$)$^6$Li radiative capture.
This work demonstrates that the NCSMC is a technique capable of addressing simultaneously bound and resonant states using the latest nuclear interaction fitted on the $A\le3$ nuclear properties. In the $s$-shell target sector, we are able to achieve convergence within the computationally feasible model space henceforth providing a stepping stone towards {\it ab initio} calculations of reaction observables of astrophysical interest together with a test bed for further approximations required to address the challenging $p$-shell target sector.

%% file: SE_scatt-p-shell.tex
\subsection{Nucleon scattering on $p$-shell targets} \label{sec:Be9}
The efficient developments for treatments and inclusions of 3N interactions in \textit{ab initio} nuclear structure and reaction calculations allow to study continuum effects also for p-shell targets. In the case of the single-nucleon projectile 3N interactions can be included explicitly via the extended kernel calculations formalism introduced in Sec.~\ref{sec:examples} (see, e.g., Eq.~\eqref{NNN-ex-express} and Refs.~\cite{Hupin2013,Langhammer2015}).
As a first application of the NCSMC with 3N effects the spectrum of \elemA{9}{Be} is studied.

The \elemA{9}{Be} spectrum is sensitive to the truncations of the localized HO model space, as shown, in the convergence problems of previous NCSM calculations where the positive-parity states were found too high in excitation energy compared to experiment~\cite{Forssen2005,Burda10}.  The splitting between the lowest $5/2^{-}$ and $1/2^{-}$ states is found overestimated in NCSM calculations using the INOY interaction model that includes 3N effects~\cite{Forssen2005}. 
This splitting is also sensitive to the 3N interactions, as shown by GFMC calculations~\cite{PhysRevC.66.044310}. The 3N contributions appeared to shift the splitting away from experiment, highlighting deficiencies of the applied 3N force models. 
The spectrum of \elemA{9}{Be} has also relevance for astrophysics by providing seed material for the \elemA{12}{C} production in the core collapse supernovae, via the ($\alpha\alpha$n,$\gamma$)\elemA{9}{Be}($\alpha$,n)\elemA{12}{C} reaction, an alternative to the triple-$\alpha$ reaction~\cite{Burda10,Efros14,Sasaqui05}. 
In particular, the accurate description of the first $1/2^{+}$ state, relevant for the cross sections and reaction rates, poses a long standing problem~\cite{Burda10,Efros14,Kuechler87}.

\begin{figure}
\centering\includegraphics[width=1.0\textwidth]{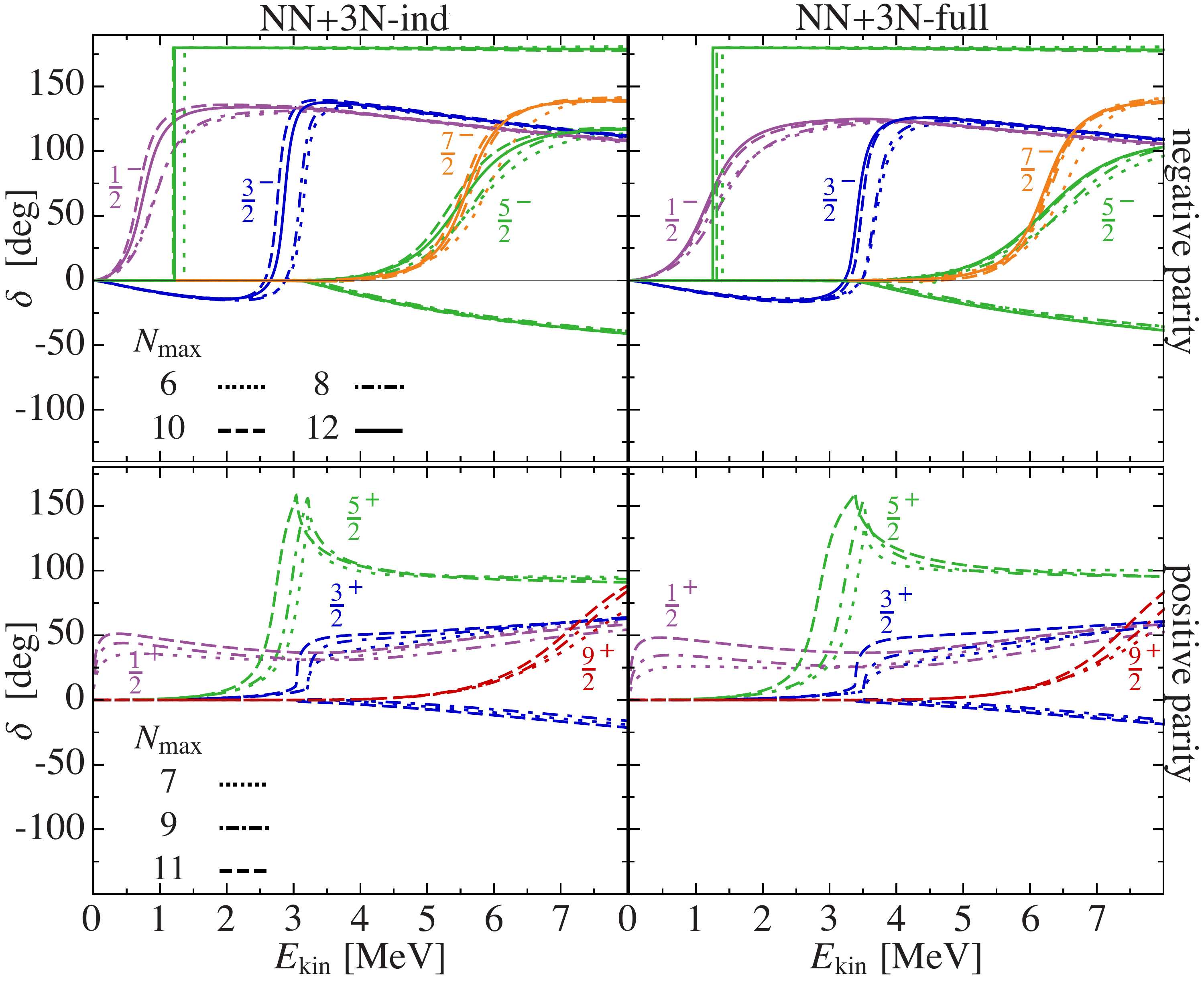}
\caption{
{\small 
NCSMC n-\elemA{8}{Be} eigenphase shifts for negative (top panels) and positive (bottom panels) parity at $N_{\text{max}} = 6-12$.
The left- and right-hand columns show the results for the NN+3N-ind and NN+3N-full Hamiltonian, respectively. Remaining parameters are $\hbar\Omega=20\mev$, $\Lambda=2.0\fm^{-1}$, and $E_{3\text{max}} = 14$. Same colors correspond to identical angular momenta. Figure taken from~\cite{Langhammer2015}.
}
\label{fig:Be9_PS_NMax}}
\end{figure}
The starting point for the \textit{ab inito} NCSMC calculations are the chiral NN interaction at N$^3$LO by Entem and Machleidt~\cite{Entem2003} combined with the local 3N interaction at N$^2$LO~\cite{Navratil2007} with a cutoff $\Lambda_{\text{cut,3N}}=400\mev$~\cite{Roth2012} (see Sec.~\ref{sec:ham}). 
This choice is motivated by the observation that for $\Lambda_{\text{cut,3N}}=500\mev$ the Hamiltonian tend to overbind the n-\elemA{8}{Be} threshold by about $800\,\text{keV}$ in calculations with the IT-NCSM at $N_{\text{max}}=12$~\cite{Langhammer2015}.
In addition, the NN and NN+3N interaction is softened by the SRG evolution, leading to the NN+3N-ind and NN+3N-full Hamiltonian, respectively, as described in Sec.~\ref{sec:srg}.    
The \elemA{9}{Be} spectrum is an ideal candidate for the NCSMC with explicit 3N interactions using a single-nucleon-projectile channel, since only the ground-state is bound, while all excited states are in the continuum above the n-\elemA{8}{Be} threshold energy, which is experimentally located at $1.665\mev$~\cite{Tilley2004155}.

\begin{figure}
\centering\includegraphics[width=1.0\textwidth]{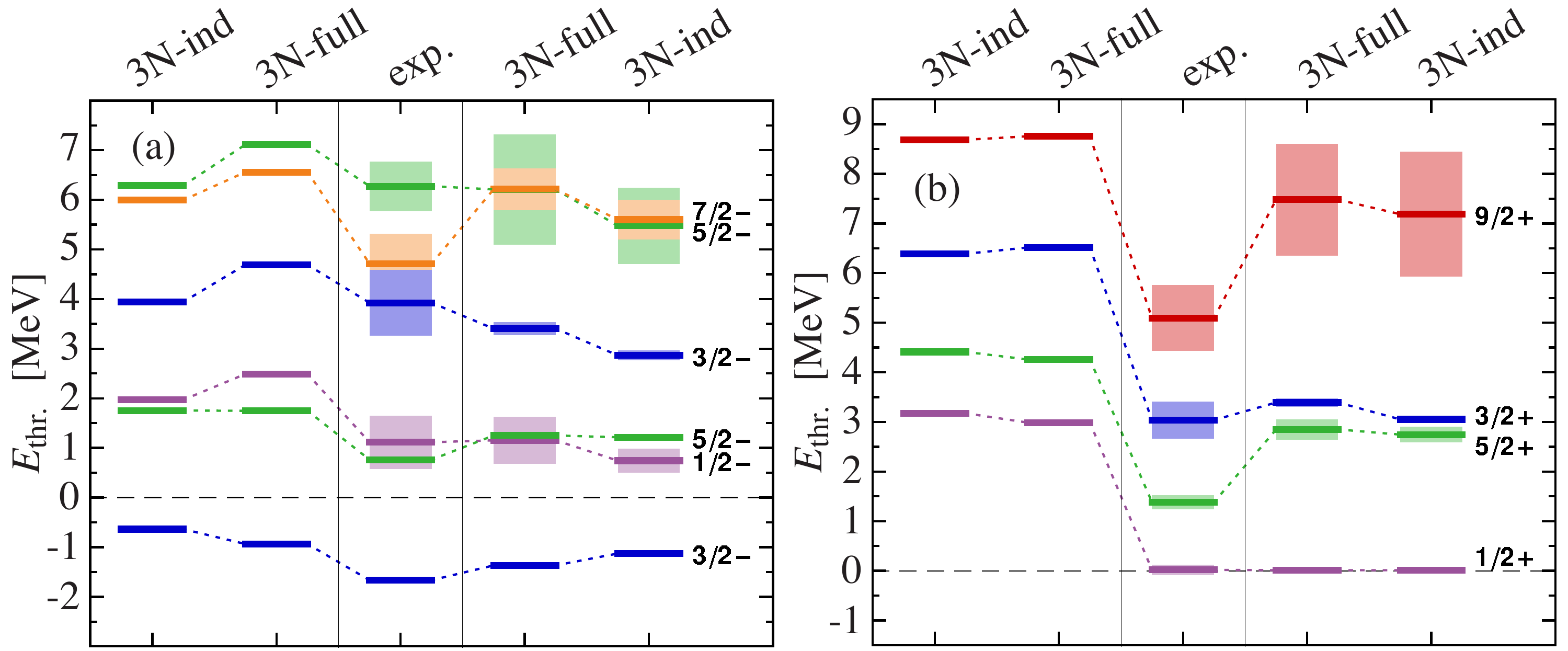}

\caption{
{\small 
Negative (a) and positive (b) parity spectrum of \elemA{9}{Be} relative to the n-\elemA{8}{Be} threshold at $N_{\text{max}}=12$ and $11$, respectively. Shown are NCSM (first two columns) and NCSMC (last two columns) results compared to experiment\cite{Tilley2004155}. First and last columns contain the energies for the NN+3N-ind and the second and fourth column for the NN+3N-full Hamiltonian, respectively. Shaded areas denote the width of the energy levels. Remaining parameters are $\hbar\Omega= 20\mev$ and $\Lambda=2.0\fm^{-1}$. Figure taken from~\cite{Langhammer2015}.
 }
\label{fig:Be9_Spec}}
\end{figure}

The NCSMC wave function~\eqref{NCSMC_wav} in Sec.~\ref{sec:ncsmc}  consist of the eigenstates of the compound system and the NCSM/RGM expansion of the cluster channel. For the first term we include the first 4 positive and 6 negative parity eigenstates of \elemA{9}{Be} obtained with the NCSM.
This selection consists of the $1/2^+$, $5/2^+$, $3/2^+$, $9/2^+$ and $3/2^-$, $5/2^-$, $1/2^-$, $3/2^-$, $7/2^-$, $5/2^-$ states and contains all excited states up to $8\mev$ above the n-\elemA{8}{Be} threshold, which is consistent with experimental data~\cite{Tilley2004155}. Moreover, in the NCSM calculations and experiment there is a gap of about $3\mev$ between the included second $5/2^{-}$ state, and the next known resonance at $11.2\mev$.
For the second term of expansion~\eqref{NCSMC_wav} we restrict ourselves to channels with a single-neutron projectile and \elemA{8}{Be} target. For the target we include the $0^+$ ground state of \elemA{8}{Be}, as well as its first excited $2^+$ state obtained from the NCSM.

To reduce the computational costs of the eigenstate calculations the importance truncated NCSM (IT-NCSM)~\cite{Roth2009,Roth2007} is used, which also simplifies the calculations of the NCSM/RGM and NCSMC coupling kernels. This is because only relevant Slater determinants are considered for the expansion of the eigenstates.
The basis reduction within the IT-NCSM has only a minor impact on the resonance positions in the \elemA{9}{Be} spectrum~\cite{Langhammer2015}.
The convergence with respect to $N_{\text{max}}$ is illustrated in Fig.~\ref{fig:Be9_PS_NMax} for the positive- and negative-parity eigenphase shifts of n-\elemA{8}{Be} using the NN+3N-ind and NN+3N-full Hamiltonian. 
The typical convergence pattern shows a shift of the resonance positions when going from $N_{\text{max}}=8$ to $10$ but only a minor change from the step to $N_{\text{max}}=12$, indicating an approach to convergence.
The eigenphase shifts are most sensitive to the model-space size near resonances with the sole exception of the $1/2^+$ eigenphase shift, which is affected at all energies.
Overall the eigenphase shifts and the spectrum resulting from the resonance centroids are reasonably converged, in particular for the negative parity states. An extensive study of the effect of incorporated truncations can be found in Ref.~\cite{Langhammer2015}. 

\begin{table*}[t]
\centering
\begin{tabular}{c || c c | c c}

                           	& \multicolumn{2}{ c |}{NCSMC} 								& \multicolumn{2}{ c }{experiment }  \\  
  \elemA{9}{Be}		& $E_{R}\,[\text{MeV}]$ 	& $\Gamma\,[\text{MeV}]$	&   $E_{R}\,[\text{MeV}]$ 	& $\Gamma\,[\text{MeV}]$ \\
         \hline          
         \hline         
 $\frac{5}{2}^{+}$   	&  3.39    						& 0.17								 & 1.38								& 0.28 \\
         \hline         
 $\frac{3}{2}^{-}$   	&  -1.367   						& -								 	& -1.66								& - \\
         \hline         
 $\frac{1}{2}^{-}$   	&  1.15   						& 0.95								 & 1.11								& 1.08 \\
         \hline     
  $\frac{5}{2}^{-}$   	&  1.25   						& 0.02$\,\text{keV}$			 & 0.76								& 0.78$\,\text{keV}$ \\
         \hline     
  $\frac{3}{2}^{-}$   	&  3.4   							& 0.26								 & 3.92								& 1.33 \\
         \hline     
  $\frac{7}{2}^{-}$   	&  6.21   						& 0.84								 & 4.71								& 1.21 \\

\end{tabular}
\caption{Energies of the bound state and resonances relative to the n-8Be threshold and resonance width in MeV for the NCSMC with the NN+3N-full Hamiltonian with $\Lambda_{cut,3N} = 400\mev$ and $N_{\text{max}} =11$ and $12$ for positive and negative parity states, respectively. The values are extracted from the procedure described in the text and compared to experiment~\cite{Tilley2004155}. 
}
\label{tab:Be9_Resonances}
\end{table*}

In Fig.~\ref{fig:Be9_Spec} we illustrate the excitation spectrum of \elemA{9}{Be}, and compare the results for the NCSM and NCSMC using the NN+3N-ind and NN+3N-full Hamiltonian.
The resonance centroid $E_{R}$ is determined by the inflection point of the eigenphase shifts and the width follows from $\Gamma=2/(\text{d}\delta(E_{\text{kin}})/\text{d}E_{\text{kin}})|_{E_{R}=E_{\text{kin}}}$ with the eigenphase shifts $\delta$ in units of radians~\cite{Thompson09}.
The resonance positions and widths are summarized in Tab.~\ref{tab:Be9_Resonances}. Note, that the applied procedure for the determination  $E_{R}$ and $\Gamma$ is generally only valid for narrow resonances that can be approximated by a Breit-Wigner shape. Thus, the $1/2^{+}$- and $3/2^{+}$-states as well as the broad resonances are not quoted in the table and require more elaborated approaches~\cite{Csoto1997}, that are currently investigated~\cite{Jeremy2015a}. 
The positive-parity states in Fig.~\ref{fig:Be9_Spec}(b) with both methods are rather insensitive to initial 3N interactions.
On the other hand,  for the negative parity states in Fig.~\ref{fig:Be9_Spec}(a), all states, except the first $5/2^-$ resonance, are sensitive to the inclusion of the initial chiral 3N interaction with effects of roughly similar size for both the NCSM and the NCSMC: the inclusion of the chiral 3N interaction increases the resonance energies relative to the threshold. For the NCSM calculations the agreement with experiment generally deteriorates when the initial 3N interaction is included, while once the continuum effects treated properly  with the NCSMC the overall agreement clearly improves when the 3N interaction is included. 

In this context it is important to note that the NCSMC spectrum is reasonably well converged, while the spectrum resulting from the pure (IT)-NCSM calculations is poorly converging at similar $N_{\text{max}}$ values~\cite{Langhammer2015}, such that conclusions about the impact of 3N interactions are only reliable for investigations that cope with continuum effects.

Although the relevance of cluster structures beyond the single-nucleon binary-cluster ansatz used here cannot be
ruled out, one might expect larger sensitivities to the NCSM model-space size if such structures were relevant. Therefore, even though one has to be also wary of some impact of SRG transformations the present deviations from experiment are likely to be connected to deficiencies of the chiral NN+3N Hamiltonian.

%
%

%% file: SE_6helium.tex
\section{Ground and continuum states of the $^6$He nucleus}
\label{sec:He6-res}

The lightest Borromean nucleus is $^6$He~\cite{Tanihata:1995yv,PhysRevLett.55.2676},
that makes it the perfect candidate to be studied within the NCSM/RGM for three-cluster
systems. Therefore the method was first used in \cite{Quaglioni:2013kma,Romero-Redondo:2014fya} 
 to study such nucleus. Both, its ground and continuum states have been studied
using a two-body interaction, namely the SRG evolved~\cite{Bogner2007,PhysRevC.77.064003} potential obtained
from the chiral N$^3$LO $NN$ interaction \cite{Entem2003} with the evolution parameter $\Lambda$=1.5 fm$^{-1}$.
Using such a soft potential has the great advantage of providing fast convergence. Furthermore,
obtaining an accurate binding energy within the NCSM is possible and therefore it provides 
a well-founded benchmark for the NCSM/RGM results.

For present calculations, only the ground state of $^4$He was included in the
cluster basis. The inclusion of some of its excited states may be necessary
in order to take into account all many-body correlations, however,
it implies an increase in the size of the problem that is not feasible 
given current computational capabilities.  
In order to overcome this limitation, it is possible to use an extension
of the NCSMC to ternary cluster, in this case the NCSM eigenstates of the   
six-body system can compensate for the missing many-body correlations. 
The extension of NCSMC to three-cluster systems and its results for $^6$He
will be presented elsewhere \cite{Romero-Redondo2016}.

\subsection{Ground state of $^6$He} 

We calculate the ground state (g.s.) of $^6$He by solving Eq. (\ref{3B_ortho}).
The convergence of this $J^{\pi}T=0^+1$ state is studied with respect
to all parameters included in the calculation. Examples of this study
are shown in section \ref{3B_conv}, where a good stability and 
convergence with respect to the maximuun hypermomentum
used in the expansion (\ref{expansionHH}) and with respect to the size of the extension
of the model space ($N_{\rm ext}$) is shown. 
In particular, the NCSM/RGM convergence of the energy with respect to the
size of the model space is shown in the third column of Table \ref{energy_6he}. While
the convergence is rather fast, the obtained energy is about
1~MeV less bound compared to the expected result with the NN potential used
(i.e. from the extrapolated value obtained through NCSM shown in the fourth column of Table~\ref{energy_6he}). This 
missing binding energy gives a measure of the effect of the 
 many-body correlations that remain unaccounted for when using
only the ground state of $^4$He in the cluster basis.    
As shown in Ref.~\cite{Romero-Redondo2016}, the
extrapolated NCSM result for the binding energy is recovered, 
already at $N_{\rm max}=10$, when working with the NCSMC.

\begin{table}
\centering
\begin{tabular}{c c c c} 
\hline
$N_{\rm max}$ & $^4$He-NCSM &$^6$He-NCSM/RGM & $^6$He-NCSM\\
\hline
6& $-27.984$ & $-28.907$ & $-27.705$ \\
8& $-28.173$ & $-28.616$ & $ -28.952$\\
10& $-28.215$ &  $-28.696$ & $-29.452$\\
12& $-28.224$ & $-28.697$ & $-29.658$ \\
\hline
Extrapolation & $-28.23(1)$ & --- & $-29.84(4)$ \\
\hline
Experimental  & $-28.296$& \multicolumn{2}{c}{$-29.268$} \\ 
\hline
\end{tabular}
\caption{In the second column we show the convergence in terms of model space size
$N_{\rm max}$ of ground state
energy of $^4$He (in MeV) within the NCSM formalism. The third column shows the same for $^6$He within the NCSM/RGM method. The last
column shows the $^6$He g.s. energies (in MeV)
for a model space size of $N_{\rm max}-2$. The extrapolated 
values for the NCSM calculations to $N_{\rm max}\rightarrow\infty$ has been obtained by an
exponential fit using $E(N_{\rm max})=E_\infty+a\; e^{-b N_{\rm max}}$. In the last row, the experimental values are shown.} 
\label{energy_6he}
\end{table}

Despite the limitation of the NCSM/RGM as to obtaining the correct binding energy 
due to the restrictions imposed in the cluster basis, i.e., the lack of inclusion
of excited states of the $\alpha$ core, the formalism gives rise to a wave function that has
the correct asymptotic behavior, which is included by construction when 
using the $R$-matrix method. This is extremely important when describing halo nuclei
such as $^6$He that exhibit an extended tail. This is a great advantage with respect to 
the NCSM that yields Gaussian asymptotic behavior due to the expansion over HO basis
states.



\begin{figure}[t]
\includegraphics[width=80mm]{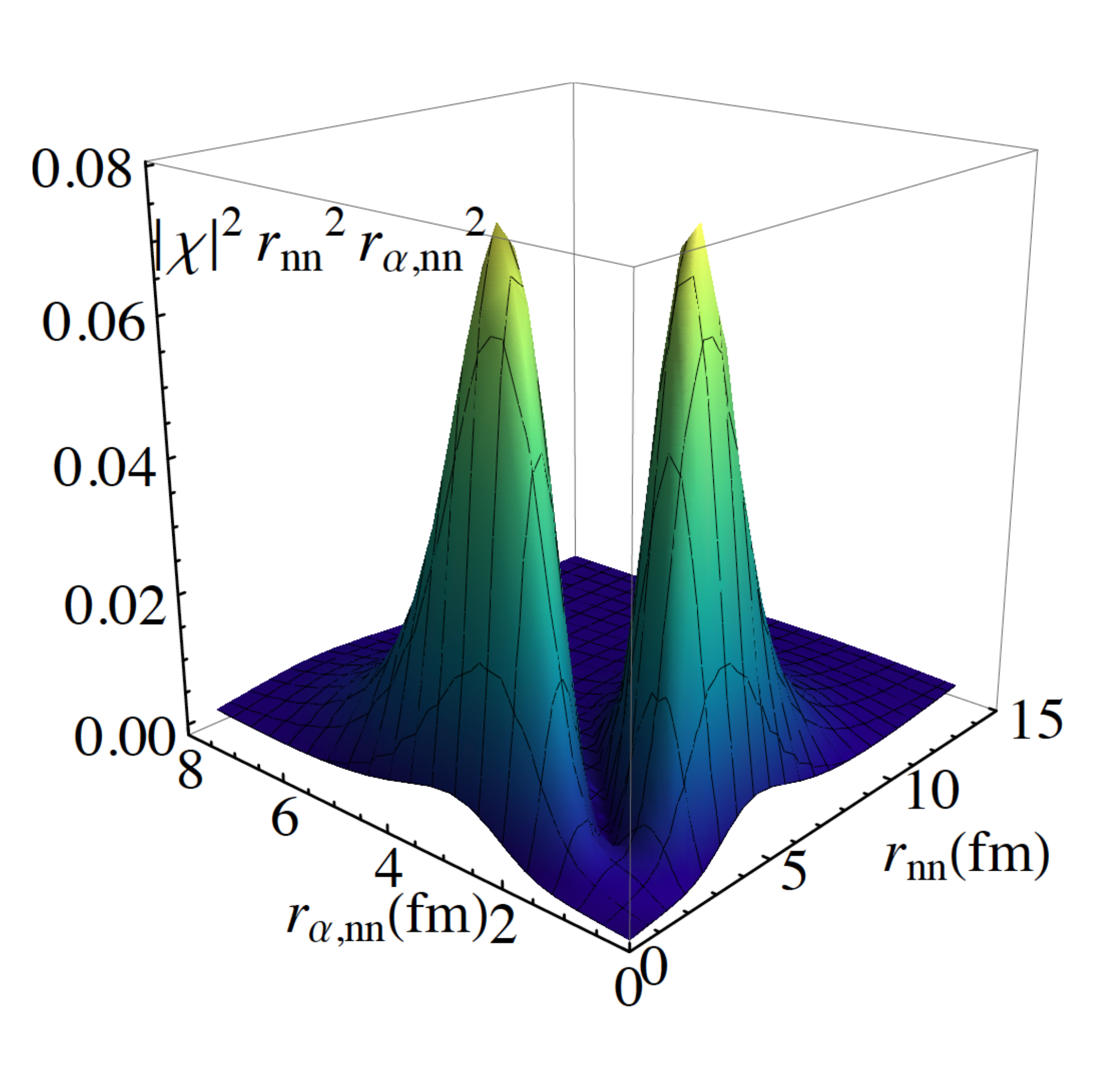} 
\includegraphics[width=85mm]{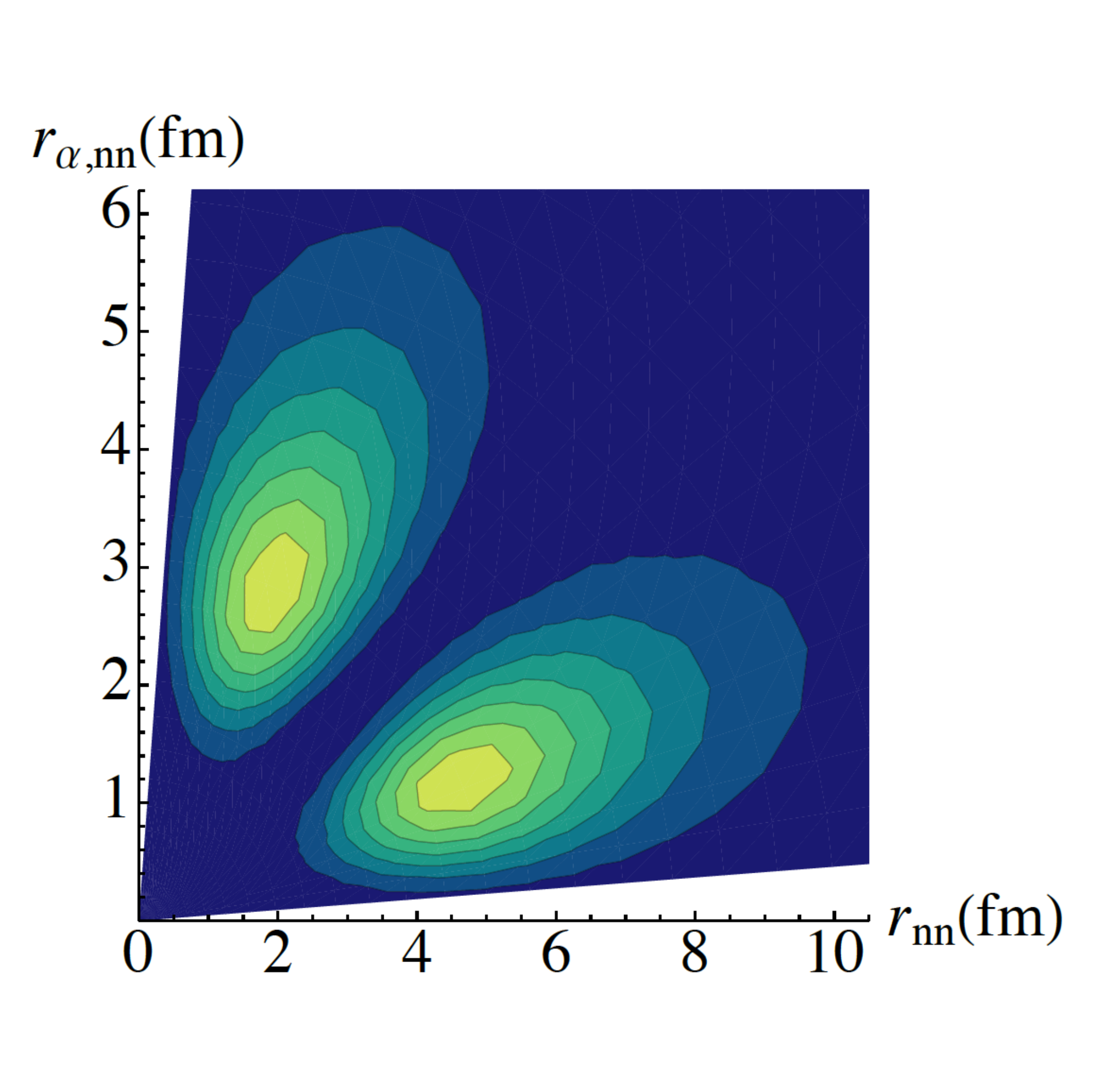}
\caption{Probability distribution (left) and its contour diagram (right) for the main component of the $^4$He+$n$+$n$
relative motion wave function for the $J^\pi T=0^+1$ ground state. The quantum numbers corresponding to this
component are $S=L=\ell_x=\ell_y=0$. Here $r_{nn} = \sqrt{2}\, \eta_{nn}$ and $r_{\alpha,nn}=\sqrt{3/4}\,\eta_{\alpha,nn}$ are respectively the distance between the two neutrons and the distance between the c.m. of $^4$He and that of the two neutrons. 
Figures first appeared in \cite{Quaglioni:2013kma}.}
\label{probability}
\end{figure}

Plotting the probability distribution (or probability density) provides a visual description
of the structure of the $^6$He g.s., in particular, it gives an idea of the distribution of the 
neutrons respect to the $\alpha$ core. In Fig. \ref{probability}, we show such distribution 
 which presents two peaks corresponding to the characteristic di-neutron (two neutrons close together)
and cigar (two neutrons far apart in opposite sides of the $\alpha$ particle) configurations
of $^6$He.

\subsection{$^4$He+$n$+$n$ Continuum}

For the study of the contiuum of $^6$He, we used the same NN potential we used in the previous section,
which allows our results to reach convergence in the HO expansions within $N_{\rm max}\sim 13$
(the largest model space currently feasible). 
We solve Eqs. (\ref{RGMrho}) using
 the corresponding asymptotic conditions (\ref{cont_asym}) in order to 
obtained the three-body phase shifts for the $J^{\pi}=0^{\pm}, 1^{\pm}$ and $2^{\pm}$ channels.
The phase shifts can be extracted either from the diagonal elements of the 
scattering matrix (diagonal phase shifts) or from its diagonalization (eigenphase shifts),
however, when large off-diagonal couplings are present, the use of eigenphase shifts is
more appropriate.      

\begin{figure}[t]
\includegraphics[height=75mm]{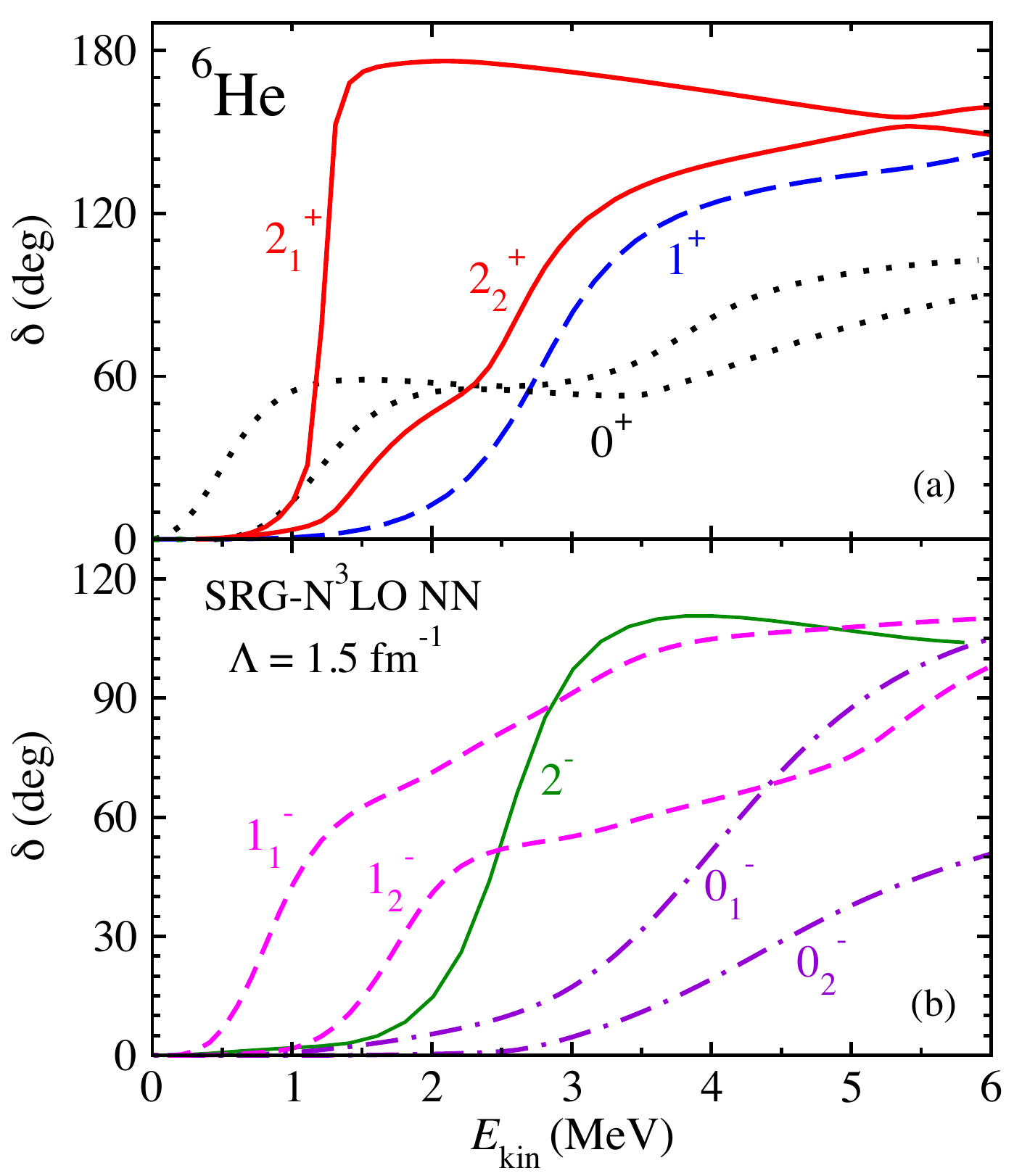} 
\includegraphics[height=75mm]{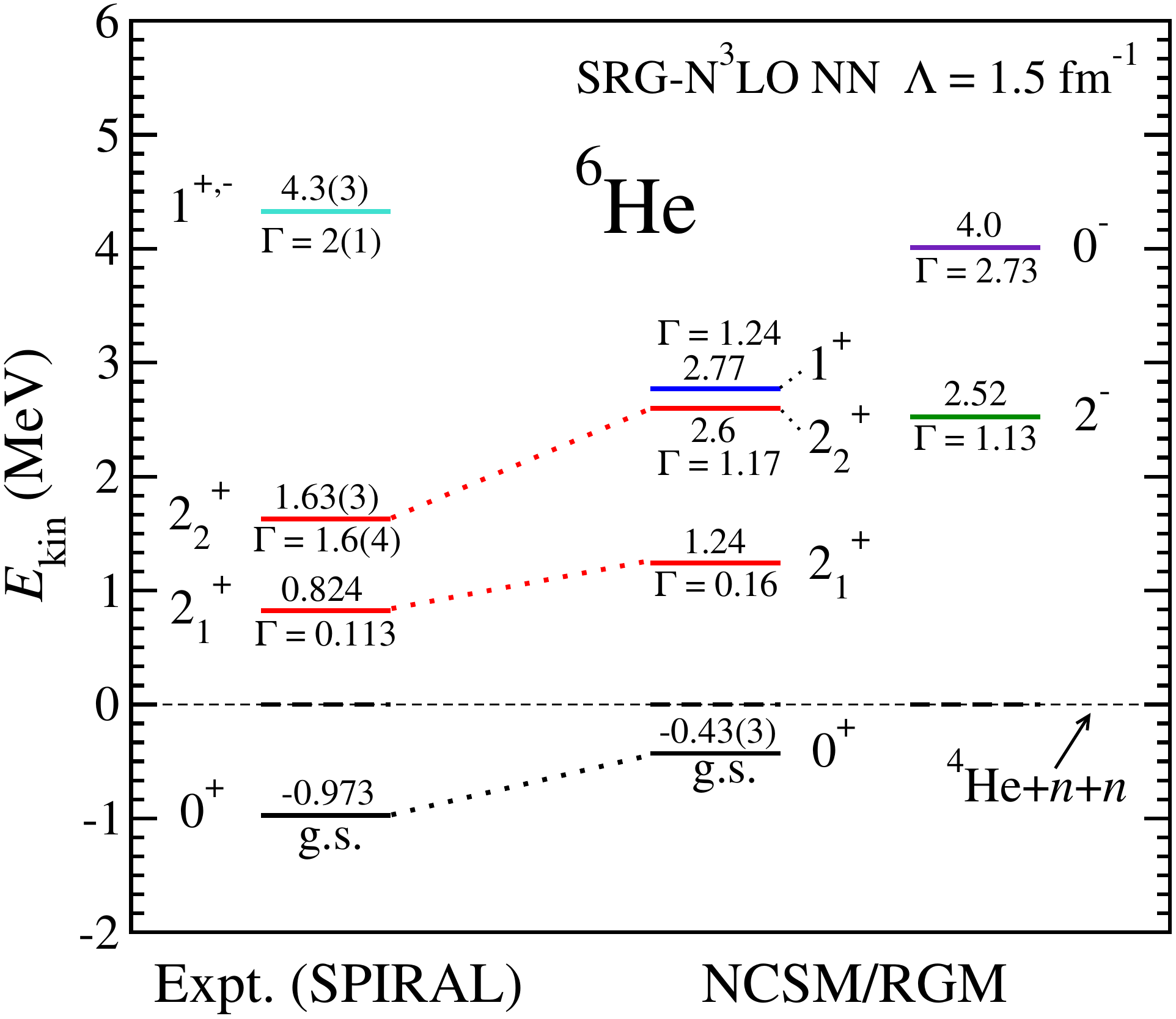}
\caption{In the left, calculated $^4{\rm He}$+$n$+$n$ (a) positive- and (b) negative-parity attractive eigenphase shifts as a function of
the kinetic energy $E_{\rm kin}$ with respect to the two-neutron emission threshold.
In the right, energy spectrum obtained from those phase shifts, compared to the most recent experimental
spectrum \cite{Mougeot:2012aq}. Figures first appeared in \cite{Romero-Redondo:2014fya}.}
\label{fig:3B_cont}
\end{figure}

From the behavior of the phase shifts it is possible to identify
the presence of resonances in the different channels. In Fig. \ref{fig:3B_cont}, we show
in the left hand panel the positive and negative parity eigenphase shifts as a function of the 
kinetic energy $E_{\rm kin}$ with respect to the two-neutron emission threshold, while in the
right hand panel, we show the energy spectrum for $^6$He, the energies and widths of the
resonances were
extracted from the phase shifts obtained for the corresponding channels.

We found several resonances, in particular we found two resonances in the $2^+$ channel, which
include the well-known narrow $2^+_1$ and the recently measured broader $2^+_2$. Additional
resonances were located in the $2^-$, $0^-$ and $1^+$ channels. However, we did not find a resonance
in the $1^-$ channel, and therefore our results do not support the idea that the accumulation of dipole strength 
at low energy is originated by a three-body resonance in this channel.

%% file: SE_radiative-capt.tex
\section{The $^7$Be$(p,\gamma)^8$B radiative capture}
\label{sec:capture}

The core temperature of the Sun can be determined with high accuracy through measurements of the $^8$B neutrino flux, currently known with a $\sim9\%$ precision~\cite{PhysRevLett.92.181301}. 
An important input in modeling this flux are the rates of the $^3$He($\alpha,\gamma$)$^7$Be and the $^7$Be($p,\gamma$)$^8$B radiative capture reactions~\cite{RevModPhys.83.195,RevModPhys.70.1265}. The $^7$Be($p,\gamma$)$^8$B reaction constitutes the final step of the nucleosynthetic chain leading to $^8$B. At solar energies this reaction proceeds by external, predominantly nonresonant $E1$, $S$- and $D$-wave capture into the weakly-bound ground state (g.s.) of $^8$B. 
Experimental determinations of the $^7$Be($p,\gamma$)$^8$B capture include direct measurements with proton beams on $^7$Be targets~\cite{PhysRevLett.50.412,PhysRevC.28.2222,PhysRevLett.90.022501,PhysRevC.68.065803} as well as indirect measurements through the breakup of a $^8$B projectile into $^7$Be and proton in the Coulomb field of a heavy target~\cite{Baur1986188,PhysRevLett.83.2910,PhysRevLett.86.2750,PhysRevC.73.015806,PhysRevLett.90.232501}. Theoretical calculations needed to extrapolate the measured S-factor to the astrophysically relevant Gamow energy were performed with several methods: the R-matrix parametrization~\cite{Barker1995693}, the potential model~\cite{PhysRevC.7.543,Typel1997147,PhysRevC.68.045802}, microscopic cluster models~\cite{Descouvemont1994341,PhysRevC.52.1130,PhysRevC.70.065802} and also using the {\it ab initio} no-core shell model wave functions for the $^8$B bound state~\cite{Navratil2006}. The most recent evaluation of the $^7$Be($p,\gamma$)$^8$B  S-factor (proportional to the cross section) at zero energy, $S_{17}(0)$, has a $\sim$10\% error dominated by the uncertainty in theory~\cite{RevModPhys.83.195,RevModPhys.70.1265}. 

We performed many-body calculations of the $^7$Be($p,\gamma$)$^8$B capture within the NCSM/RGM starting from a NN interaction that describes two-nucleon properties with a high accuracy ~\cite{Navratil2011a}. In particular, we used an SRG evolved chiral N$^3$LO NN~\cite{Entem2003} and chose the SRG evolution parameter $\Lambda$ so that the experimental separation energy (s.e.) of the $^8$B weakly bound $2^+$ ground-state with respect to the $^7$Be+$p$ is reproduced in the largest-space calculation that we were able to reach. We note that for the calculation of the low-energy behavior of the $S_{17}$ S-factor, a correct s.e. is crucial. Using the five lowest eigenstates of $^7$Be (i.e., $3/2^-$ g.s. and $1/2^-, 7/2^-, 5/2^-_1$ and $5/2^-_2$ excited states) in the $N_{\rm max}{=}10$ model space and solving the NCSM/RGM equations with bound-state boundary conditions we were able to reproduce experimental s.e. for $\Lambda=1.86$ fm$^{-1}$. 

\begin{figure}
\begin{minipage}{0.60\columnwidth}
\includegraphics*[width=\columnwidth]{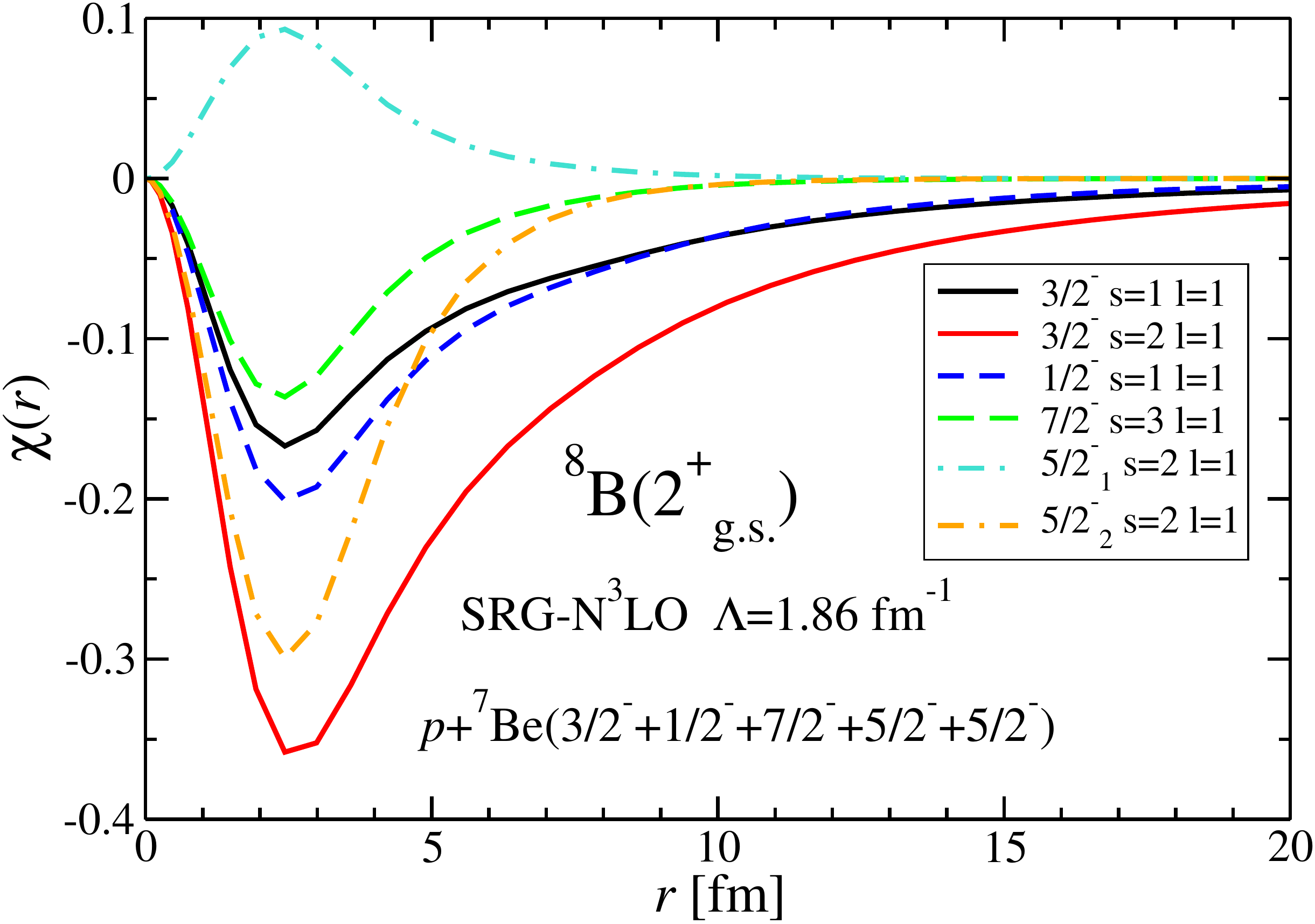}
\end{minipage}
\begin{minipage}{0.40\columnwidth}
\centering
\caption{Dominant $P$-wave components of the $2^+$ $^8$B g.s.\ wave function for $N_{\rm max}=10$ and $\hbar\Omega=18$ MeV, using the SRG-N$^3$LO $NN$ potential with $\Lambda=1.86$ fm$^{-1}$. The NCSM/RGM calculation includes $^7$Be g.s.\ and $1/2^-$, $7/2^-$, $5/2^-_1$ and  $5/2^-_2$ excited states. Figure from Ref. \cite{Navratil2011a}.}
\label{fig:B8_gs}
\end{minipage}
\end{figure}
In Fig.~\ref{fig:B8_gs}, we plot the most significant components of the radial wave functions $\chi(r)$ for the $2^+$ g.s. of $^8$B. The dominant component is clearly the channel-spin $s{=}2$ $P$-wave of the $^7$Be(g.s.)-$p$ that extends to a distance far beyond the plotted range. Remarkably, we notice a substantial contribution from the $^7$Be$(5/2^-_2){-}p$ $P$-wave in the channel spin $s{=}2$. (The other possible $s{=}3$ $P$-wave configuration is negligible). At the same time, the $^7$Be $5/2^-_2$ state is dominated by a $^6$Li-$p$ channel-spin $s{=}3/2$ $P$-wave configuration. Within the NCSM framework relevant to the present calculations this was shown (for the mirror $^7$Li-$n$ system) in Ref.~\cite{PhysRevC.70.054324}. Therefore, such a large contribution of the $s{=}2$ $^7$Be$(5/2^-_2){-}p$ $P$-wave to the $^8$B ground state seems to indicate the presence of two antiparallel protons outside of a $^6$Li core, and that their exchanges are important. Clearly, for a realistic description of the $^8$B g.s., this state must be taken into account.

\begin{figure}
\begin{minipage}{0.60\columnwidth}
\includegraphics*[width=\columnwidth]{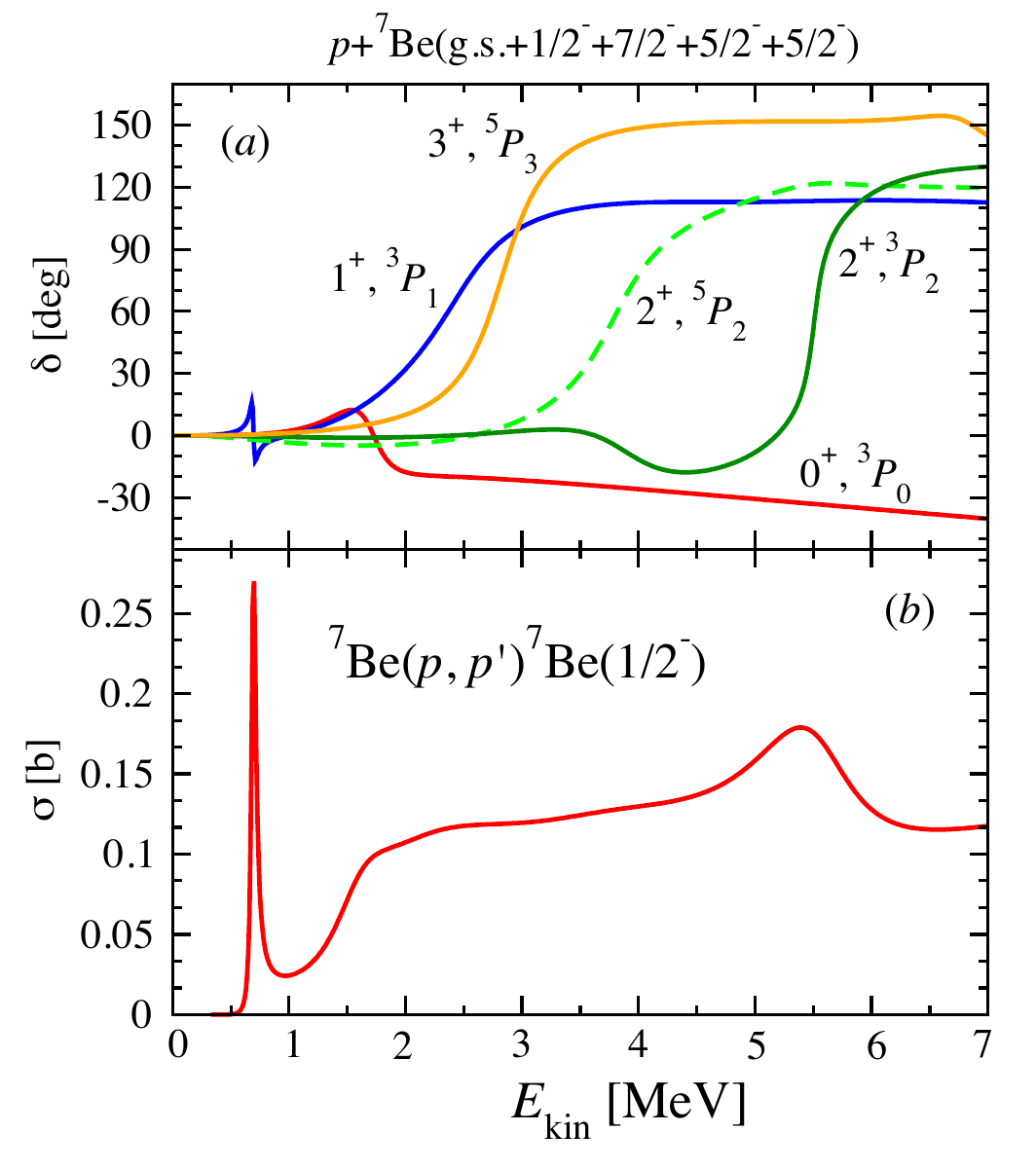}
\end{minipage}
\begin{minipage}{0.40\columnwidth}
\centering
\caption{$P$-wave $(a)$ diagonal phase shifts of $p$-$^7$Be elastic scattering and the inelastic $^7$Be($p$,$p'$)$^7$Be(1/2$^-$) cross section $(b)$. Calculations as described in Fig.~\protect\ref{fig:B8_gs}. Figure adapted from Ref. \cite{Navratil2011a}.}
\label{fig:Be7_31755_Pwaves}
\end{minipage}
\end{figure}

Next, using the same NN interaction, we solve the NCSM/RGM equations with scattering-state boundary conditions for a chosen range of energies and obtain scattering wave functions and the scattering matrix. The resulting phase shifts and cross sections are displayed in Fig.~\ref{fig:Be7_31755_Pwaves}. All energies are in the center of mass (c.m.). We find several $P$-wave resonances in the considered energy range. The first $1^+$ resonance, manifested in both the elastic and inelastic cross sections, corresponds to the experimental $^8$B $1^+$ state at $E_x{=}0.77$ MeV (0.63 MeV above the $p$-$^7$Be threshold)~\cite{Tilley2004155}. The $3^+$ resonance, responsible for the peak in the elastic cross section, corresponds to the experimental $^8$B $3^+$ state at $E_x{=}2.32$ MeV. However, we also find a low-lying $0^+$ and additional $1^+$ and $2^+$ resonances that can be distinguished in the inelastic cross section. In particular, the $s{=}1$ $P$-wave $2^+$ resonance is clearly visible. There is also an $s{=}2$ $P$-wave $2^+$ resonance with some impact on the elastic cross section. These resonances are not included in the current $A{=}8$ evaluation~\cite{Tilley2004155}. We note, however, that the authors of the recent Ref.~\cite{PhysRevC.82.011601} do claim observation of  low-lying $0^+$ and $2^+$ resonances based on an R-matrix analysis of their $p$-$^7$Be scattering experiment. Their suggested $0^+$ resonance at 1.9 MeV is quite close to the calculated $0^+$ energy of the present work.

\begin{figure}[b]
\begin{minipage}{18pc}
\includegraphics[width=18pc]{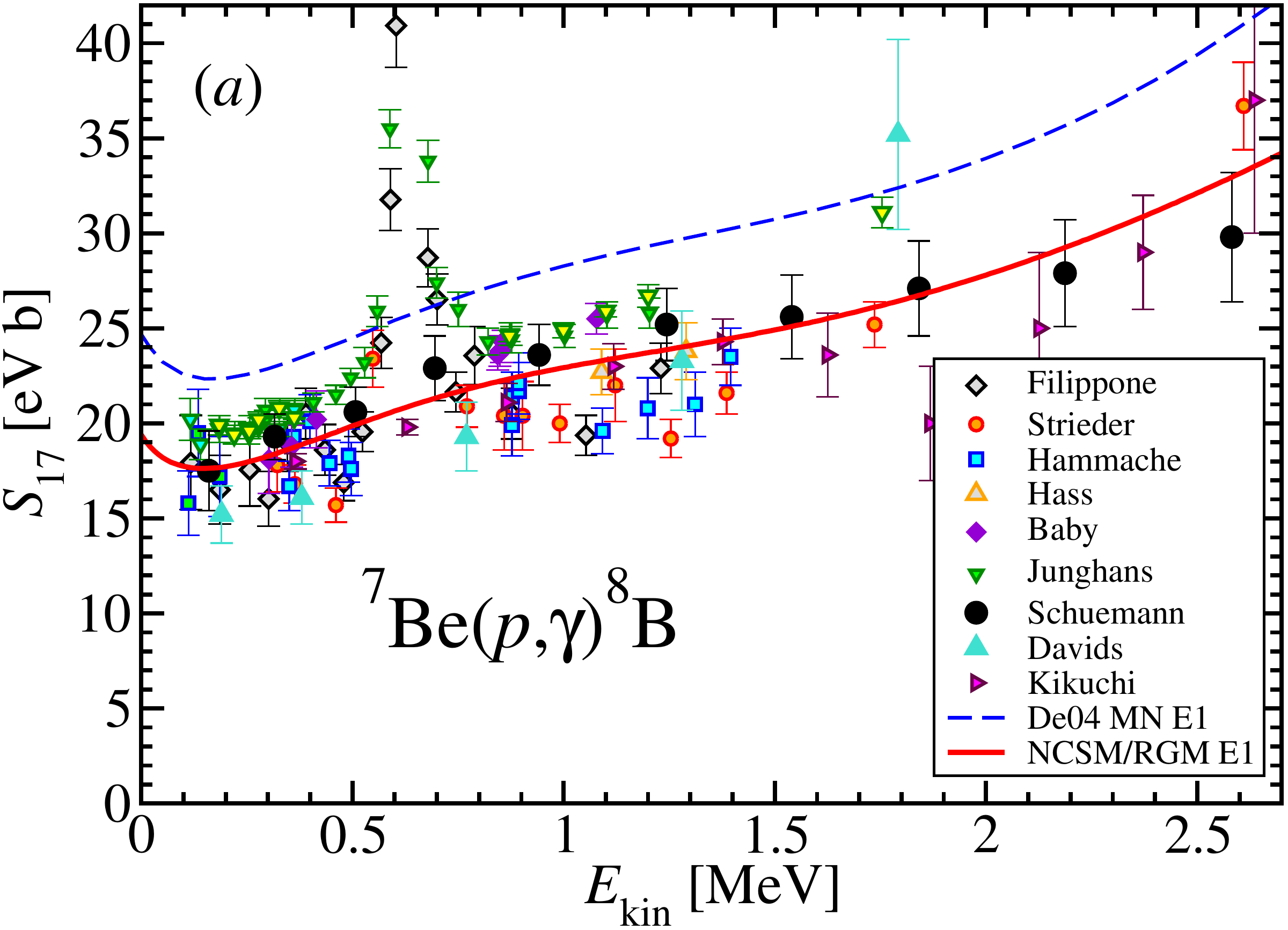}
\end{minipage}\hspace{1pc}%
\begin{minipage}{18pc}
\includegraphics[width=18pc]{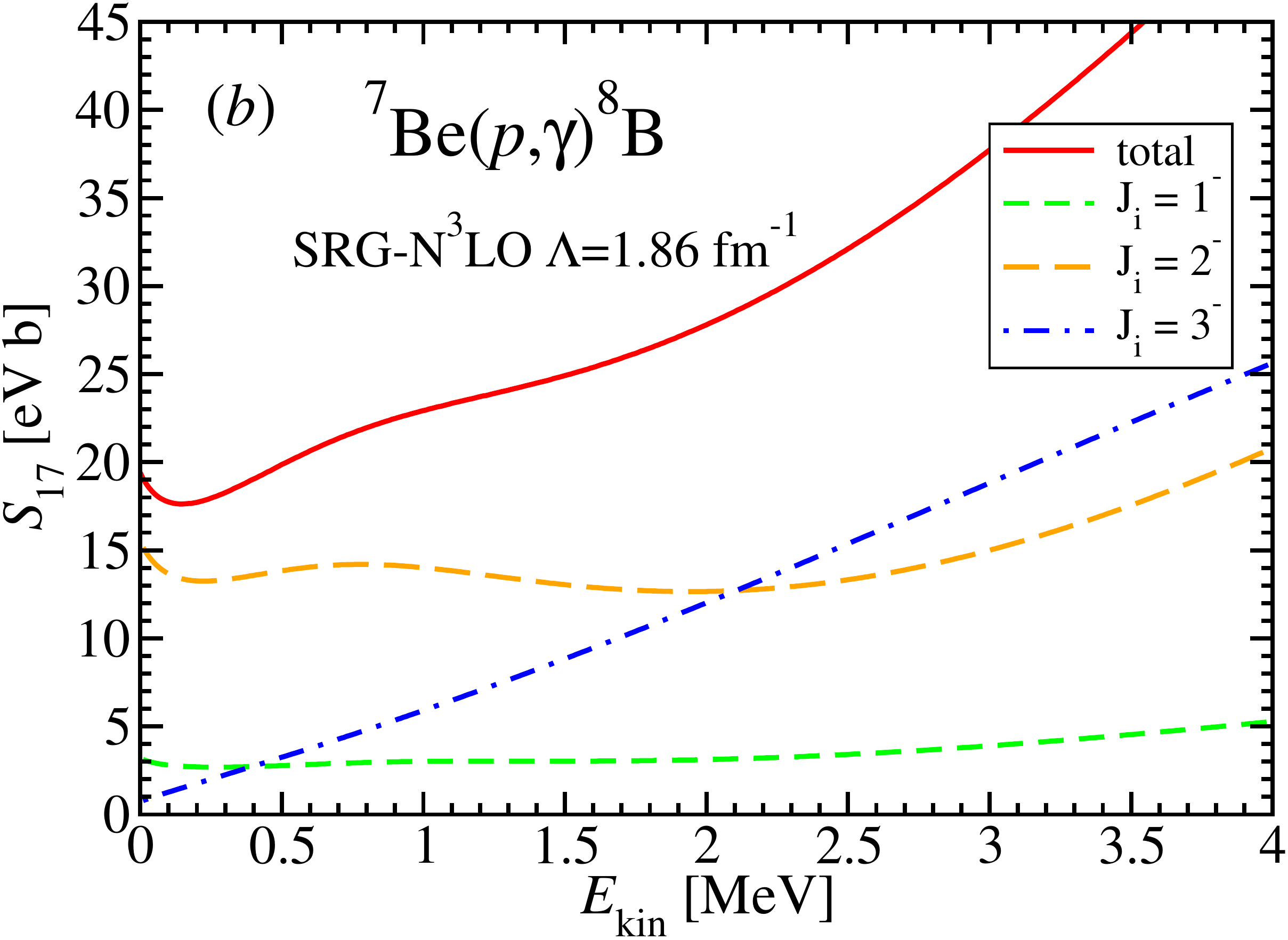}
\end{minipage} 
\caption{Calculated $^7$Be($p$,$\gamma$)$^8$B S-factor as function of the energy in the c.m.\ compared to data and the microscopic cluster-model calculations of Ref.~\cite{PhysRevC.70.065802} (blue dashed line) with the Minnesota (MN) interaction $(a)$. Only $E1$ transitions were considered. Initial-state partial wave contributions are shown in panel $(b)$. Calculation as described in Fig.~\protect\ref{fig:B8_gs}. Figures from Ref. \cite{Navratil2011a}.}
\label{fig:p7Be1} 
\end{figure}

With the resulting bound- and scattering-state wave functions that are properly orthonormalized and antisymmetrized, we calculate the $^7$Be($p$,$\gamma$)$^8$B radiative capture using a one-body $E1$ transition operator. We use the one-body $E1$ operator defined in Eq.~(\ref{E1op}) that includes the leading effects of the meson-exchange currents through the Siegert's theorem. The resulting $S_{17}$ astrophysical factor is compared to several experimental data sets in Figure~\ref{fig:p7Be1}.  In the data, one can see also the contribution from the $1^+$ resonance due to the $M1$ capture that does not contribute to a theoretical calculation outside of the resonance and is negligible at astrophysical energies~\cite{RevModPhys.83.195,RevModPhys.70.1265}. Our calculated S-factor is somewhat lower than the recent Junghans data~\cite{PhysRevC.68.065803} but the shape reproduces closely the trend of the GSI data~\cite{PhysRevC.73.015806,PhysRevLett.90.232501}, which were extracted from Coulomb breakup. The shape is also quite similar to that obtained within the microscopic three-cluster model~\cite{PhysRevC.70.065802} (see the dashed line in Fig.~\ref{fig:p7Be1} $(a)$) used, after scaling to the data, in the most recent $S_{17}$ evaluation~\cite{RevModPhys.83.195}. The contributions from the initial $1^-$, $2^-$ and $3^-$ partial waves are shown in panel $(b)$ of Fig.~\ref{fig:p7Be1}. 

An interesting feature of the S-factor is its flattening around 1.5 MeV. As seen in Fig.~\ref{fig:p7Be1} $(b)$, this phenomenon is due to the $S$-wave contribution that dominates the $J_i{=}2^-$  and $1^-$ partial waves at low energies. The increase of flattening with the number of $^7$Be  eigenstates included in the calculation, see Fig. 5 in Ref.~\cite{Navratil2011a}, indicates that this is an effect due to the many-body correlations. This finding corroborates the observations of Ref.~\cite{PhysRevC.70.065802}, where the flattening was attributed to the deformation of the $^7$Be core. We also note that the flattening found in the present work is slightly larger than that obtained in the microscopic three-cluster model of Ref.~\cite{PhysRevC.70.065802}. Presumably, this is because in the three-cluster model the $^7$Be structure was assumed to be of $^3$He-$^4$He nature only, while the NCSM wave functions include in addition $^6$Li-$p$ configurations, particularly for the $5/2^-_2$ $^7$Be state, as discussed earlier.

The convergence of our results with respect to the size of the HO model space was assessed by means of calculations up to $N_{\rm max}{=}12$ within the importance-truncated NCSM with (due to computational limitations) only the first three-eigenstates of $^7$Be. The $N_{\rm max}{=}10$  and 12 S-factors are very close. As for the convergence in the number of $^7$Be states, we explored it by means of calculations including up to 8 $^7$Be eigenstates in a $N_{\rm max}{=}8$ basis (larger $N_{\rm max}$ values were out of reach with more then five $^7$Be states). Based on this analysis, we estimated the uncertainty of the obtained S-factor. Finally, our calculated $S_{17}(0){=} 19.4(7)$~eV~b is on the lower side, but consistent with the latest evaluation $20.8\pm 0.7({\rm expt})\pm 1.4({\rm theory})$~eV~b~\cite{RevModPhys.83.195,RevModPhys.70.1265}.

The calculations discussed in this section and published in Ref.~\cite{Navratil2011a} can and will be further improved. In particular, we should include the 3N interaction, both the chiral 3N as well as the induced 3N from the SRG transformation. This is a neccessary step towards an {\it ab initio} description. As demonstrated in Sec.~\ref{sec:Be9} in calculations for $^9$Be, we have now developed the capability to do that. Also, to further improve the convergence of the capture calculations, we should utilize the NCSMC expansion of the wave functions rather than just the NCSM/RGM. Again, this has been developed including the capability to calculate the $E1$ and $M1$ contributions to the capture S-factor as discussed in Sections~\ref{sec:ncsmc} and \ref{sec:E1}.

Concerning the related reaction, $^3$He($\alpha,\gamma$)$^7$Be, we have already performed calculations of its S-factor using similarly the SRG evolved chiral EFT NN interaction this time, however, within the NCSMC formalism~\cite{Jeremy2015a}. Actually, for the $^3$He-$^4$He system, we would be unable to achieve any reasonable convergence within the NCSM/RGM alone for technical reasons. The application of the NCSMC becomes unavoidable in this case. We plan to include the 3N interactions also for this reaction in the future, most likely using the normal-ordering approximation.

%% file: SE_fusion.tex
\section{The $^3$H$(d,n)^4$He fusion}
\label{sec:fusion}

The $^3$H($d$,$n$)$^4$He and $^3$He($d$,$p$)$^4$He reactions are
leading processes in the primordial formation of the very light
elements (mass number, $A\le7$), affecting the predictions of Big Bang
nuleosynthesis for light nucleus abundances~\cite{1475-7516-2004-12-010}. With its
low activation energy and high yield, $^3$H($d$,$n$)$^4$He is also the
easiest reaction to achieve on Earth, and is pursued by research
facilities directed toward developing fusion power by either magnetic
({\em e.g.}\ ITER) or inertial ({\em e.g.}\ NIF) confinement.   
The cross section for the $d+^3$H fusion is well known experimentally,
while more uncertain is the situation for the branch of this reaction,
$^3$H$(d, \gamma n)^4$He that produces $17.9$ MeV $\gamma$-rays~\cite{PhysRevLett.53.767,PhysRevC.47.29} and that is being considered as a possible plasma diagnostics in modern fusion experiments. Larger uncertainties dominate also
the $^3$He($d$,$p$)$^4$He reaction that is known for presenting considerable electron-screening effects
at energies accessible by beam-target experiments. Here, the electrons bound to the target, usually a neutral atom or molecule,
lead to increasing values for the reaction-rate with decreasing energy, 
effectively preventing direct access to the astrophysically relevant bare-nucleus cross section. Consensus on the physics mechanism behind this enhancement is not been reached yet~\cite{Kimura2005229}, largely because of the difficulty of determining the absolute value of the bare cross section.
 
Past  theoretical investigations of these fusion reactions include various $R$-matrix analyses of experimental data at higher energies~\cite{PhysRevLett.59.763,PhysRevC.56.2646,PhysRevC.75.027601,Descouvemont2004203}  as well as microscopic calculations with phenomenological interactions~\cite{PhysRevC.41.1191,Langanke1991,PhysRevC.55.536}. However, in view of remaining experimental challenges and the large role played by theory in extracting the astrophysically important information, it is highly desirable to achieve a microscopic description of the $^3$H($d$,$n$)$^4$He and $^3$He($d$,$p$)$^4$He fusion reactions that encompasses the dynamic of all five nucleons and is based on the
fundamental underlying physics: the realistic interactions among nucleons and the structure of the fusing nuclei.

We made the first step in this direction by performing NCSM/RGM calculations using a realistic NN interaction~\cite{Navratil2012}. We started from the SRG-evolved chiral N$^3$LO $NN$ interaction~\cite{Entem2003} with $\Lambda{=}1.5$ fm$^{-1}$, for which we reproduce the experimental $Q$-values of both reactions within $1\%$. This interaction, at the same time, provides an accurate description of the two-nucleon scattering data and of the deuteron properties. The NCSM/RGM calculations were performed starting from eigenstates of the interacting nuclei, {\em i.e.} $^2$H, $^3$H, $^3$He and $^4$He, calculated within the NCSM with the above NN interaction. Important for determining the magnitude of the fusion reactions considered here is the Coulomb interaction. The NCSM/RGM (and the NCSMC) allows for a proper handling of such interaction (particularly its long-range component, which is treated exactly), as described in Sec.~\ref{sec:rgm2} (see Eqs.~(\ref{Hamiltonian})-(\ref{locCoul})). Further, even though the fusion proceeds at very low energies, the deformation and virtual breakup of the reacting nuclei cannot be disregarded, particularly for the weakly-bound deuteron. A proper treatment of deuteron-breakup effects requires the inclusion of three-body continuum states (neutron-proton-nucleus) and is very challenging. In the first fusion application we limited ourselves to binary-cluster channels and approximated virtual three-body breakup effects by discretizing the continuum with excited deuteron pseudostates, strategy that proved successful in our $d$-$^4$He calculations as demonstrated in Sec.~\ref{sec:s-shell}. 

\begin{figure}
\begin{minipage}{0.55\columnwidth}
\includegraphics*[width=1.1\columnwidth]{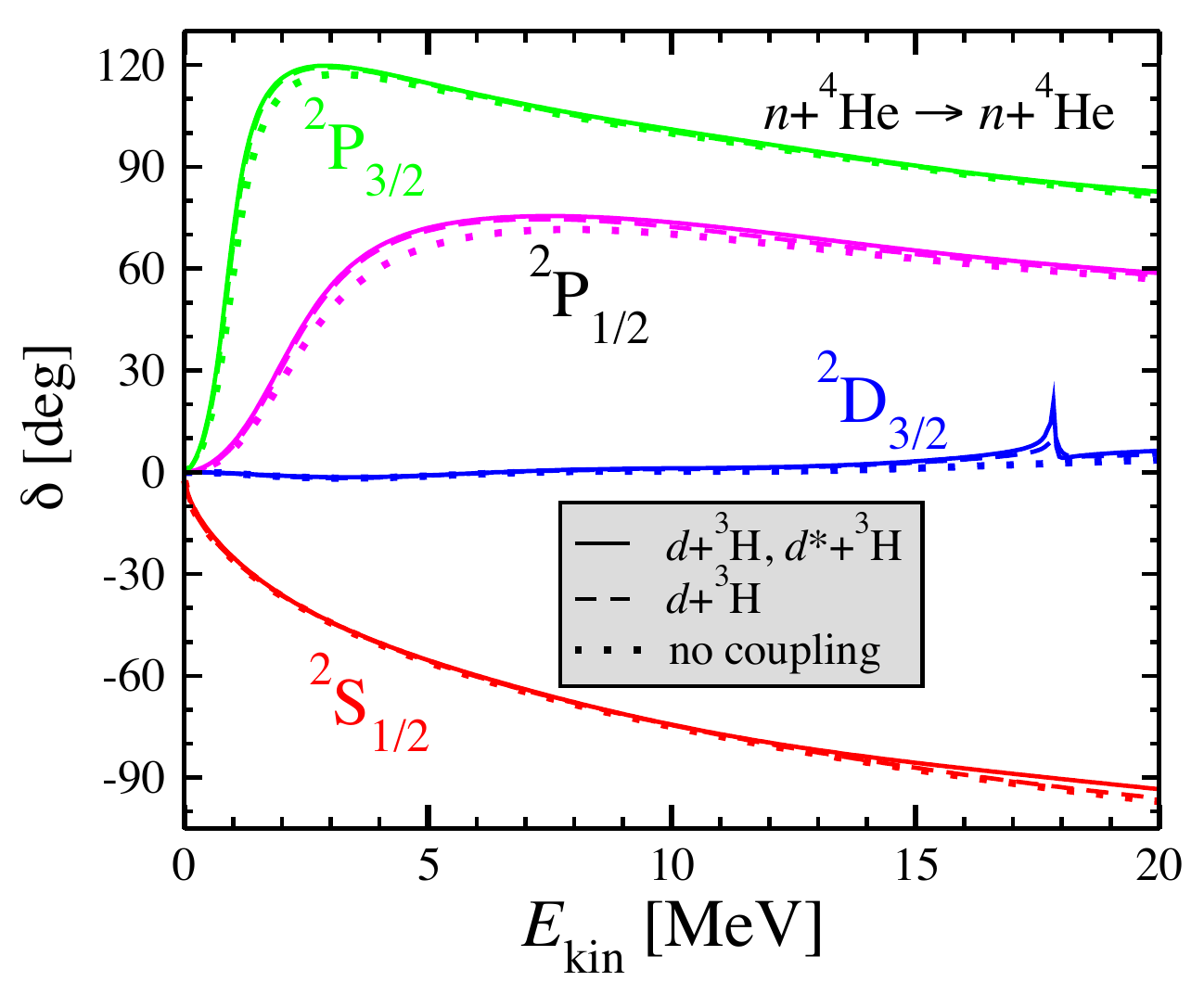}
\end{minipage}
\begin{minipage}{0.45\columnwidth}
\centering
\caption{Calculated elastic $n$-$^4$He phase shifts. The dashed (dotted) lines are obtained 
with (without) coupling of the $n$($p$)-$^4$He and $d$-$^3$H($^3$He) channels and all nuclei in their g.s. The full lines represent calculations that further couple channels with one $^3S_1{-}^3D_1$ deuteron pseudostate. The SRG-N$^3$LO $NN$ potential with $\Lambda{=}1.5$ fm$^{-1}$ and the HO space with $N_{\rm max}{=}12$ ($N_{\rm max}{=}13$ for the negative parity) and $\hbar\Omega{=}14$ MeV were used. Figure adapted from Ref. \cite{Navratil2012}.}
\label{fig:nHe4_phase_shifts}
\end{minipage}
\end{figure}
Also in Sec.~\ref{sec:s-shell}, we discussed the nucleon-$^4$He scattering phase shifts calculated by considering only the $n(p)$-$^4$He binary cluster channels. Here, we extend those calculations by including the coupling to the 
$d$-$^3$H ($d$-$^3$He) channels. The impact of this coupling can be judged (in the $n$-$^4$He case) from Fig.~\ref{fig:nHe4_phase_shifts}. 
Besides a slight shift of the $P$-wave resonances to lower energies,  
the most striking feature is the appearance of a resonance in the $^2D_{3/2}$ partial wave, just above the $d$-$^3$H ($d$-$^3$He) threshold. The further inclusion of distortions of the deuteron via an $^2$H $^3S_1$-$^3D_1$ pseudostate $(d^*)$, enhances this resonance. By investigating also the $d$-$^3$H($^3$He) scattering, we find a resonance in the $^4S_{3/2}$ channel, i.e., an $S$-wave between the $d$ and $^3$H($^3$He) with their spins aligned, at the same energy where we observe a resonance in the $^2D_{3/2}$ $n(p)$-$^4$He phase shift. This resonance is further enhanced by distortions of the deuteron. On the contrary, when the spins of the $d$ and $^3$H($^3$He) are opposite, i.e., in the $^2S_{1/2}$ channel, the Pauli blocking causes a repulsion between the two nuclei.

The $^3$H($d$,$n$)$^4$He and $^3$He($d$,$p$)$^4$He fusion (or more accurately transfer) reactions cross sections are then strongly enhanced near the resonance energy (experimentally at 50 keV and 200 keV, respectively). The fusion at the resonance proceeds from the $^4S_{3/2}$  $d$-$^3$H($^3$He) channel to the $^2D_{3/2}$ $n(p)$-$^4$He channel with a realease of large amount of energy due to the dramatic difference in the threshold energies of the two binary-cluster systems (17.6 MeV and 18.35 MeV, respectively).  
\begin{figure}[b]
\begin{minipage}{18pc}
\includegraphics[width=18pc]{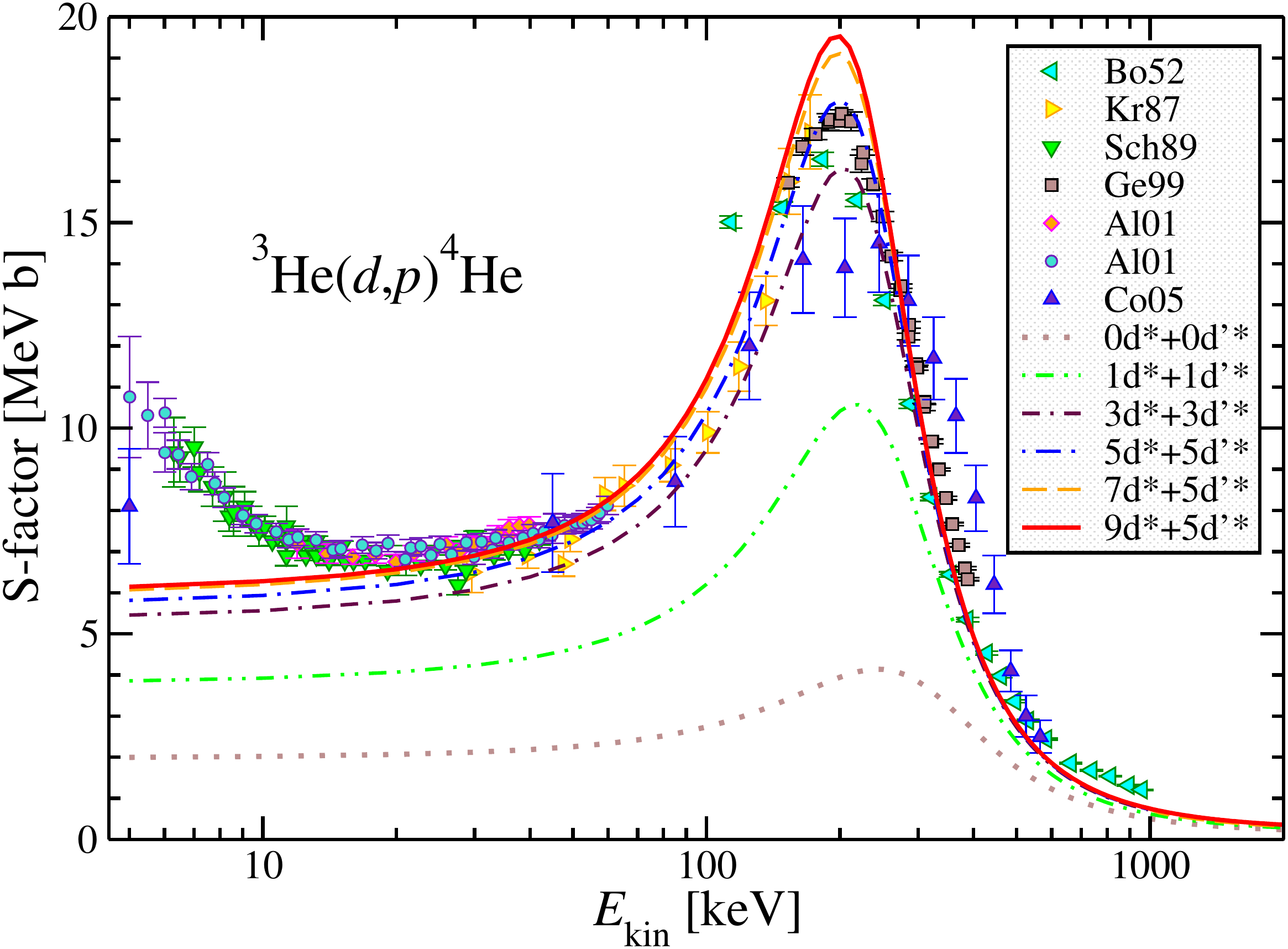}
\caption{\label{fig:d3He}Calculated S-factor of the $^3$He$(d,p)^4$He reaction compared to experimental data. Convergence with the number of $^2$H pseudostates in the $^3S_1$-$^3D_1$ ($d^*$) and $^3D_2$ ($d^{\prime *}$) channels. Parameters as in Fig.~\ref{fig:nHe4_phase_shifts}. Figure from Ref. \cite{Navratil2012}.}
\end{minipage}\hspace{1pc}%
\begin{minipage}{18pc}
\includegraphics[width=18pc]{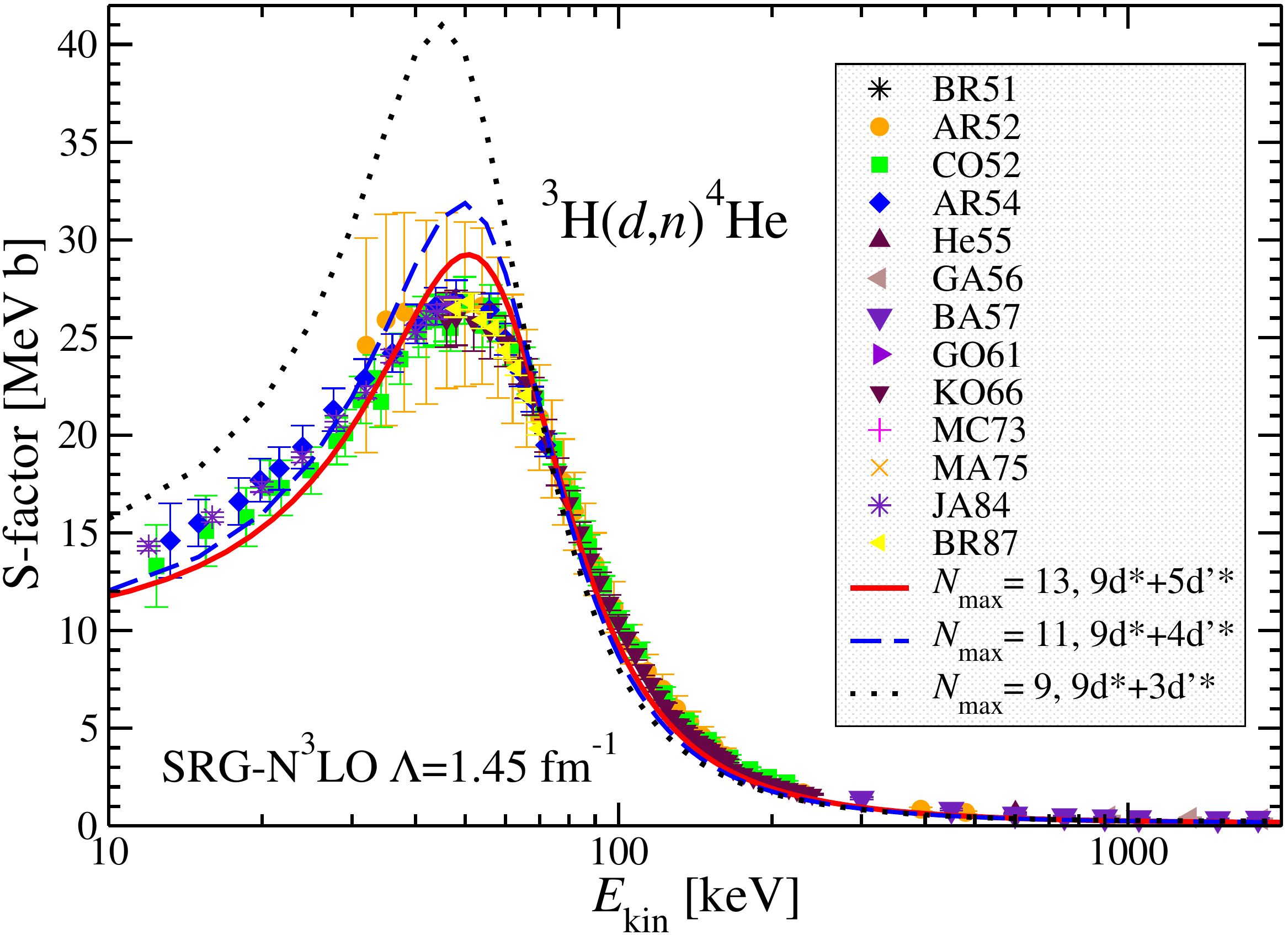}
\caption{\label{fig:d3H}Calculated $^3$H$(d,n)^4$He S-factor compared to experimental data. Convergence with $N_{\rm max}$ obtained for the SRG-N$^3$LO $NN$ potential with $\Lambda=1.45$ fm$^{-1}$ at $\hbar\Omega=14$ MeV. Figure from Ref. \cite{Navratil2012}.}
\end{minipage} 
\end{figure}
Our calculated S-factors (proportional to the cross sections) are shown in Figs.~\ref{fig:d3He} and \ref{fig:d3H}. In paticular, Fig.~\ref{fig:d3He} presents results obtained for the $^3$He$(d,p)^4$He S-factor. The deuteron deformation and its virtual breakup, approximated by means of $d$ pseudostates, play a crucial role. The S-factor increases dramatically with the number of pseudostates until convergence is reached for $9d^*+5d^{\prime*}$. The dependence upon the HO basis size is illustrated by the $^3$H$(d,n)^4$He results of Fig.~\ref{fig:d3H}. The convergence is satisfactory and we expect that an $N_{\rm max}=15$ calculation, which is currently out of reach, would not yield significantly different results. The experimental position of the $^3$He$(d,p)^4$He S-factor is reproduced within few tens of keV. Correspondingly, we find an overall fair agreement with experiment for this reaction, if we exclude the region at very low energy, where the accelerator data are enhanced by laboratory electron screening.
The $^3$H$(d,n)^4$He S-factor is not described as well with  $\Lambda=1.5$ fm$^{-1}$, see Fig. 2(a) in Ref.~\cite{Navratil2012}. Due to the very low activation energy of this reaction, the S-factor (particularly peak position and height) is extremely sensitive to higher-order effects in the nuclear interaction, such as three-nucleon force (not included in this calculation) and missing isospin-breaking effects in the integration kernels (which are obtained in the isospin formalism).  To compensate for these missing higher-order effects in the interaction and reproduce the position of the  $^3$H$(d,n)^4$He S-factor, we performed additional calculations using lower $\Lambda$ values. This led to the theoretical S-factor of Fig.~\ref{fig:d3H} (obtained for $\Lambda=1.45$ fm$^{-1}$), that is in overall better agreement with data, although it presents a slightly narrower and somewhat overestimated peak. This calculation would suggest that some electron-screening enhancement could also be present in the $^3$H$(d,n)^4$He measured S factor below $\approx$10 keV c.m.\ energy. However, these results cannot be considered conclusive until more accurate calculations using a complete nuclear interaction are performed. 

Overall, however, the results discussed above and published in Ref.~\cite{Navratil2012} are promising and pave the way for microscopic investigations of polarization and electron screening effects, of the $^3$H($d$,$\gamma n$)$^4$He bremsstrahlung and other reactions relevant to fusion research that are less well understood or hard to measure. Due to the rapid progress in the formulation and implementation of our formalism, we are in a position to perform significantly improved calculations for these reactions within the NCSMC formalism outlined in the previous sections of this paper. By coupling the NCSM/RGM binary-cluster basis with the NCSM eigenstates for $^5$He ($^5$Li) will take into account more five-nucleon correlations and polarization of not just the deuteron but also of the $^3$H($^3$He) and the $^4$He at and near the resonance energy and further improve the converence of the calculations compared to those discussed in this section. Further, since we developed the capability to include the 3N interaction in the NCSMC (for both the single-nucleon and the deuteron projectiles) we are in a position to calculate the $^3$H($d$,$n$)$^4$He and $^3$He($d$,$p$)$^4$He fusion with a realistic chiral EFT NN+3N Hamiltonian.

\begin{figure}
\begin{minipage}{0.55\columnwidth}
\includegraphics*[width=1.1\columnwidth]{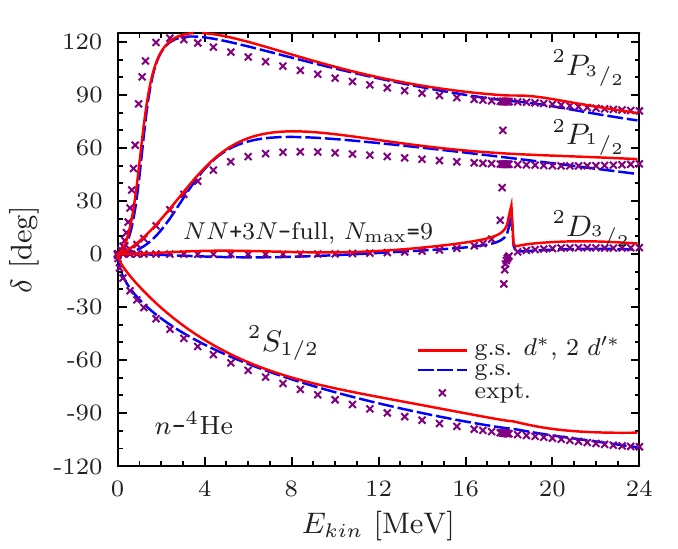}
\end{minipage}
\begin{minipage}{0.45\columnwidth}
\centering
\caption{Preliminary results (lines) for the $n$-$^4$He scattering phase shifts from zero to 24 MeV in the center-of-mass energy compared to an R-matrix anaysis of experiment (crosses). All calculations were performed at $N_{\rm max}{=}9$ within the NCSMC including $n$+$^4$He(g.s.) and $d$+$^3$H(g.s.) continuous basis states with up to two $^2$H pseudostates in the $^3S_1$-$^3D_1$ ($d^*$) and $^3D_2$ ($d^{\prime *}$) channels, as well as square-integrable discrete eigenstates of the compound $^5$He nucleus. The chiral NN+3N interaction SRG evolved with $\Lambda{=}2$ fm$^{-1}$ and $\hbar\Omega{=}20$ MeV were used.}
\label{fig:nHe4_phase_shifts_NN3N}
\end{minipage}
\end{figure}
Work in this direction is under way with the first preliminary results published in Ref.~\cite{Quaglioni:2015via}. In Fig.~\ref{fig:nHe4_phase_shifts_NN3N}, we present the $n$-$^4$He scattering phase shifts obtained within the NCSMC wih the chiral NN+3N interaction with $\Lambda_{\rm cut,3N}{=}500$~MeV. The phase shifts are comparable to those shown in Fig.~\ref{fig:nHe4_phase_shifts}. The difference is, however, that now we are not fine-tuning the SRG parameter $\Lambda$. We selected a standard value, $\Lambda{=}2$~fm$^{-1}$, and checked that phase shift and resonance position results are less sensitive to variations of the $\Lambda$ compared to the NN-only case of Fig.~\ref{fig:nHe4_phase_shifts}. This was done in particular by repeating calculations using $\Lambda{=}1.7$ fm$^{-1}$. More on the SRG $\Lambda$ sensitivity of nucleon-$^4$He phase shifts, see Fig.~\ref{fig:srg-convergence} and the discussion in Ref. \cite{Hupin2013}. Despite the fairly small size of the HO basis, the calculation is in close agreement with experiment. In particular, besides a slight shift of the $P$-wave resonances to lower energies, the inclusion of $d$+$^3$H channels leads to the appearance of a resonance in the $^2D_{3/2}$ partial wave, just above the $d$+$^3$H threshold. This is the exit channel of the deuterium-tritium fusion. What is in particular encouraging is the fact that the resonance appears close to the experimental resonance energy. It is a consequnce of the chiral NN+3N interaction rather than of an SRG fine-tuning.

It should be noted that the ($d$,$p$), ($d$,$n$) transfer reaction formalism can be readily generalized for $A{>}5$ masses. We took the first steps in that direction and investigated the $^7$Li($d$,$p$)$^8$Li reaction witin the NCSM/RGM~\cite{Raimondi:2016njp}.

%% file: SE_conclusions.tex
\section{Conclusions and outlook}
\label{sec:concl}
{\it Ab initio} theory of light and medium mass nuclei is a rapidly
evolving field with many exciting advances in the past few
years. Several new methods have been introduced capable to describe
bound-state properties of nuclei as heavy as nickel. Similarly, there
has been a significant progress in calculations of unbound states,
nuclear scattering and reactions, mostly in light nuclei so far.

In this contribution, we reviewed the recently introduced unified
approach to nuclear bound and continuum states based on the coupling
of a square-integrable basis ($A$-nucleon NCSM eigenstates), suitable for
the description of many-nucleon correlations, and a continuous basis
(NCSM/RGM cluster states) suitable for a description of long-range
correlations, cluster correlations and scattering. 
This {\it ab initio} method, the No-Core Shell Model with Continuum, 
is capable of describing efficiently: $i)$ short- and medium-range
nucleon-nucleon correlations thanks to the large HO basis expansions
used to obtain the NCSM eigenstates, and $ii)$ long-range cluster
correlations thanks to the NCSM/RGM cluster-basis expansion. 

We demonstrated the potential of the NCSMC in calculations of nucleon
scattering on $^4$He, in highlighting the connection between the
deuteron scattering on $^4$He and the structure of $^6$Li, and in studying
the continuum and 3N effects in the structure of $^9$Be. We further
presented the extension of the formalism to three-body cluster systems
and discussed in detail calculations of bound and resonance states of
the Borromean halo nucleus $^6$He. 

We introduced in this paper the formalism for electromagnetic transition
calculations within the NCSMC and reviewed our first application to
reactions important for astrophysics, the $^7$Be($p,\gamma$)$^8$B
radiative capture. We also discussed our past and ongoing calculations
of the $^3$H($d,n$)$^4$He transfer reaction relevant to future energy
generation on Earth. 

The NCSMC is a versatile method with many applications. We are
developing the formalism needed to study transfer ($d,p$),
($d,n$) and ($p,t$) reactions frequently used in radioactive beam
experiments. We will extend the calculations throughout the
$p$-shell and light $sd$-shell nuclei and investigate 
($p,\gamma$), ($\alpha,\gamma$) and ($n,\gamma$) capture reactions
relevant to nuclear astrophysics. We will investigate the bremsstrahlung process
$^3$H($d,n \gamma$)$^4$He relevant to the fusion research. We will calculate 
weak decays relevant to testing of fundamental symmetries such
as the $^6$He beta decay that is being measured with high precision at
present.

Our long-term goals are then studies of systems with three-body
clusters, in particular the Borromean exotic nucleus $^{11}$Li, and in
general reactions with three-body final states such as
$^3$He($^3$He,2$p$)$^4$He. Ultimate goal for the forseeable future is
to study alpha clustering, e.g., in $^{12}$C and $^{16}$O, and
reactions involving $^4$He, e.g., $^8$Be($\alpha,\gamma$)$^{12}$C,
$^{12}$C($\alpha,\gamma$)$^{16}$O important for stellar burning, the
$^{11}$B($p,\alpha$)$^8$Be aneutron reaction explored as a candidate
for the future fusion energy generation as well as the
$^{13}$C($\alpha,n$)$^{16}$O relevant to the i- and s-processes. The
first two of these reactions were identified as one of the drivers of exascale computing \cite{exa09}.

\ack

This work was supported in part by Natural Sciences and Engineering Research Council of Canada (NSERC) under Grant No. 401945-2011.
TRIUMF receives funding via a contribution through the National Research Council Canada. 
Prepared in part by LLNL under Contract DE-AC52-07NA27344. This material is based upon work supported 
by the U.S.\ Department of Energy, Office of Science, Office of Nuclear Physics, under Work Proposal Number SCW1158.
Computing support for this work came from the LLNL institutional Computing Grand Challenge Program and from 
an INCITE Award on the Titan supercomputer of the Oak Ridge Leadership Computing Facility (OLCF) at ORNL. 
Further, computing support for this work came in part from the National Energy Research Scientific Computing 
Center (edison) supported by the Office of Science of the U.S. Department of Energy under Contract No. DE-AC02-05CH11231, 
from the LOEWE-CSC Frankfurt, and from the computing center of the TU Darmstadt (lichtenberg).